\documentclass[onecolumn,aps,prd,superscriptaddress,nofootinbib,amsmath,amssymb,floats,floatfix,showkeys,notitlepage,longbibliography]{revtex4-1}

\usepackage[dvips]{graphicx,color}
\usepackage{subfig}
\usepackage{subcaption}
\usepackage{times}
\usepackage{mathrsfs}  
\usepackage{xcolor}
\usepackage[%
  colorlinks=true,
  urlcolor=blue,
  linkcolor=red,
  citecolor=blue
]{hyperref}
\usepackage{orcidlink}

\begin{document}
\title{\Large Evolution of Realistic Neutron star in the framework of $f(Q)$ gravity}

\author{Samprity Das \orcidlink{0000-0002-2661-9659}}
\email{samprity.das@s.amity.edu}
\affiliation{ Department of Mathematics, Shibpur Dinobundhoo Institution (College),\\
Shibpur, Howrah 711102; \\
Department of Mathematics, Amity University, Kolkata, New Town, Rajarhat, Kolkata 700135,
India.}

\author{Surajit Chattopadhyay \orcidlink{0000-0002-5175-2873}}
\email{schattopadhyay1@kol.amity.edu; surajitchatto@outlook.com}
\affiliation{ Department of Mathematics, Amity University, Major
Arterial Road, Action Area II, Rajarhat, New Town, Kolkata 700135,
India.(Communicating author)}

\date{\today}

\begin{abstract}
This work analyses and evaluates a few realistic compact objects in the presence of a gravitational interaction between two particles with a nonmetricity $Q$. In the $f(Q)$ gravity framework, we have selected the anisotropic equation of motion and have determined $f(Q)$ to be a linear function of nonmetricity $Q$. To evaluate the field equations in our work, we have opted to employ the Krori-Barua metric. We calculated the anisotropic factor for each of the four compact objects and found that the anisotropic component is positive and increases monotonically and interpreted that the nuclear force can oppose the gravitational attraction. At last, the relationship between mass and radius has been determined and illustrated visually. We have noted that the compactness of the pulsars LMC X-4, SMC X-4, Cen X-3, and Vela X-1 is inside the Buchdahl's limit for varying values of $a$. This has led to the interpretation that these pulsars are neutron stars in a modified gravity background of $f(Q)$. In addition, we calculated the model mass and, using thirty distinct choices of $a$, ran the Chi-Square test to see if there was a noticeable difference between the observed and model-generated masses. We have also looked at how the surface redshift has changed over time and whether the compact objects in our model that were previously described are compact.
\end{abstract}

\maketitle
\section{Introduction}

Over the past few decades, scientists have developed modified gravitational theories other than General Relativity(GR), such as $f(R)$ gravity \cite{new1, new2, bib31,bib33}, $f(T)$ gravity \cite{bib29,bib32,new3}, $f(G)$ gravity\cite{bib32}, $f(Q)$ gravity \cite{bib34,bib35}, $f(R,T)$ gravity \cite{bib28,bib30}, and so on to find a more precise and meaningful accelerated explanation for our expanding universe. Among these, we have chosen to study the compact star in the background of the $f(Q)$ gravity theory proposed by Jimnez et al. \cite{bib36,bib37}, where $Q$ is the nonmetricity, starts receiving a gravitational pull. By defining the $f(Q)$ Lagrangian as polynomial equations of the redshift $z$, the authors developed an intriguing set of constraints on $f(Q)$ gravity \cite{bib38}. This suggests that feasible $f(Q)$ models have coefficients comparable to the GR model, specifically the $\Lambda$CDM model. To determine whether this new formalism provides any workable alternatives to explain the universe's late-time acceleration, an innovative approach has been constructed at the background level. Authors \cite{bib39,bib40} have developed a nonlinear form of $f(Q)$ as a function of nonmetricity $Q$ and looked into several fascinating baseline cosmologies including accelerating solutions related to dark energy and inflation and how the universe would react to cosmic disturbances as well as its behaviour. The terms "GR's teleparallel equivalent (TEGR) \cite{bib41,bib42} and symmetric teleparallel GR (STGR)" \cite{bib43,bib44} refer to two different but equivalent representations of gravity that result from choosing either the torsion ($T$) or the nonmetricity ($Q$) as the geometric basis. Torsion, for example, is the field that characterizes gravity according to TEGR. Here curvature and non-metricity are both zero, and the affine connection is the Weitzenböck connection \cite{bib45,bib46}. Here, the affine connection, torsion invariants, and ultimately the field equations can be obtained from tetrads, which are the fundamental objects. Conversely, STEGR is predicated on non-metricity. In this instance, a flat spacetime can be considered, and the gravitational field is linked to non-metricity. In STEGR, geometry has a non-metric connection and a metric tensor but the total curvature and torsion are diminishing. $f(Q)$ gravity has been verified in numerous theoretical investigations including the bouncing model \cite{bib47}, wormhole solution \cite{bib48,bib49,bib50}, energy conditions \cite{bib51}, development of stellar objects without the existence of singularity \cite{bib53,bib54}, the Newtonian limit \cite{bib52} and so on.\\
Astrophysicists have recently developed an avid curiosity in examining stellar objects against the backdrop of modified gravity, such as black holes, gravastars, neutron stars, quark stars, and so on. These are resources with a restricted radius and a high density. Although the genesis of these formations is still being investigated, there is proof that they exist. The presence of (hyper) nuclear matter in compact stars could display some exceptional properties, such as superfluidity and superconductivity, a contained neutrino component in the early stages of formation, or extraordinarily high magnetic fields \cite{bib56,bib57}. The five main regions of a neutron star are the atmosphere, the outer and inner crust, and the outer and inner core. A graphical preview is available in Sedrakian et al. \cite{bib55}. The densest known type of stuff in the universe, its structure and attributes of which remain unknown theoretically, makes up neutron stars. The neutron star matter equation of state can be highly constrained and theoretical predictions of their composition can be ruled out by measurements of their masses or radii. A white dwarf can be less dense than a neutron star because of the neutrons' degeneration pressure, which permits a higher density than in the event of the electrons, which are the main component of white dwarfs. The density of a star is determined by the type of matter that makes up the star \cite{bib58,bib59}. In addition to considering aspects of particle physics and nuclear physics, the models that represent compact stars enable us to analyze or suggest some state equations as well as identify the magnitude levels of the density inside the star \cite{bib60,bib61}. Neutron star models in $f(R)$ gravity framework were demonstrated in \cite{Odi5,Odi6,Odi7,Odi8}.
\\
This paper presents a detailed analysis of the field equations in $f(Q)$-gravity theory with a vanishing complexity factor. In section II, we have developed $f(Q)$ as a linear function of nonmetricity $Q$. In Section III, the density, radial pressure, and transversal pressure of a neutron star have been determined using the Krori-Barua metric potential. We have obtained a boundary condition for the integrating constants $a$ and $b$ by applying the matching condition from Schwarzschild solution and the non-sigularity condition of neutron star density. Four realistic compact objects—LMC X-4, SMC X-4, Cen X-3, and Vela X-1—have been selected for a thorough investigation against the backdrop of our model. The evolution of density and pressure, validation of the energy conditions for the neutron star, validation of the equation of state parameters, and study of causality and stability are all covered in Section III. Using several values of $a$ and adhering to the boundary requirements, we examined the mass-radius relation of the aforementioned realistic compact objects in section IV. To see if there is a discrepancy between the estimated mass and the mass that is produced by the model, we additionally conducted hypothesis testing. We have also examined the evolution of the surface redshift and the compactness of the earlier-mentioned compact objects in our model. We have finally concluded the results.

\section{The $f(Q)$ gravity theory}

In this work we choose $f(Q)$ gravity background proposed by Jiménez et al. \cite{bib5}, where the Einstein-Hilbert action is represented as
\begin{equation}
S=\int \sqrt{-g}d^4x\left[\frac{1}{2}f(Q)+\mathcal{L}_m\right],
\label{E1}
\end{equation}
where $g$ is the determinant of the metric $g_{\mu\nu}$, $f(Q)$ denotes the function of $Q$ and $\mathcal{L}_{m}$ denotes the matter Lagrangian density. The nonmetricity tensor in $f(Q)$ gravity defined as
\begin{equation}
Q_{{xyz}}=\nabla \, _xg_{{yz}}=-L^w{}_{{xy}}g_{{wz}}-L^w{}_{{xz}}g_{{wy}}.
\label{E2}
\end{equation}
Two independent traces of nonmetricity tensor computed as
\begin{equation}
Q_a=Q_a{}^b{}_{ b},\bar{Q}_a=Q^b{}_{{ab}}.
\end{equation}
The deformation term in $f(Q)$ gravity is
\begin{equation}
L_{{yz}}^x=\frac{1}{2}Q_{{yz}}^x-Q_{({yz})}{}^x,
\end{equation}
and the nonmetricity scalar is defined by the nonmetricity conjugate $P^{{xbc}}$
\begin{equation}
Q=-g^{{yz}}\left(L_{{bz}}^xL_{{yx}}^b-L_{{xb}}^bL_{{yz}}^x\right)=-P^{{xbc}}Q_{{xbc}},
\label{E3}
\end{equation}
and the corresponding tensor is
\begin{equation}
P^x{}_{{yz}}=\frac{1}{4}\left[-Q_{{yz}}^x+2Q_{({yz})}^x-Q^xg_{{yz}}-\tilde{Q}^xg_{{yz}}-\delta _y^xQ_y\right].
\end{equation}
By varying the metric coefficient $g_{\mu\nu}$ in Eq. (\ref{E1}) the field equations of $f(Q)$ gravity obtained as
\begin{equation}
\frac{-2}{\sqrt{-g}}\nabla _a\left(\sqrt{-g}f_QP_{{yz}}^x\right)+f_Q\left(P_z^{{xb}}Q_{{yab}}-2P^{{xb}}{}_yQ_{{xbz}}\right)+\frac{1}{2}g_{{yz}}f=\kappa T_{{yz}},
\end{equation}
where $f_Q=\frac{ df}{dQ}$. 
The energy-momentum tensor for fluid description spacetime can be described as
\begin{equation}
T_{{yz}}=\frac{-2}{\sqrt{-g}}\frac{\delta \sqrt{-g}}{\delta \sqrt{g_{{yz}}}}\mathcal{L}_m.
\end{equation}
Varying Eq. (\ref{E1}) with respect to the connection the following relation can be obtained
\begin{equation}
\nabla _y\nabla _z\left(\sqrt{-g}f_QP^{{yz}}{}_x\right)=0.
\end{equation}
The affine connection established with the curvature less and the torsion less as the following
\begin{equation}
\Gamma^{a}_{yz}=(\frac{\delta x^{a}}{\delta \eta^{\lambda}})\delta_{y}\delta_{z}\eta^{\lambda}.
\end{equation}
The nonmetricity reduced as the following by a special choice of the coordinate system, for the coincide gauge
\begin{equation}
Q_{xyz}=\delta_{x}g_{yz}.
\end{equation}
The spherically symmetric metric for GR with spherically symmetric coordinates $(t,r,\theta,\phi)$ is of the form \cite{bib3}
\begin{equation}
    ds^2 = e^{\mu(r)} dt^2 - e^{\phi(r)} dr^2 - r^{2} (d \theta ^{2} + sin ^{2} \theta d r^{2}),
  \label{E023}
\end{equation}

From Eq. (\ref{E3}) and Eq. (\ref{E023}) the nonmetricity scalar $Q$ defined as
\begin{equation}
Q(r)=- \frac{2 e^{-\mu (r)}}{r}\left(\frac{1}{r}+ \phi '(r)\right),
\label{E001}
\end{equation}
here $'$ denotes the derivative with respect to radial co-ordinate $r$. From anisotropic fluid equation and $f(Q)$ field equation, we get the equations of motions as \cite{bib1}
\begin{equation}
\rho =-\frac{f}{2}+f_Q\left(Q+\frac{1}{r^2}+\frac{e^{-\mu }}{r}(\phi '+\mu ')\right),
\label{E002}
\end{equation}
\begin{equation}
p_r=\frac{f}{2}-f_Q\left(Q+\frac{1}{r^2}\right),
\label{E003}
\end{equation}
\begin{equation}
p_t=\frac{f}{2}-f_Q\left(\frac{Q}{2}-e^{-\mu }\left(\frac{\phi {''}}{2}+\left(\frac{\phi '}{4}+\frac{1}{2 r}\right)(\phi '-\mu ')\right)\right),
\label{E004}
\end{equation}
where $\rho$ is matter density and $p_r$ and $p_t$ are radial pressure and tangential pressure respectively.
\begin{equation}
\frac{{cot\theta}}{2}Q'f_{{QQ}}=0.
\label{E005}
\end{equation}
Wang et al.'s \cite{bib2} work established that the Schwarzschild (anti-) de Sitter solution is the exact solution obtained from the exterior solution of field equations if and only if $f_{QQ}=0$. Consequently, in order to deduce the functional form of $f (Q)$ and find the solution for self-gravitating compact star models, one must assume that $f_{QQ}=0$.
\begin{equation}
f_{{QQ}}=0.
\label{E006}
\end{equation}
Hence reformulated $f$ attain the form
\begin{equation}
f=a  Q+ b,
\label{E007}
\end{equation}
where $a$ and $b$ are integrating constant, the ratio of which ($\frac{b}{a}$) has a corresponds to the cosmological constant $\Lambda$.

\section{The Krori-Barua Spacetime}

Inspired by the Krori-Barua metric potential we assume the following metric potential in Eq. (\ref{E023}) \cite{bib6} as
\begin{equation}
\mu (r)=A r^2,
\label{E008}
\end{equation}
\begin{equation}
\phi (r)=B r^2+C,
\label{E009}
\end{equation}
where $A$, $B$, and $C$ are constants with dimensions of $A$ and $B$ are $(Km^{-2})$ and $(km^{-2})$ respectively and $C$ is a dimensionless quantity. The gravitational potentials and their derivatives are guaranteed to be limited at the center by the KB-spacetime in Eq. (\ref{E008}) and Eq. (\ref{E009}).
Reconstructed density, radial pressure and tangential pressure from Eq. (\ref{E002}), Eq. (\ref{E003}), and Eq. (\ref{E004}) are obtained using Eq. (\ref{E001}), Eq. (\ref{E007}), Eq. (\ref{E008}), and Eq. (\ref{E009})
\begin{equation}
\rho =\frac{2 a e^{-A r^2} \left(-1+e^{A r^2}+2 A r^2\right)}{r^2}-b
\label{E010}
\end{equation}
\begin{equation}
p_r=\frac{b}{2}+\frac{a e^{-A r^2} \left(1-e^{A r^2}+2 B r^2\right)}{r^2}
\label{E011}
\end{equation}
\begin{equation}
p_t=\frac{1}{2} e^{-A r^2} \left(b e^{A r^2}-2 a \left(A+A B r^2-B \left(2+B r^2\right)\right)\right)
\label{E012}
\end{equation}
From the non-singularity condition of density we derive
\begin{equation}
\frac{a e^{-A r^2} \left(-1+e^{A r^2}+2 A r^2\right)}{r^2}>\frac{b}{2}
\label{E013}
\end{equation}
It is important to note that in order for compact objects to behave appropriately, the following requirements must be achieved:
\begin{itemize}
\item Density, radial pressure and tangential pressure must have finite and maximum value near the centre.
\item $d\rho/dr$, $dp_{r}/dr$, $dp_{t}/dr$ should have negative graphical representation \cite{bib11}.
\item Null energy conditions(NEC)
 $NEC : \rho + p_{r} \geq 0 $ , $\rho + p_{t} \geq 0$, 
 Strong energy conditions(SEC) 
 $SEC : \rho + p_{r} \geq 0 $ , $\rho + p_{t} \geq 0$ , $\rho + p_{r} + 2 p_{t} \geq 0$ must be satisfied \cite{bib10}.
\item Corresponding equation of State for radial pressure($\omega_r = \frac{p_r}{\rho}$) and tangential pressure($\omega_t=\frac{p_t}{\rho}$) must be positive and lies within (0,1) for well behaved compact star model .
\item The radial and transverse square speed of sound must be greater zero and lies within $0< v^2 _{sr}, v^2 _{st} <1$, where $v^2 _{sr}= \frac{dp_r}{d\rho}$ and $v^2 _{st}= \frac{dp_t}{d\rho}$ \cite{bib12}.
\item Compactness of the objects i.e. $u(r)= \frac{m}{r}$, must be less than Buchdahl’s limit, which is  $u(r) < 4/9$ \cite{bib8}. Here $m$ represents mass of the object and $r$ radius of the object.
\item The surface redshift $z_{s}=(1-2 u) ^{-1/2} -1$, lies within permissible limit i.e. $z(r) \leq 5.211$ \cite{bib9}.
\item For the stability of the model the adiabatic index $\Gamma$ must satisfy the condition $\Gamma(=\frac{p_r +\rho}{p_r} v^2 _{sr}) > \frac{4}{3}$ \cite{bib13}.
\end{itemize}

\subsection{Matching Condition from Schwarzschild solution}

When selecting from the many matching restrictions for analyzing compact objects in the GR backdrop, the Schwarzschild approach is seen to be the best option. According to the Jebsen-Birkhoff theorem, for spherically symmetric spacetime, the solution of the Einstein field equations (EFE) needs to be asymptotically flat and static.
The unknown parameters can be found by matching the interior spacetime and the exterior Schwarzschild line element\cite{bib7}

\begin{equation}
 ds^2 = F(r) dt^2 - F(r)^{-1} dr^2 - r^{2} (d \theta ^{2} + sin ^{2} \theta d r^{2}),
 \label{E014}   
\end{equation}
for $F(r)=(1-\frac{2m}{r})$,  with the interior line element in Eq.(\ref{E023}). 

The continuity of the first and second fundamental forms over the surface $\sum$, is the process of connecting both space-time at the boundary. According to the first fundamental form, intrinsic geometry of the interface, which is represented by the metric tensor $g_{ij}$ generated by $\mathcal{M^-}$, the external one and $\mathcal{M}^+$, the internal one, meets \cite{revi1}
\begin{equation}
    \begin{array}{cc}
         g_{rr}^- |_ { r = R}=g_{rr}^+ |_{r=R}, \\
         g_{tt}^- |_{r=R}=g_{tt}^+ |_{r=R},
    \end{array}
    \label{E051}
\end{equation}
 
The metric corresponds to the interior line element in Eq. (\ref{E023}) provides the boundary condition 
 \begin{equation}
     \left[{ds}^2\right]{}_{\sum }=0.
     \label{E024}
 \end{equation}
 Eq. (\ref{E024}) provides the boundary condition connecting from  Eq. (\ref{E051})
 \begin{equation}
     \begin{array}{cc}
        e^{\beta -} |_ { r = R}=e^{\beta +}|_{r=R}, \\
         e^{\alpha -}|_{r=R}=e^{\alpha +}|_{r=R},
     \end{array}
     \label{E025}
 \end{equation}
 and $\frac{\partial g_{tt}}{\partial r}$ across the boundary surface $r=R$ between the exterior and the interior region gives
 \begin{equation}
    \left(\frac{\partial e^{\alpha -}}{\partial r}\right){}_{r=R}=\left(\frac{\partial e^{\alpha +}}{\partial r}\right){}_{r=R}.
    \label{E026}
 \end{equation}
 Here $(+)$ denotes the exterior geometry and $(-)$ denotes the interior geometry. Using metric Eq. (\ref{E014}) and boundary conditions from Eq. (\ref{E025}) and Eq. (\ref{E026}) in Eq. (\ref{E004}) we get explicit expressions of $A$, $B$ and $C$ of the form 
 \begin{equation}
 \begin{array}{cc}
     e^{{BR}^2+C}=1-\frac{2M}{R} , \\
     e^{{AR}^2}=\left(1-\frac{2M}{R}\right)^{-1} , \\
     B R e^{{BR}^2+C}=\frac{M}{R^2},
 \end{array}
 \label{E028}
 \end{equation}
 where $M$= total mass for radius $R$.
 
 Solving the system of equations in Eq. (\ref{E028}) we get
 \begin{equation}
     \begin{array}{cc}
          A=\frac{1}{r^2}{Log}\left[\frac{R}{R-2 M}\right], \\
         B=\frac{M}{R^2(R-2 M)}, \\
         C={Log}\left[1-\frac{2M}{R}\right]-\frac{M}{R}\left(1-\frac{2M}{R}\right)^{-1}.
         \label{E029}
     \end{array}
 \end{equation}
 Here $A$ and $B$ has dimension of per square distance and $C$ is a dimensionless quantity. 
 The condition between integrating constants $a$ and $b$ has been evaluated for the estimated mass $M(M_{\odot})$ and the chosen estimated radius $r(km)$, for the specific compact objects with particular values of $A$, $B$, and $C$ \cite{bib4}.
 
\begin{table}[!h]
\centering
\begin{tabular}{||c c c c c c c c||}
\hline
Sl. No. & Star Model & $A$ & $B$ & $C$ & Estimated Radius & Estimated Mass & Derived Condition \\[0.5ex]
 &  & $(\mathrm{km}^{-2})$ & $(\mathrm{km}^{-2})$ &  & $R~(\mathrm{km})$ & $M(M_{\odot})$ &  \\
\hline\hline

1 & LMC X-4 
& 0.00674276 
& 0.00429139 
& $-0.760325$ 
& $8.301 \pm 0.2$ 
& $1.04 \pm 0.09$ 
& $\dfrac{b}{a} < 0.0272426$ \\
\hline

2 & SMC X-4 
& 0.00728223 
& 0.00490204 
& $-0.950209$ 
& $8.831 \pm 0.09$ 
& $1.29 \pm 0.05$ 
& $\dfrac{b}{a} < 0.0273703$ \\
\hline

3 & Cen X-3 
& 0.00779851 
& 0.00551327 
& $-1.12133$ 
& $9.178 \pm 0.13$ 
& $1.49 \pm 0.08$ 
& $\dfrac{b}{a} < 0.0272107$ \\
\hline

4 & Vela X-1 
& 0.00871723 
& 0.00666462 
& $-1.4058$ 
& $9.56 \pm 0.08$ 
& $1.77 \pm 0.08$ 
& $\dfrac{b}{a} < 0.0274588$ \\
[1ex]

\hline
\end{tabular}

\caption{\label{Table:1} Derived condition on $a$ and $b$ from the non-singularity condition of density.}
\end{table}
Values of $A$, $B$, $C$, the estimated mass $M$ for the estimated radius $R$ of the specific compact objects has been collected from observational data and the condition on the parameter $\frac{b}{a}$ is derived from the non-singularity condition of density, which we have elaborately explained in the next subsection.

\subsubsection{Density, Pressure and Anisotropic factor}

In this section we have analyzed physical attributes such as density, radial pressure, and tangential pressure within the compact stars in the background of $f(Q)$ gravity theory. By carrying out some analytical calculations and creating profiles of the various model parameters of the obtained solutions, we hope to learn more specifics about the various properties of the stellar configuration in this section. Lastly, we compare our findings with both the $f(Q)$ gravity model and the observational constraints.

\begin{figure}[!ht]
    \centering

    \begin{minipage}{0.24\textwidth}
        \centering
        \includegraphics[width=\linewidth]{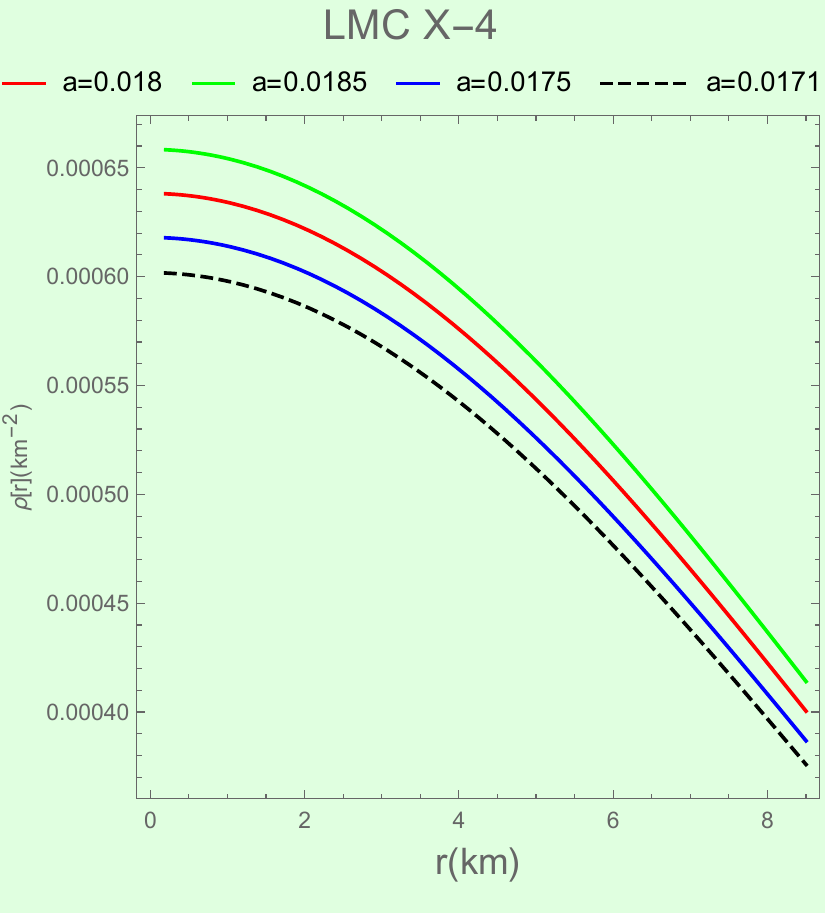}
        \small (a) LMC X-4
    \end{minipage}
    \hfill
    \begin{minipage}{0.24\textwidth}
        \centering
        \includegraphics[width=\linewidth]{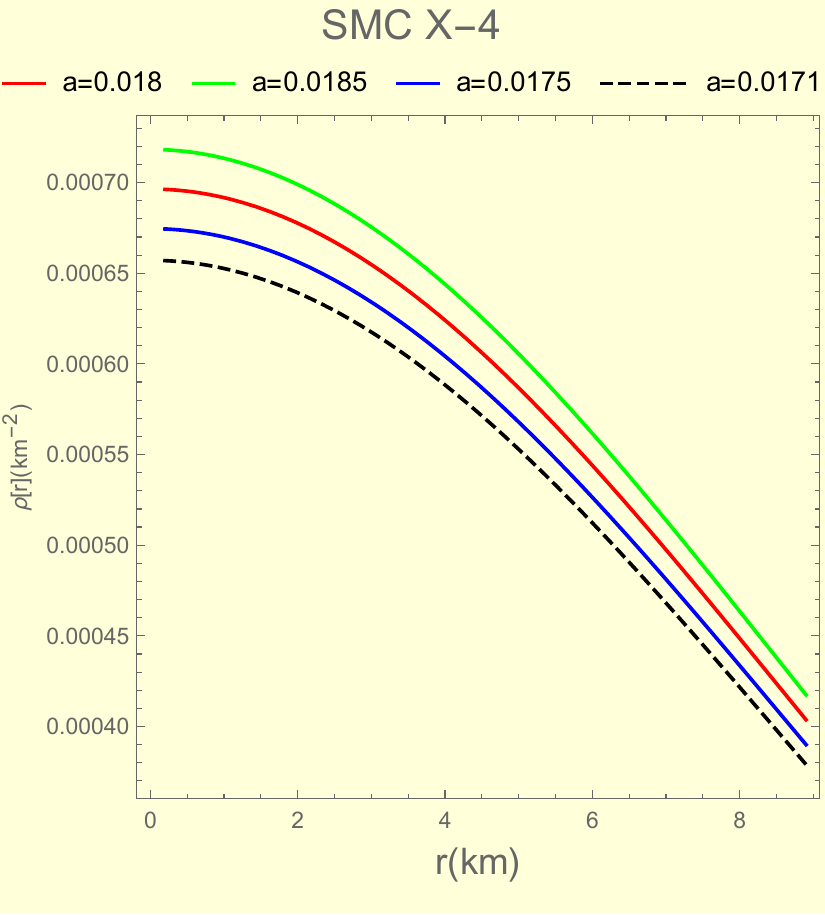}
        \small (b) SMC X-4
    \end{minipage}
    \hfill
    \begin{minipage}{0.24\textwidth}
        \centering
        \includegraphics[width=\linewidth]{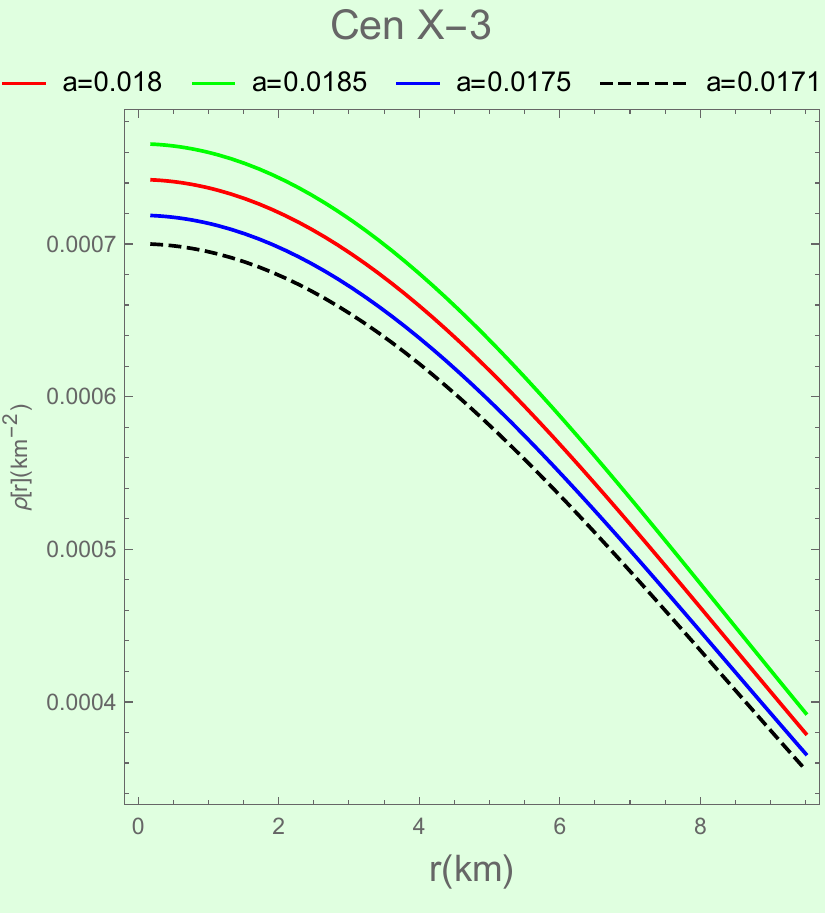}
        \small (c) Cen X-3
    \end{minipage}
    \hfill
    \begin{minipage}{0.24\textwidth}
        \centering
        \includegraphics[width=\linewidth]{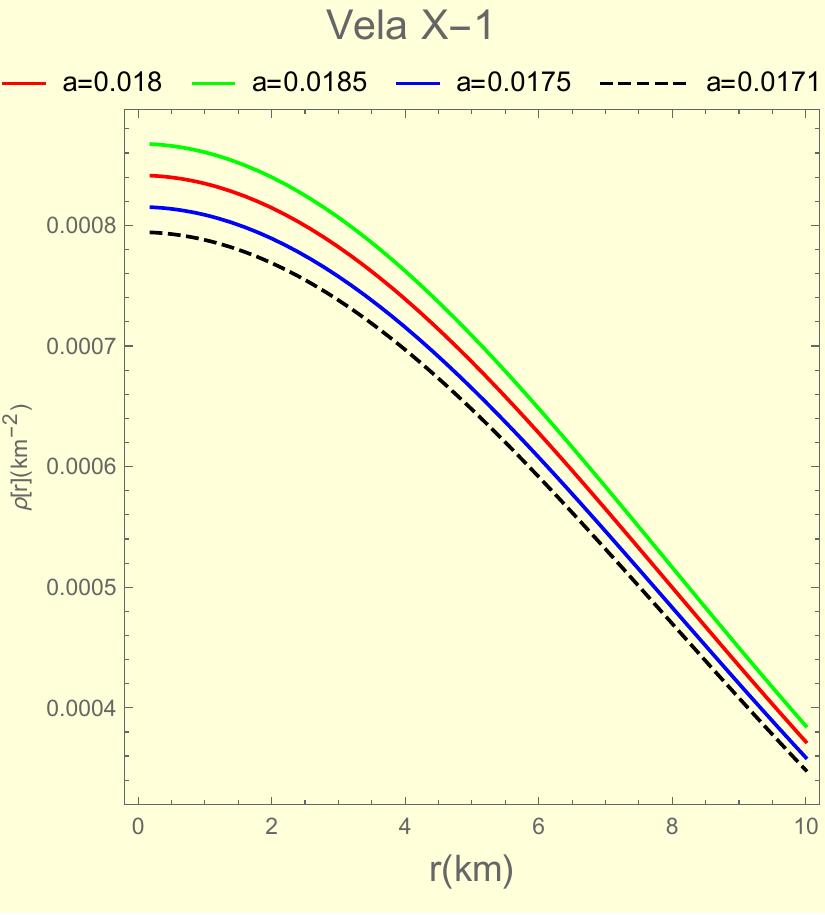}
        \small (d) Vela X-1
    \end{minipage}

    \caption{Density evolution with respect to radial coordinate $r$ for (a) LMC X-4, (b) SMC X-4, (c) Cen X-3, and (d) Vela X-1.}
    \label{fig:density}
\end{figure}

At this juncture let us have some discussion on the choice of $a$ and $b$ as presented in Table \ref{Table:1}. If we consider the requirement of $\rho >0$ we must have $\frac{a e^{-A r^2} \left(-1+e^{A r^2}+2 A r^2\right)}{r^2}>\frac{b}{2}$ and consequently $\frac{2 e^{-A R^2} \left(-1+e^{A R^2}+2 A R^2\right)}{R^2}>\frac{b}{a}$, where $R$ is the radius of the aforementioned compact objects. Here the values of $A$ and $R$ are known from the observational data mentioned in Table 1. From here, we can derive the condition on the parameter $\frac{b}{a}$. Further more if we consider the expression $\rho=2 a e^{-A r^2} \left(\frac{-1+e^{A r^2}}{r^2}+2 A \right)-b$, taking limit $r \rightarrow 0$, $\rho$ comes out to be $(4Aa-b)$, which is finite. Therefore for $a>0$, we get non-singular and finite value near the core and throughout the interior of the star. In view of this to depict the evolution of density, radial pressure, and tangential pressure pictorially we have chosen $a$= 0.018,0.0185,0.0175, 0.0171 and observed the evolution of the density and pressure within the compact objects. Therefore $\frac{b}{a} < 4A$ for $r \rightarrow 0$ i.e. near the core of the star and this condition is satisfied in all the cases in Table \ref{Table:1}. Graphical representation density of Eq. (\ref{E010}) is depicted in Fig \ref{fig:density}.

\begin{figure}[!ht]
    \centering

    \begin{minipage}{0.24\textwidth}
        \centering
        \includegraphics[width=\linewidth]{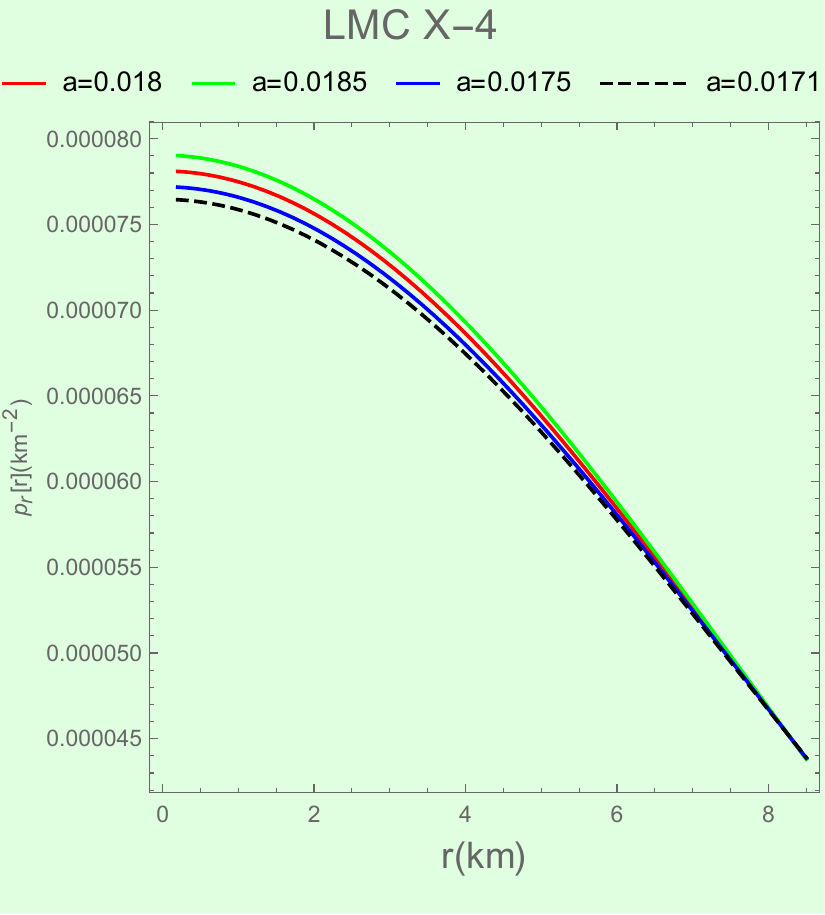}
        \small (a) LMC X-4
    \end{minipage}
    \hfill
    \begin{minipage}{0.24\textwidth}
        \centering
        \includegraphics[width=\linewidth]{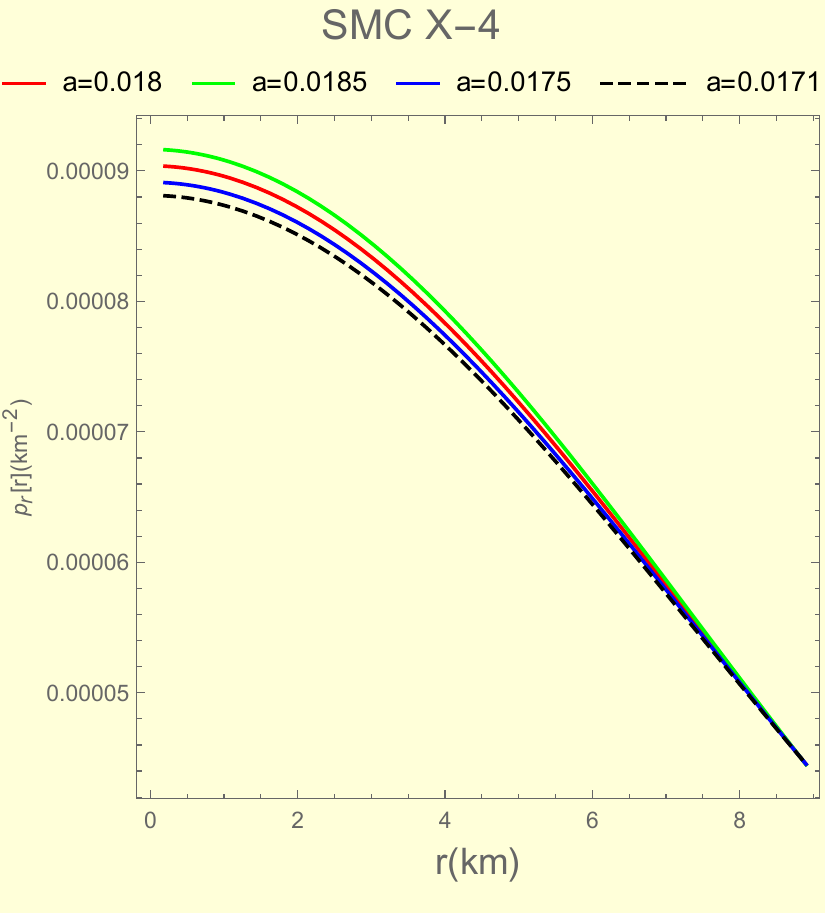}
        \small (b) SMC X-4
    \end{minipage}
    \hfill
    \begin{minipage}{0.24\textwidth}
        \centering
        \includegraphics[width=\linewidth]{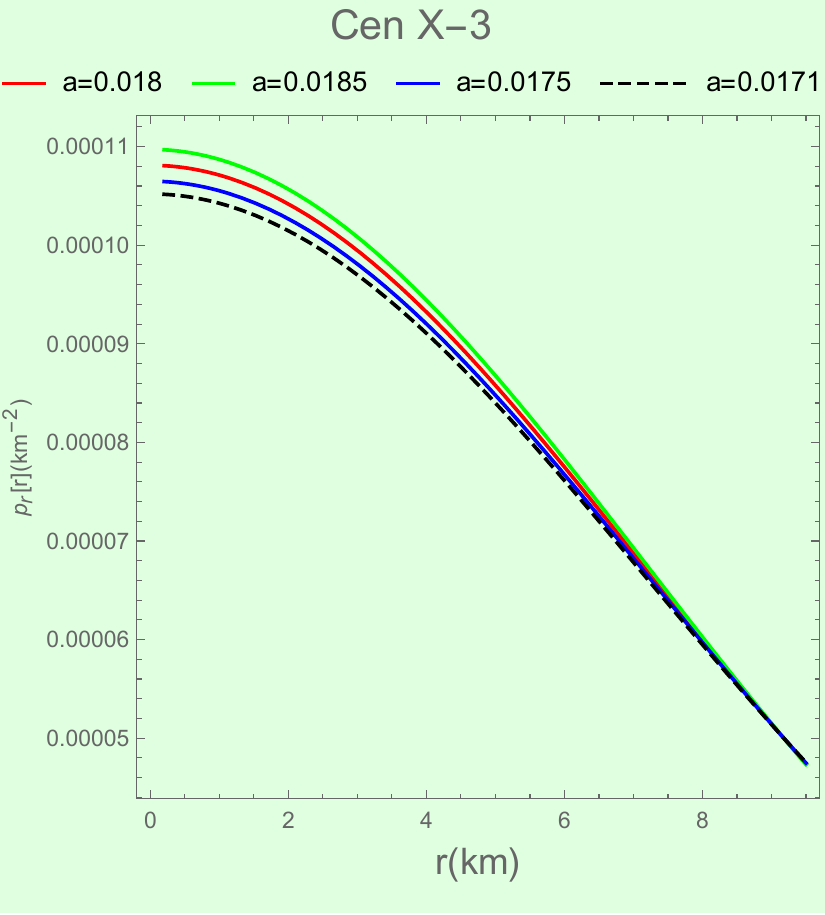}
        \small (c) Cen X-3
    \end{minipage}
    \hfill
    \begin{minipage}{0.24\textwidth}
        \centering
        \includegraphics[width=\linewidth]{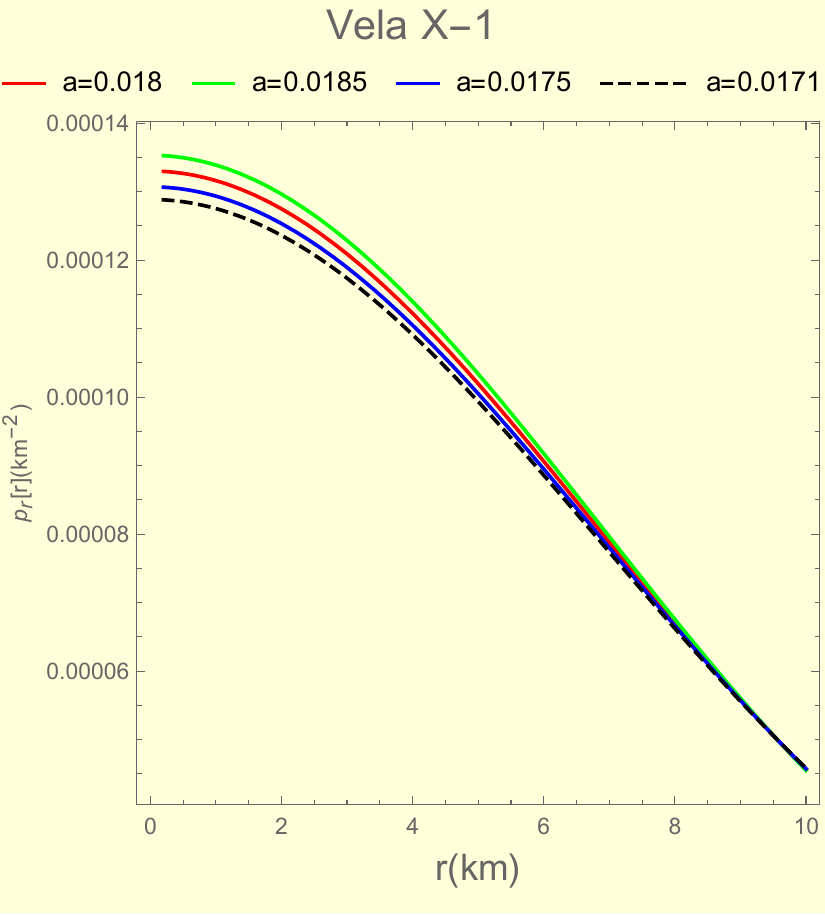}
        \small (d) Vela X-1
    \end{minipage}

    \caption{Radial pressure evolution with respect to the radial coordinate $r$ for (a) LMC X-4, (b) SMC X-4, (c) Cen X-3, and (d) Vela X-1.}
    
    \label{fig:radial}
\end{figure}

Graphical representation of radial pressure Eq. (\ref{E011}) is depicted in Fig \ref{fig:radial}.

\begin{figure}[!ht]
    \centering

    \begin{minipage}{0.24\textwidth}
        \centering
        \includegraphics[width=\linewidth]{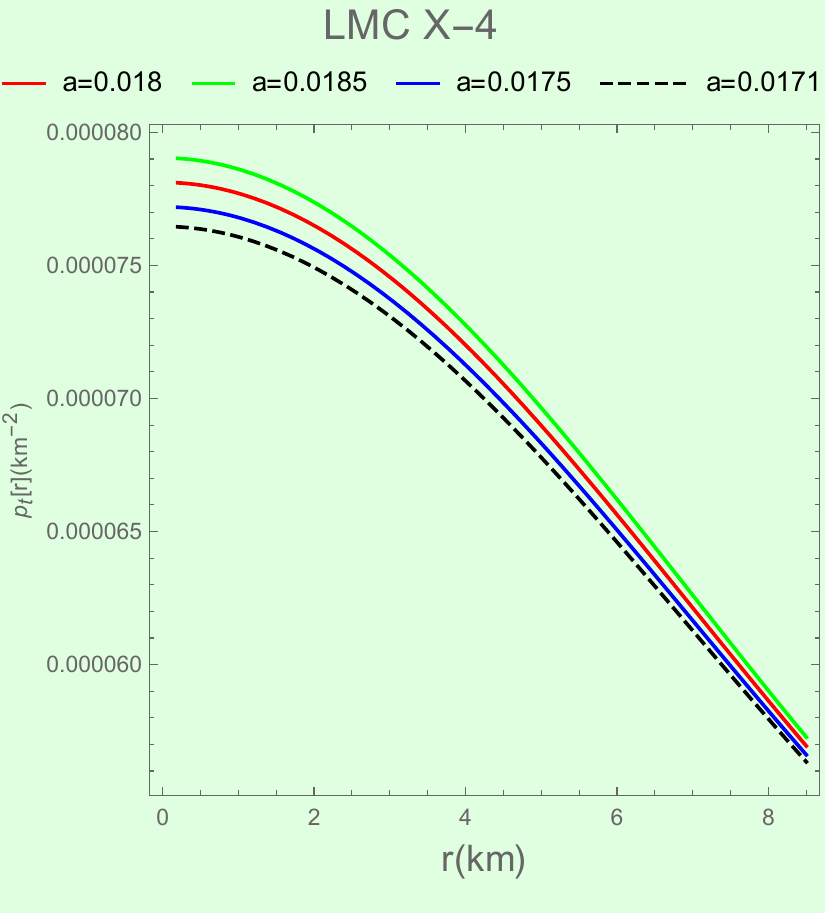}
        \small (a) LMC X-4
    \end{minipage}
    \hfill
    \begin{minipage}{0.24\textwidth}
        \centering
        \includegraphics[width=\linewidth]{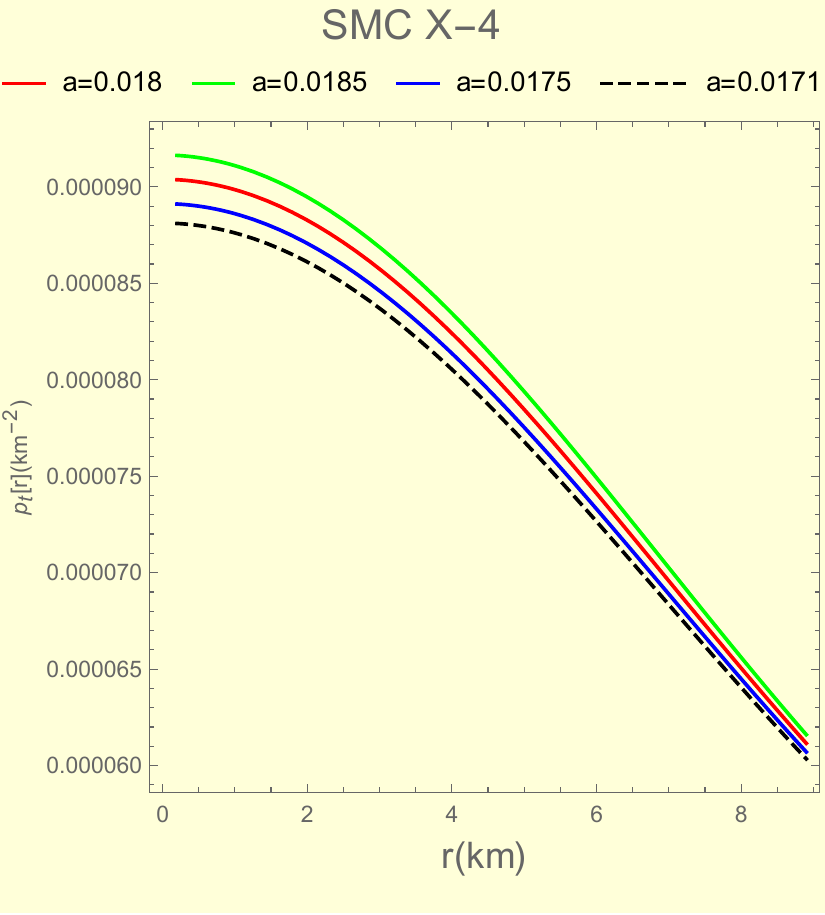}
        \small (b) SMC X-4
    \end{minipage}
    \hfill
    \begin{minipage}{0.24\textwidth}
        \centering
        \includegraphics[width=\linewidth]{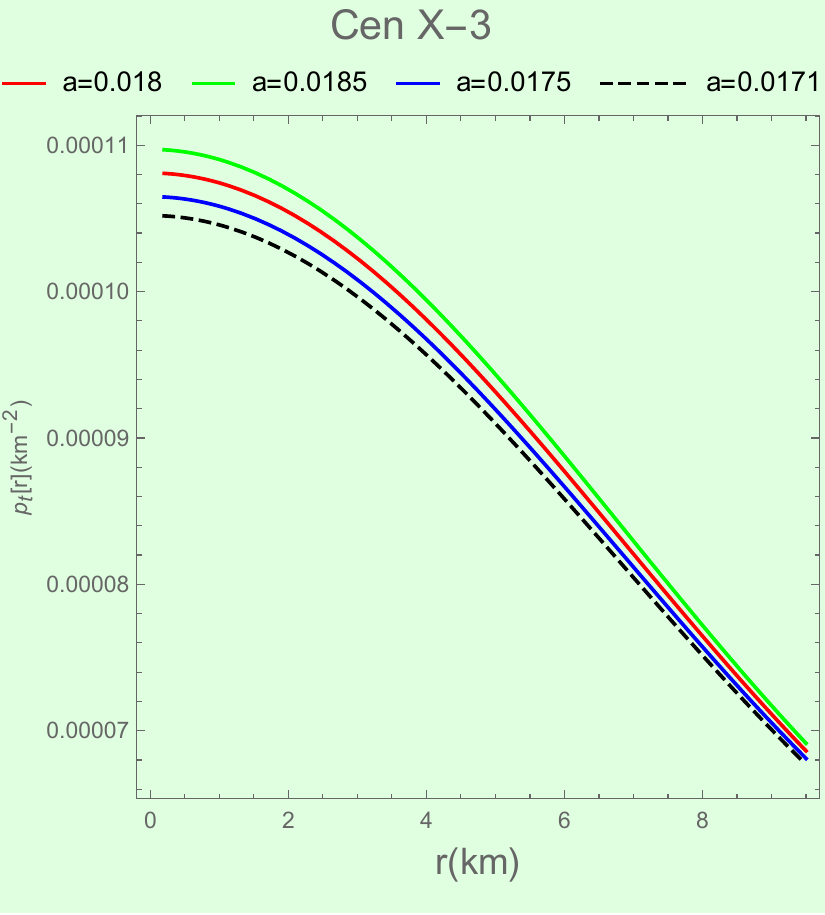}
        \small (c) Cen X-3
    \end{minipage}
    \hfill
    \begin{minipage}{0.24\textwidth}
        \centering
        \includegraphics[width=\linewidth]{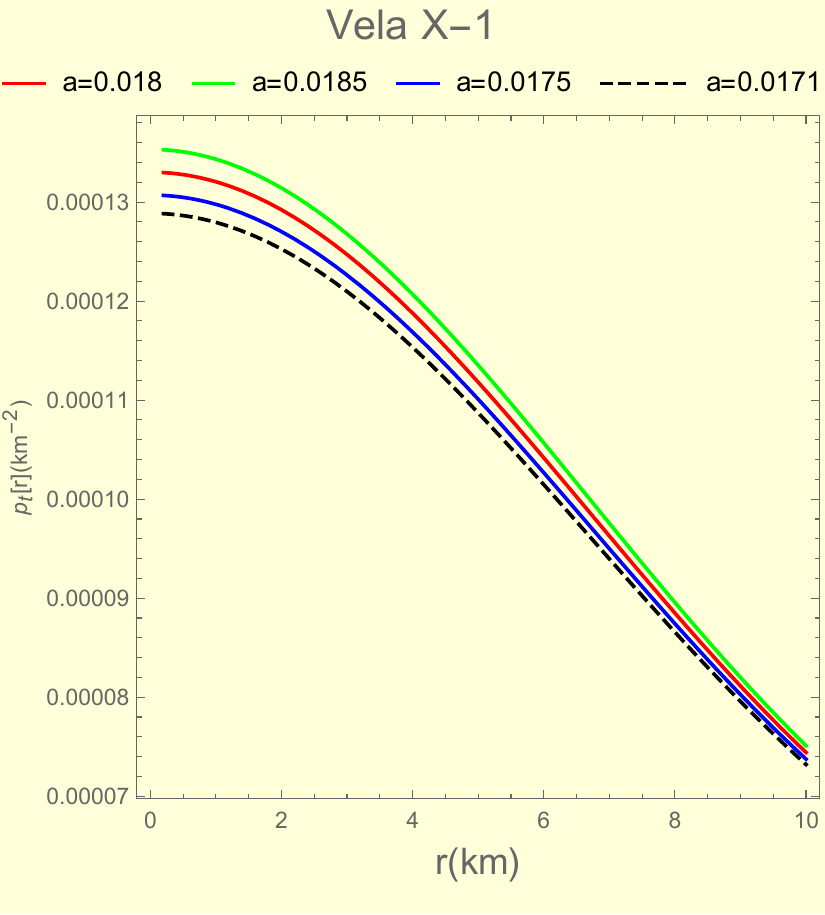}
        \small (d) Vela X-1
    \end{minipage}

    \caption{Tangential pressure evolution with respect to the radial coordinate $r$ for (a) LMC X-4, (b) SMC X-4, (c) Cen X-3, and (d) Vela X-1.}
    
    \label{fig:tangential}
\end{figure}

Graphical representation of tangential pressure Eq. (\ref{E012}) is depicted in Fig \ref{fig:tangential}.

Fig \ref{fig:density}, Fig \ref{fig:radial}, and Fig \ref{fig:tangential} show the density, radial pressure and tangential pressure profiles for the pulsar LMC X-4, SMC X-4, Cen X-3, and Vela X-1. Here we can observe a decaying pattern in the density, radial, and tangential pressure with respect to $r$ but they are positive at the boundary of the stars. Also, the radial and tangential pressure vanishes at the surface of the star and has maximum value near the core of the star, which indicate highly compact core for each star. Hence the density, radial pressure, and tangential pressure demonstrate decreasing function graphically.\\

Understanding the nature of the compact object requires an understanding of density and pressure in both radial and transverse directions. The density and pressure of the compact objects should reach their highest value close to their center and then monotonically drop over their surface, meaning that their physical properties shouldn't have any singularities. Furthermore, the anisotropic factor, denoted by $\Delta=p_{t}-p_{r}$, indicates that the pressure in both the transversal and radial directions is not equal for any compact object i.e. $p_{t} > p_{r}$. 

\begin{figure}[!ht]
    \centering

    \begin{minipage}{0.24\textwidth}
        \centering
        \includegraphics[width=\linewidth]{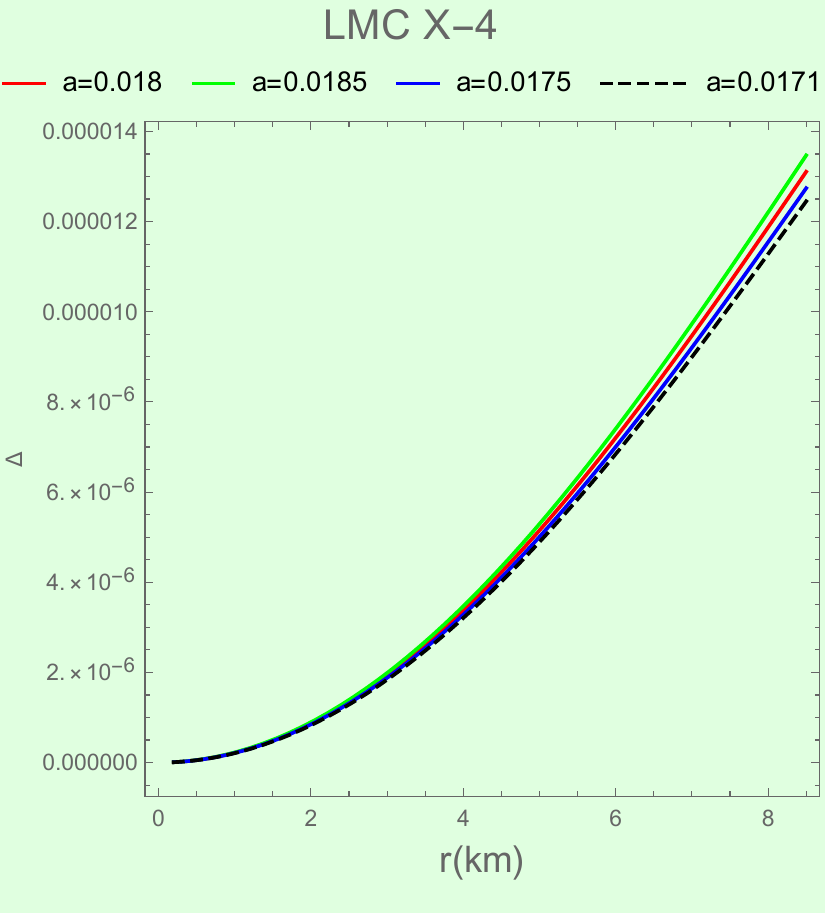}
        \small (a) LMC X-4
    \end{minipage}
    \hfill
    \begin{minipage}{0.24\textwidth}
        \centering
        \includegraphics[width=\linewidth]{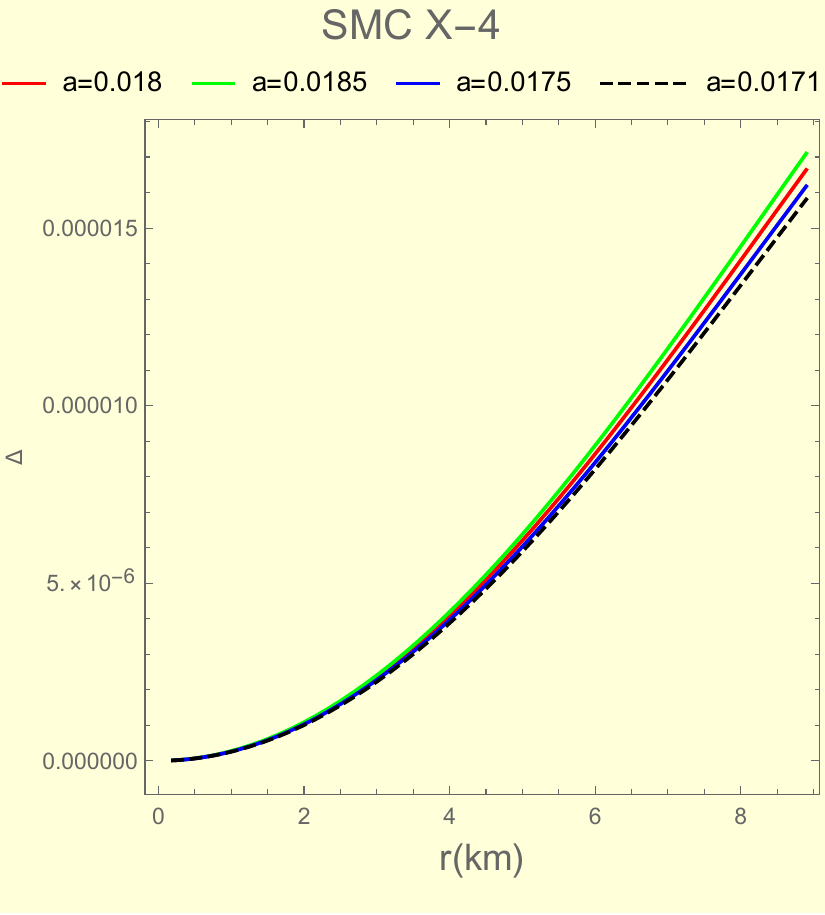}
        \small (b) SMC X-4
    \end{minipage}
    \hfill
    \begin{minipage}{0.24\textwidth}
        \centering
        \includegraphics[width=\linewidth]{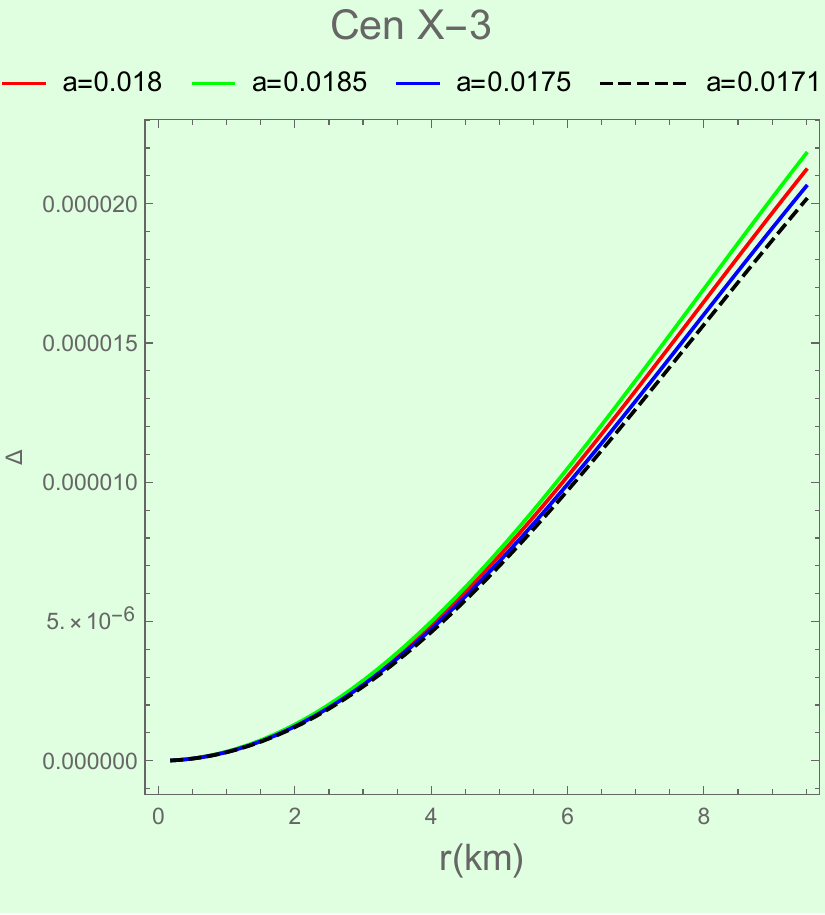}
        \small (c) Cen X-3
    \end{minipage}
    \hfill
    \begin{minipage}{0.24\textwidth}
        \centering
        \includegraphics[width=\linewidth]{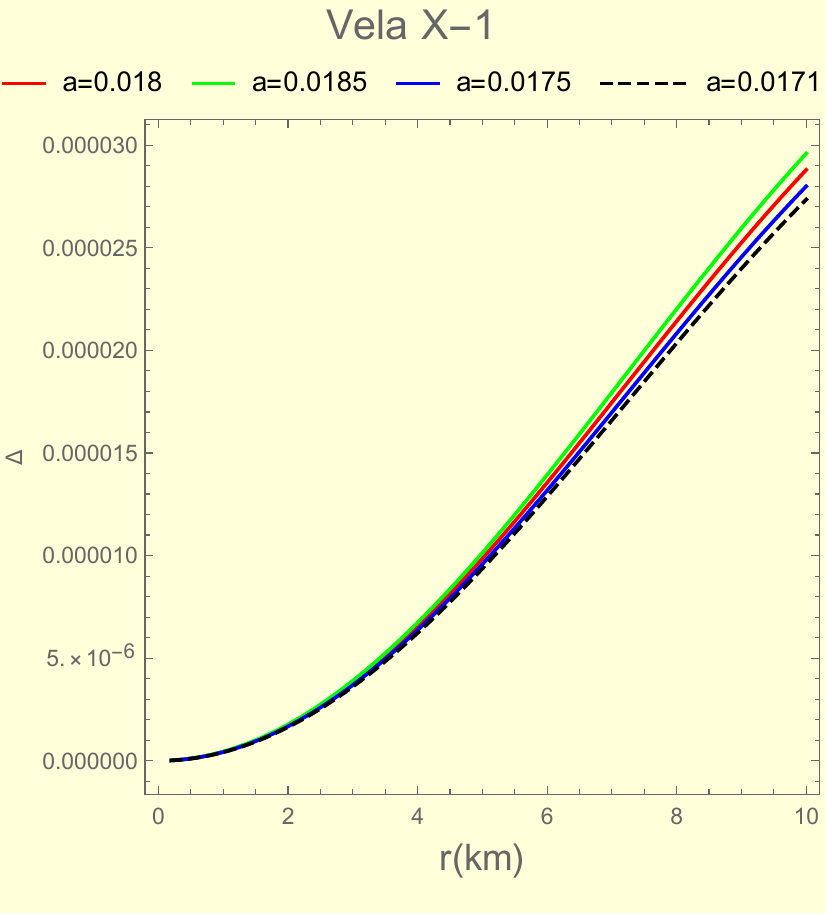}
        \small (d) Vela X-1
    \end{minipage}

    \caption{Anisotropic factor evolution with respect to the radial coordinate $r$ for (a) LMC X-4, (b) SMC X-4, (c) Cen X-3, and (d) Vela X-1.}
    
    \label{fig:ani}
\end{figure}

Anisotropic factor has been depicted graphically in Fig \ref{fig:ani} and show positive value and monotonically increases along the surface. Hence, we conclude $p_{t} > p_{r}$ and indicate that anisotropic factor has a outward direction.\\

 To further consolidate our result we have derived the gradient of the density, radial pressure and tangential pressure.
Gradient of density from Eq. (\ref{E010})
\begin{equation}
\frac{{d\rho }}{{dr}}=-\frac{4 a e^{-A r^2} \left(-1+e^{A r^2}-A r^2+2 A^2 r^4\right)}{r^3}
\label{E032}
\end{equation}
Gradient of radial pressure from Eq. (\ref{E011})
\begin{equation}
\frac{{dp}_r}{{dr}}=\frac{2 a e^{-A r^2} \left(-1+e^{A r^2}-A \left(r^2+2 B r^4\right)\right)}{r^3}
\label{E033}
\end{equation}
Gradient of radial pressure from Eq. (\ref{E012})
\begin{equation}
\frac{{dp}_t}{{dr}}=2 a e^{-A r^2} r \left(B^2+A^2 \left(1+B r^2\right)-A B \left(3+B r^2\right)\right)
\label{E034}
\end{equation}

\begin{figure}[!ht]
    \centering

    \begin{minipage}{0.24\textwidth}
        \centering
        \includegraphics[width=\linewidth]{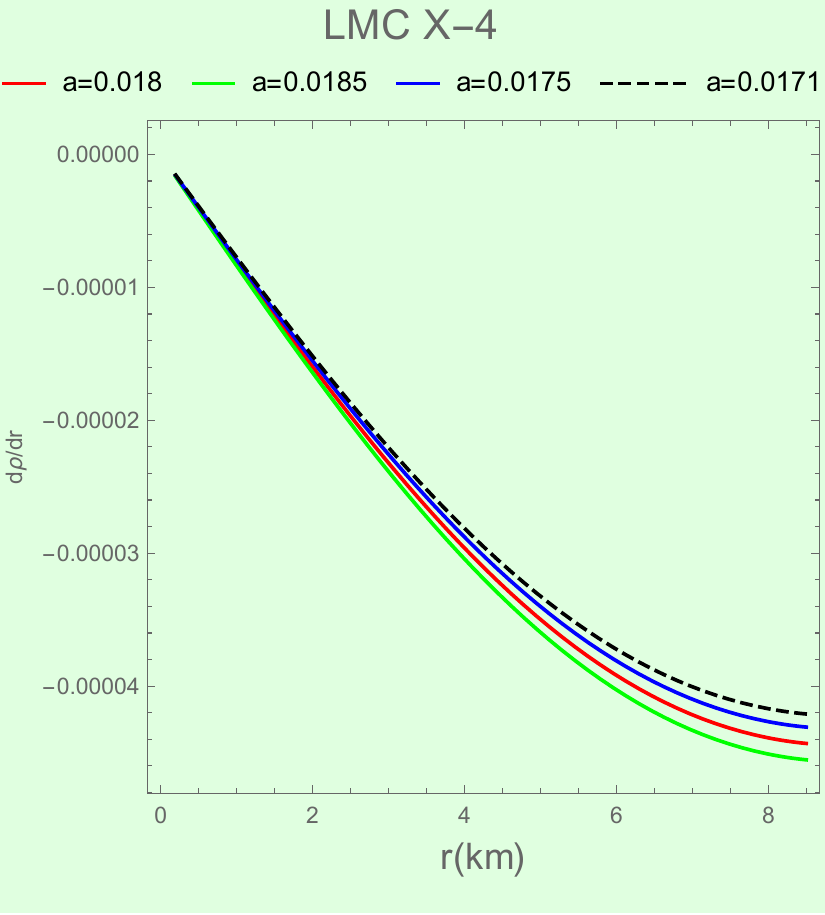}
        \small (a) LMC X-4
    \end{minipage}
    \hfill
    \begin{minipage}{0.24\textwidth}
        \centering
        \includegraphics[width=\linewidth]{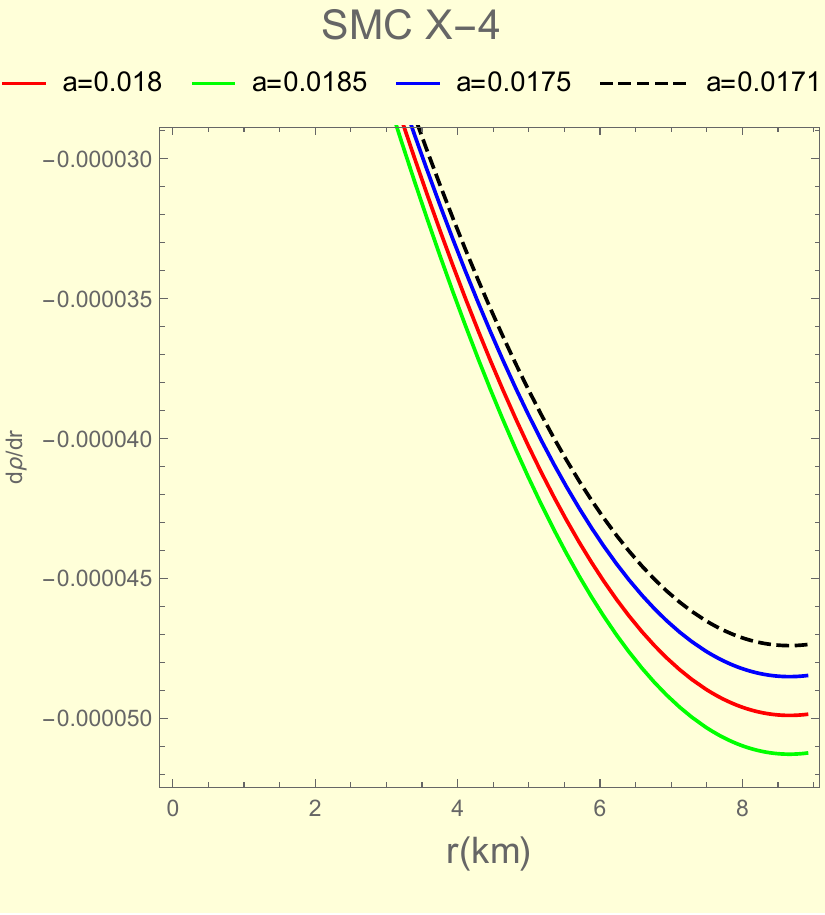}
        \small (b) SMC X-4
    \end{minipage}
    \hfill
    \begin{minipage}{0.24\textwidth}
        \centering
        \includegraphics[width=\linewidth]{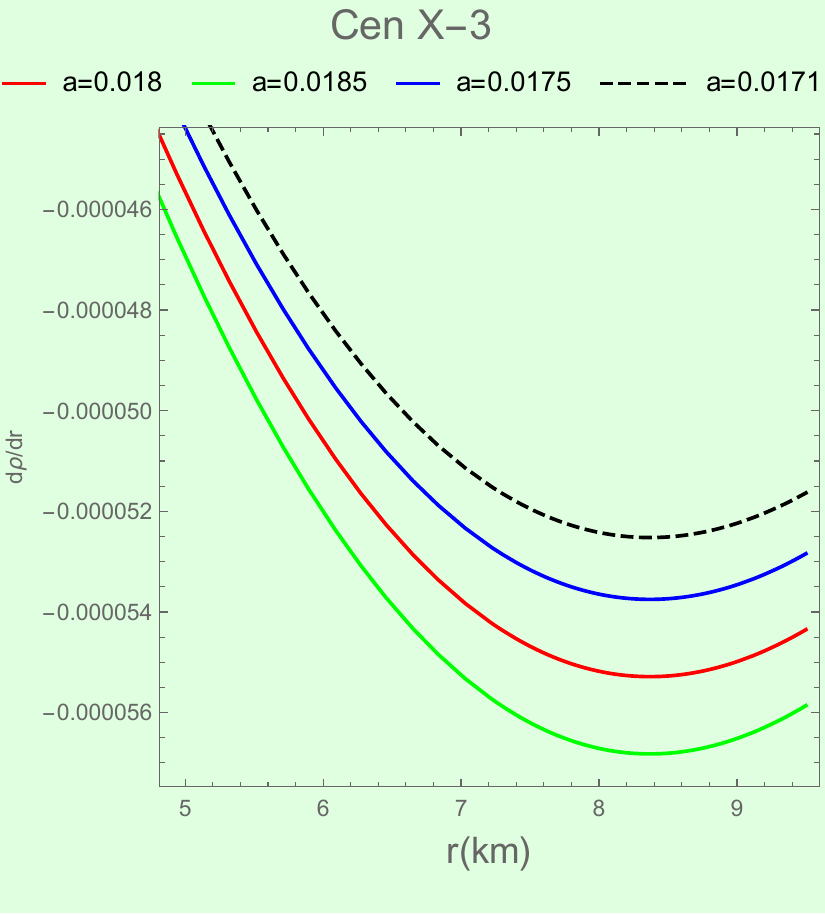}
        \small (c) Cen X-3
    \end{minipage}
    \hfill
    \begin{minipage}{0.24\textwidth}
        \centering
        \includegraphics[width=\linewidth]{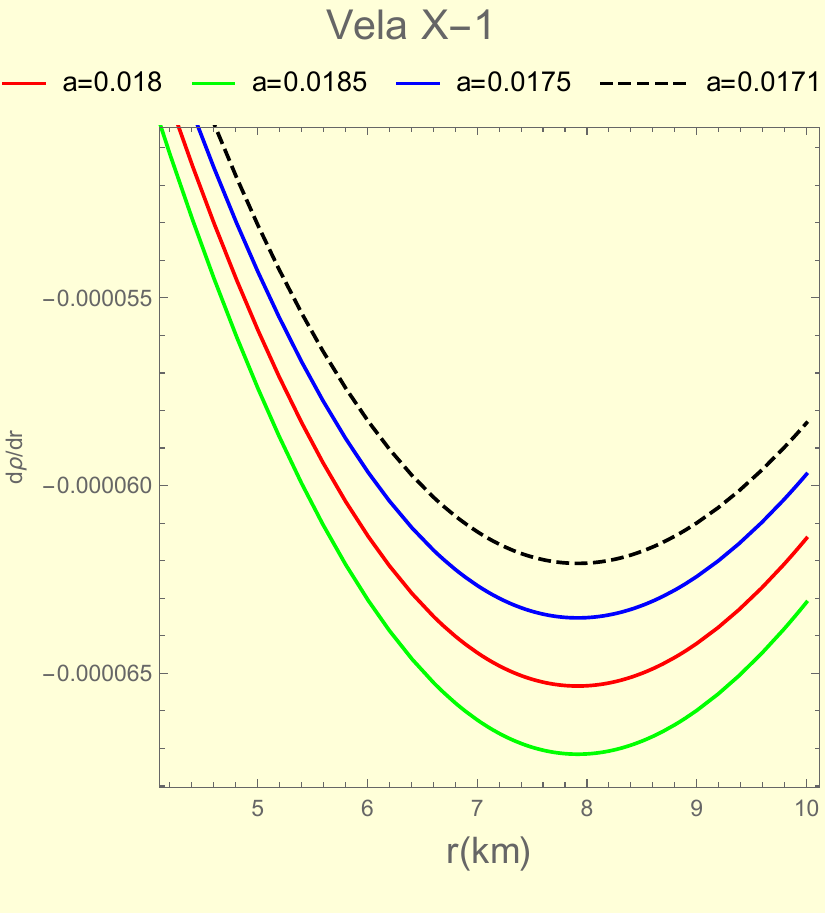}
        \small (d) Vela X-1
    \end{minipage}

    \caption{Density gradient evolution with respect to the radial coordinate $r$ for (a) LMC X-4, (b) SMC X-4, (c) Cen X-3, and (d) Vela X-1.}
    
    \label{fig:drho}
\end{figure}

Graphical representation of density gradient from Eq. (\ref{E032}) is depicted in Fig \ref{fig:drho}.

\begin{figure}[!ht]
    \centering

    \begin{minipage}{0.24\textwidth}
        \centering
        \includegraphics[width=\linewidth]{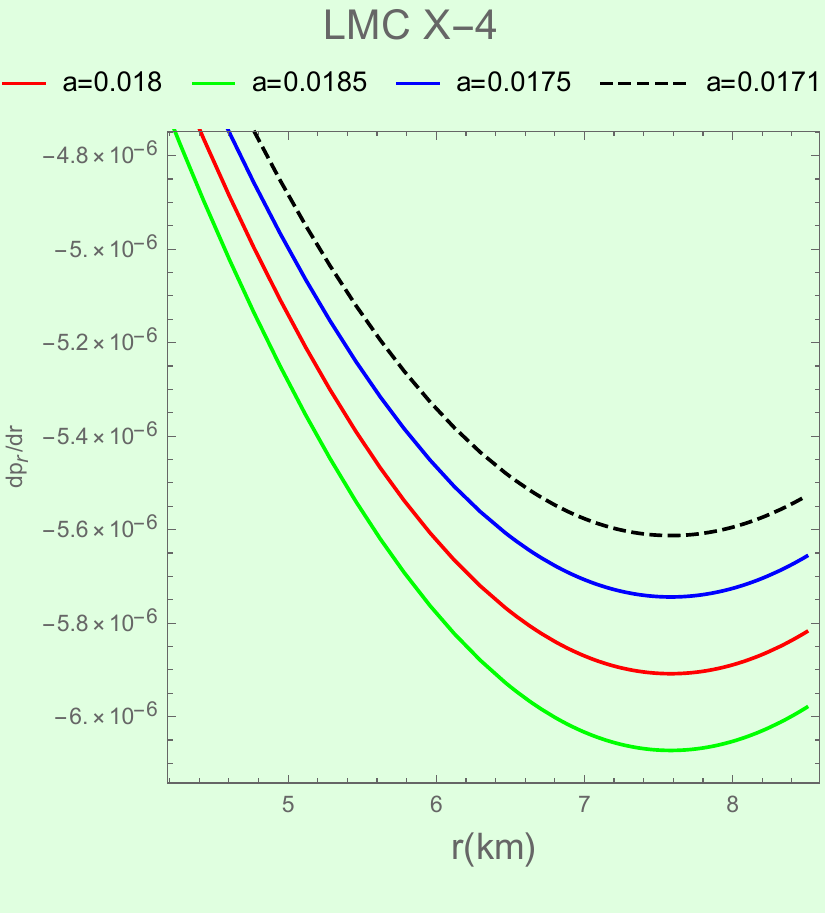}
        \small (a) LMC X-4
    \end{minipage}
    \hfill
    \begin{minipage}{0.24\textwidth}
        \centering
        \includegraphics[width=\linewidth]{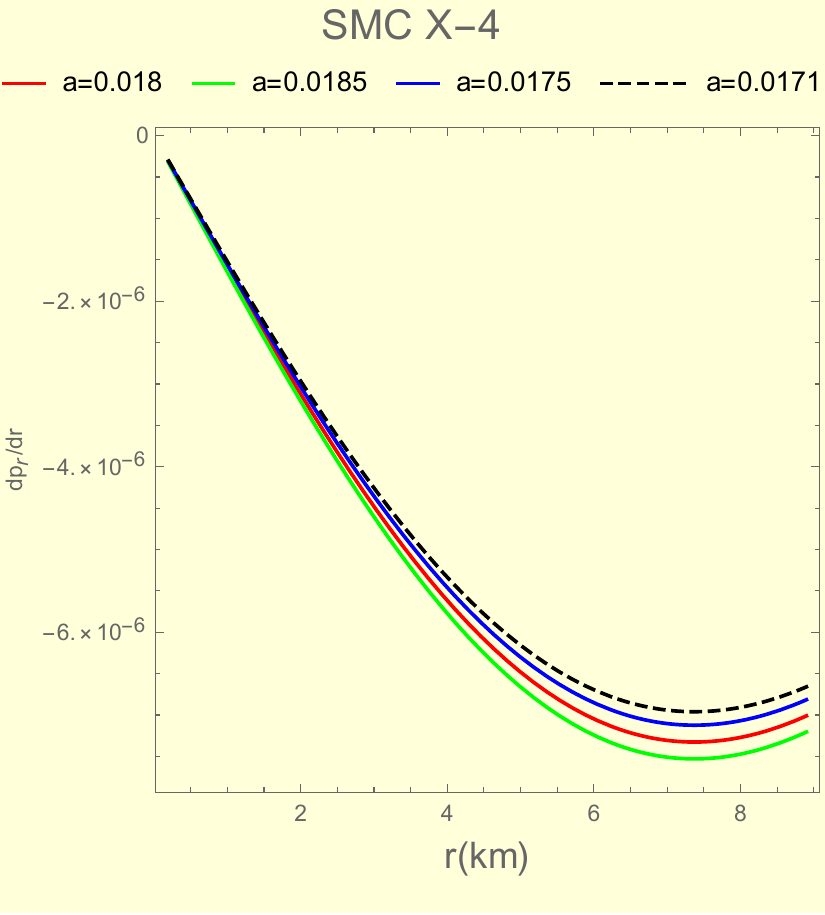}
        \small (b) SMC X-4
    \end{minipage}
    \hfill
    \begin{minipage}{0.24\textwidth}
        \centering
        \includegraphics[width=\linewidth]{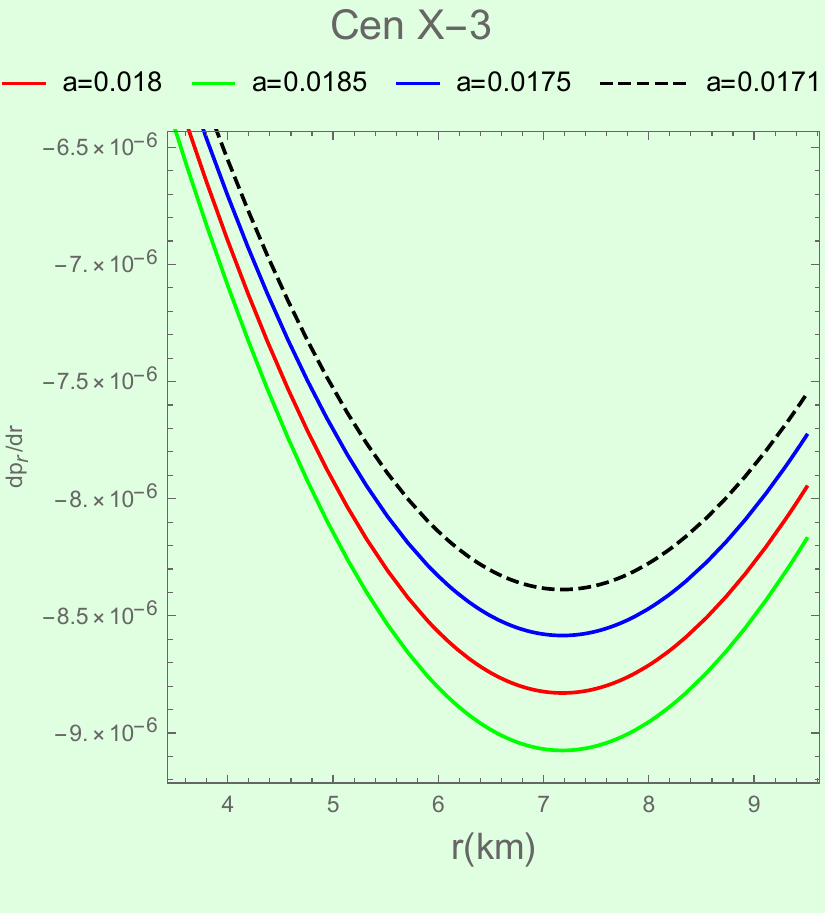}
        \small (c) Cen X-3
    \end{minipage}
    \hfill
    \begin{minipage}{0.24\textwidth}
        \centering
        \includegraphics[width=\linewidth]{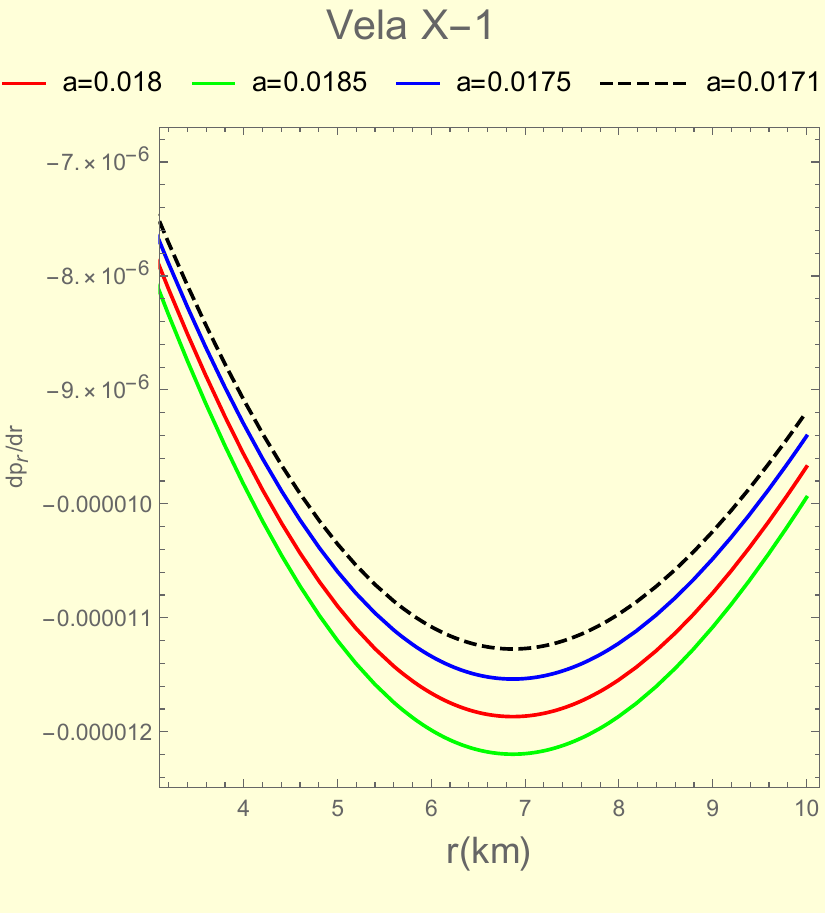}
        \small (d) Vela X-1
    \end{minipage}

    \caption{Radial pressure gradient evolution with respect to the radial coordinate $r$ for (a) LMC X-4, (b) SMC X-4, (c) Cen X-3, and (d) Vela X-1.}
    
    \label{fig:dradial}
\end{figure}

Graphical representation of radial pressure gradient from Eq. (\ref{E033}) is depicted in Fig \ref{fig:dradial}.

\begin{figure}[!ht]
    \centering

    \begin{minipage}{0.24\textwidth}
        \centering
        \includegraphics[width=\linewidth]{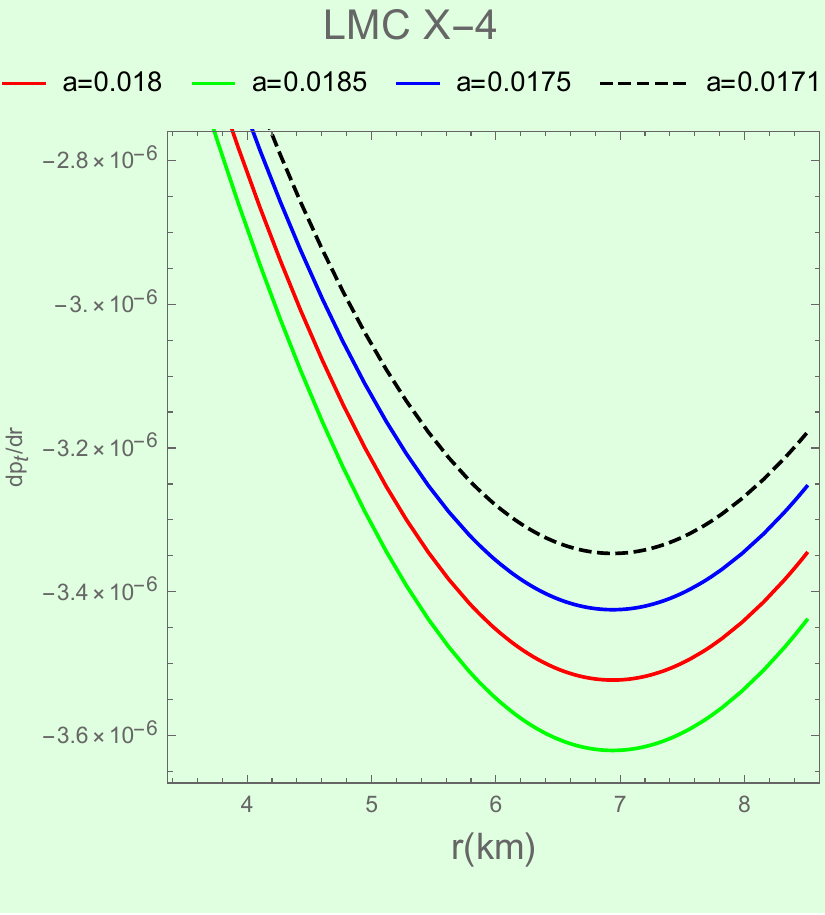}
        \small (a) LMC X-4
    \end{minipage}
    \hfill
    \begin{minipage}{0.24\textwidth}
        \centering
        \includegraphics[width=\linewidth]{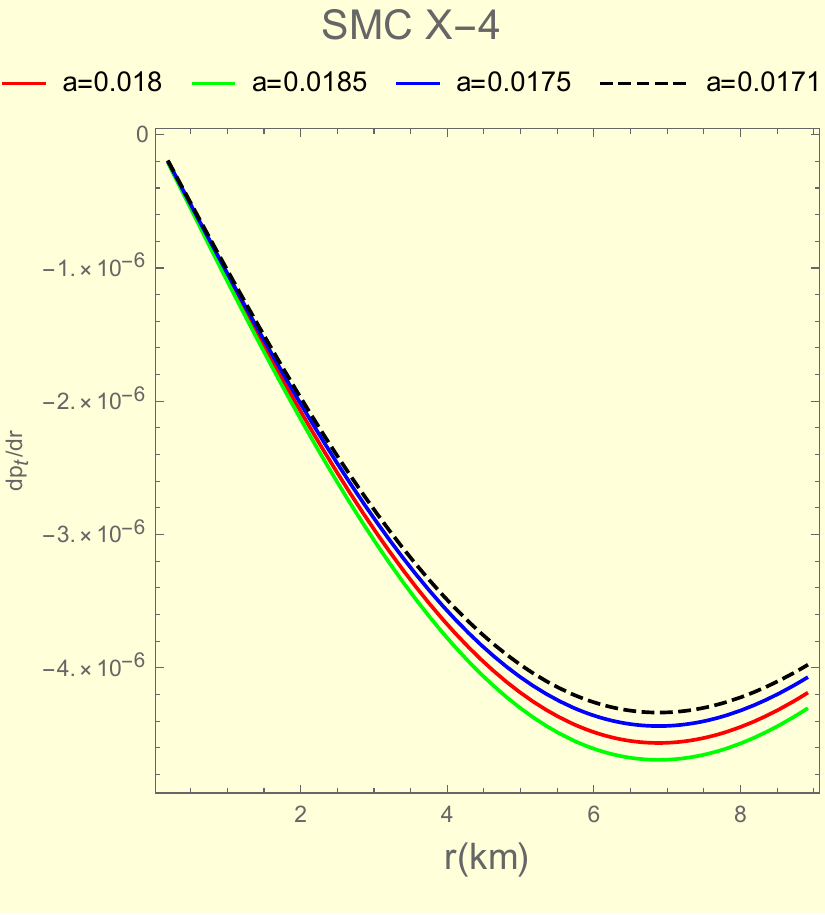}
        \small (b) SMC X-4
    \end{minipage}
    \hfill
    \begin{minipage}{0.24\textwidth}
        \centering
        \includegraphics[width=\linewidth]{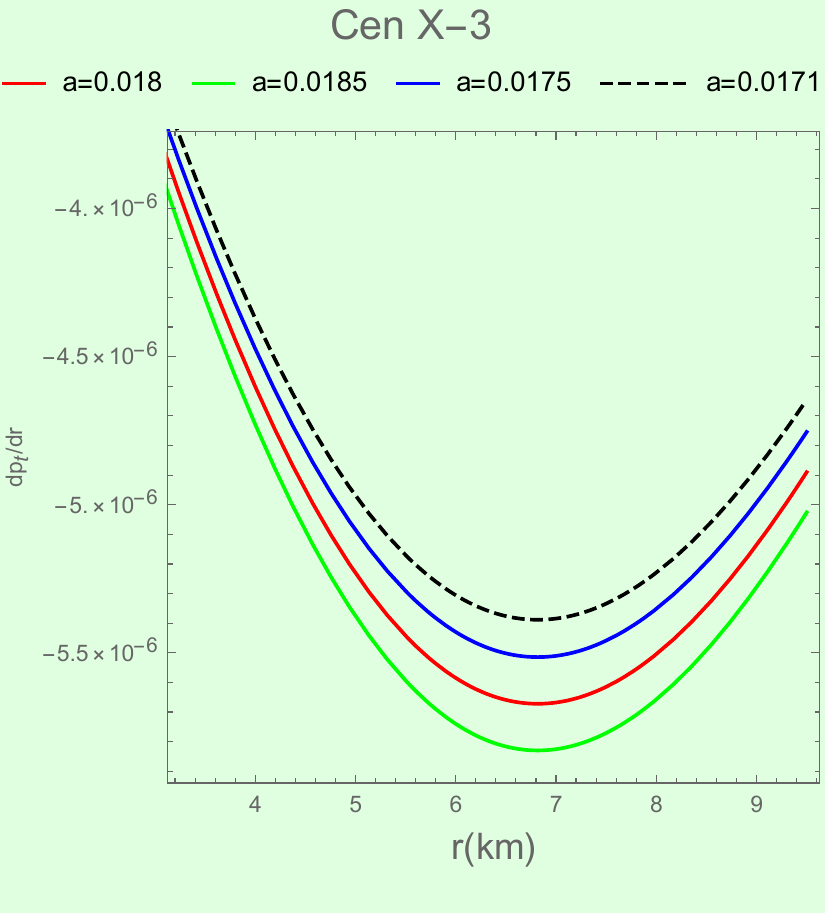}
        \small (c) Cen X-3
    \end{minipage}
    \hfill
    \begin{minipage}{0.24\textwidth}
        \centering
        \includegraphics[width=\linewidth]{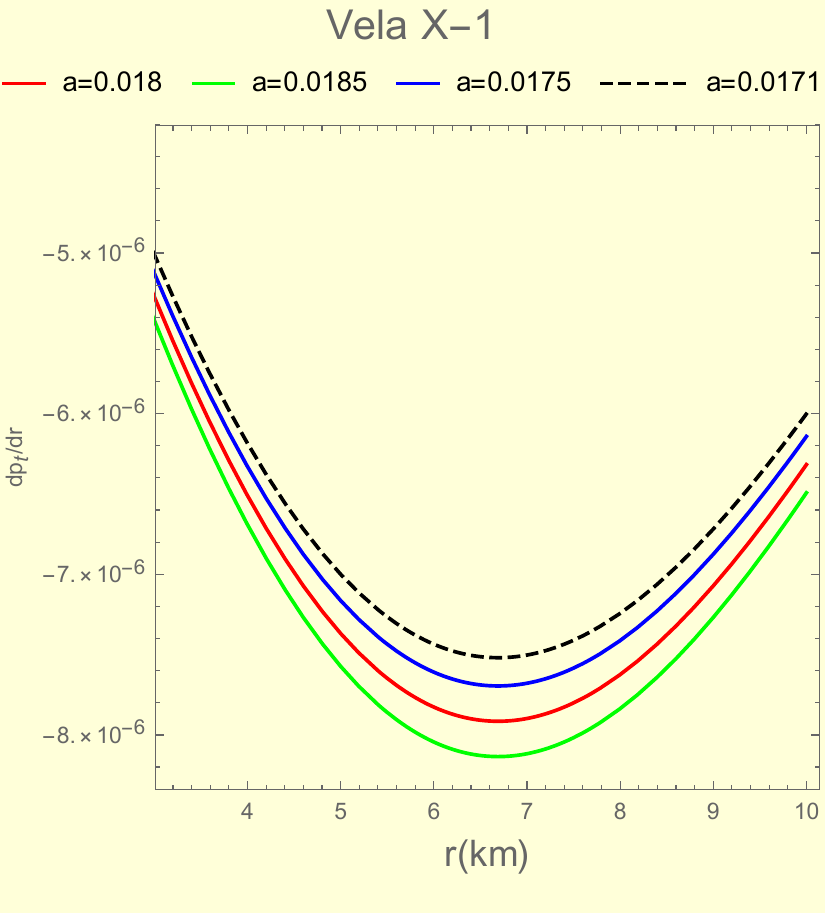}
        \small (d) Vela X-1
    \end{minipage}

    \caption{Tangential pressure gradient evolution with respect to the radial coordinate $r$ for (a) LMC X-4, (b) SMC X-4, (c) Cen X-3, and (d) Vela X-1.}
    
    \label{fig:dtangential}
\end{figure}

Graphical representation of radial pressure gradient from Eq. (\ref{E034}) is depicted in Fig \ref{fig:dtangential}.

Fig \ref{fig:drho}, Fig \ref{fig:dradial}, and Fig \ref{fig:dtangential} show the gradient of density, radial pressure and tangential pressure profiles for the pulsar LMC X-4, SMC X-4, Cen X-3, and Vela X-1. Three of the graphical representation show negative profiles. Hence our model is viable for further study. We would like to mention that density, radial pressure, and tangential pressure decrease with radius of the star and hence they are monotone deceasing function of $r$. As a result the gradients are negative.

\subsection{Energy Conditions}

We will talk about the energy state of the compact object in this section. These energy conditions are useful for assessing the matter's unusual behaviour inside the compact object.
Null Energy Condition (NEC) and Strong Energy Condition (SEC) are the main two categories into which energy states are divided, and their corresponding boundaries are \cite{bib14} \\
$(i)$ $NEC : \rho + p_{r} \geq 0 $ , $\rho + p_{t} \geq 0$ \\
$(ii)$ $SEC : \rho + p_{r} \geq 0 $ , $\rho + p_{t} \geq 0$ , $\rho + p_{r} + 2 p_{t} \geq 0$. \\
The Hawking-Penrose singularity hypothesis and the second law of black hole thermodynamics \cite{bib14} can be effectively examined by means of the energy conditions. Energy conditions have been used to explain several interesting phenomena in cosmology \cite{bib15,bib16}. Energy conditions have an impact on both understanding the DE phase and future singularity research. Furthermore, fulfilment of these prerequisites is necessary to comprehend the peculiar behaviour of matter in the compact structure model. We can draw the conclusion that our star models are workable if the energy conditions are met.

Hence the simplified expressions for energy conditions are
\begin{equation}
\rho +p_r=-\frac{b}{2}+\frac{a e^{-A r^2} \left(-1+e^{A r^2}+4 A r^2+2 B r^2\right)}{r^2}
\label{E030}
\end{equation}
\begin{equation}
\rho +p_t=\frac{e^{-A r^2} \left(-b e^{A r^2} r^2+2 a \left(-2+2 e^{A r^2}+3 A r^2+2 B r^2-A B r^4+B^2 r^4\right)\right)}{2 r^2}
\label{E031}
\end{equation}
\begin{equation}
\rho +p_r+2 p_t=\frac{e^{-A r^2} \left(b e^{A r^2} r^2+2 a \left(-1+e^{A r^2}+2 A r^2+6 B r^2-2 A B r^4+2 B^2 r^4\right)\right)}{2 r^2}
\label{E035}
\end{equation}

\begin{figure}[!ht]
    \centering

    \begin{minipage}{0.24\textwidth}
        \centering
        \includegraphics[width=\linewidth]{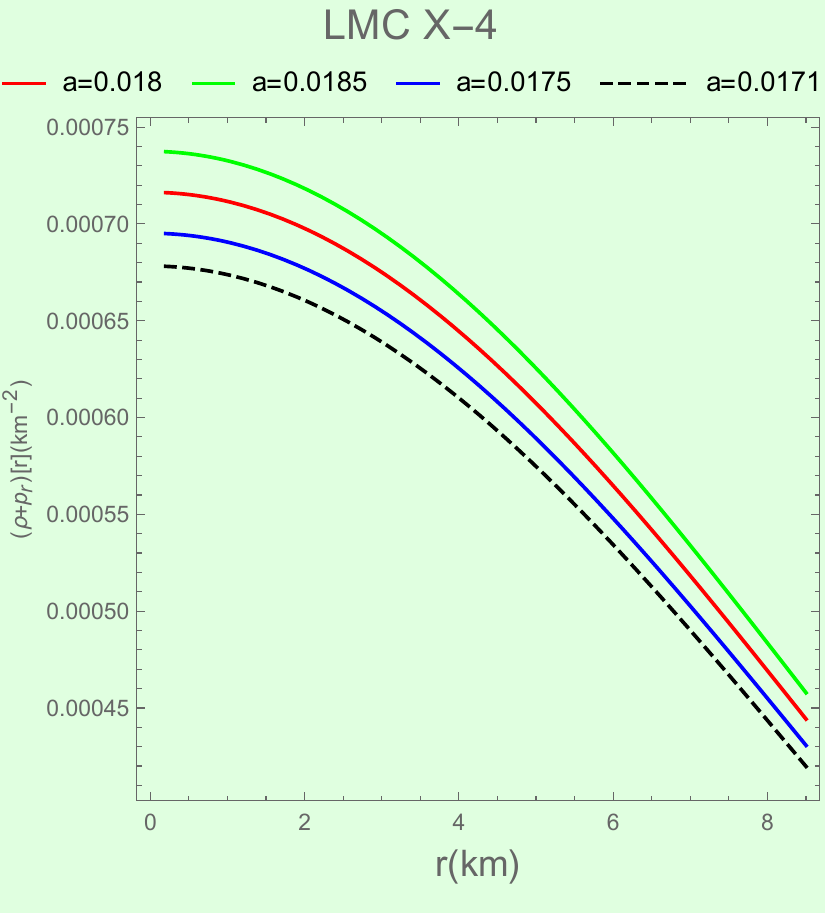}
        \small (a) LMC X-4
    \end{minipage}
    \hfill
    \begin{minipage}{0.24\textwidth}
        \centering
        \includegraphics[width=\linewidth]{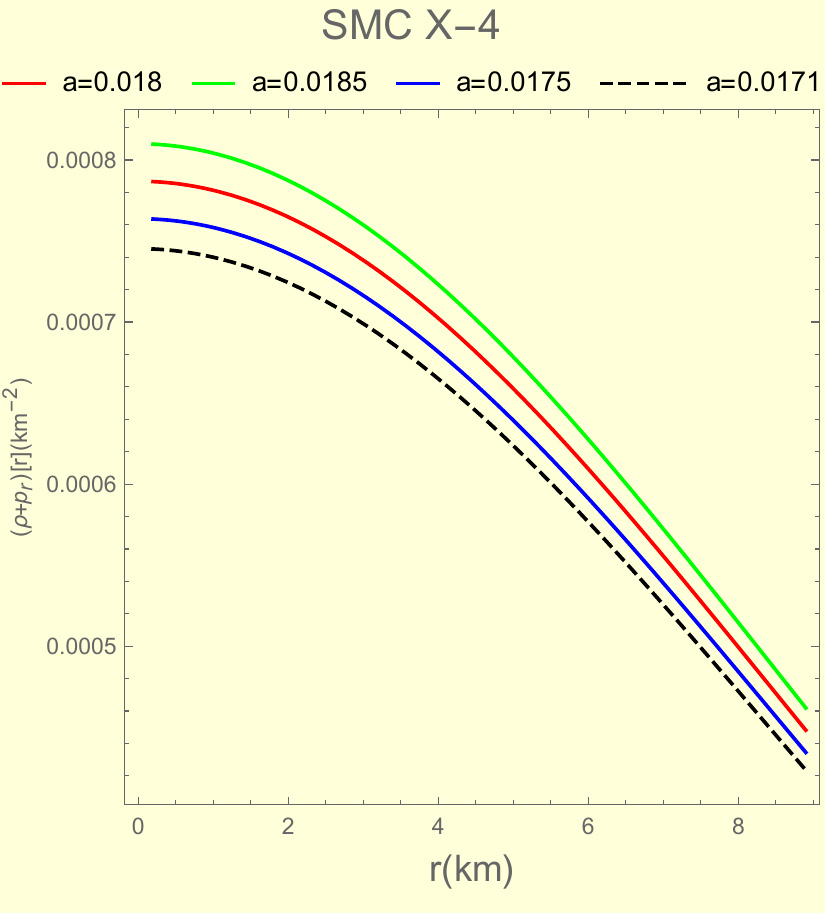}
        \small (b) SMC X-4
    \end{minipage}
    \hfill
    \begin{minipage}{0.24\textwidth}
        \centering
        \includegraphics[width=\linewidth]{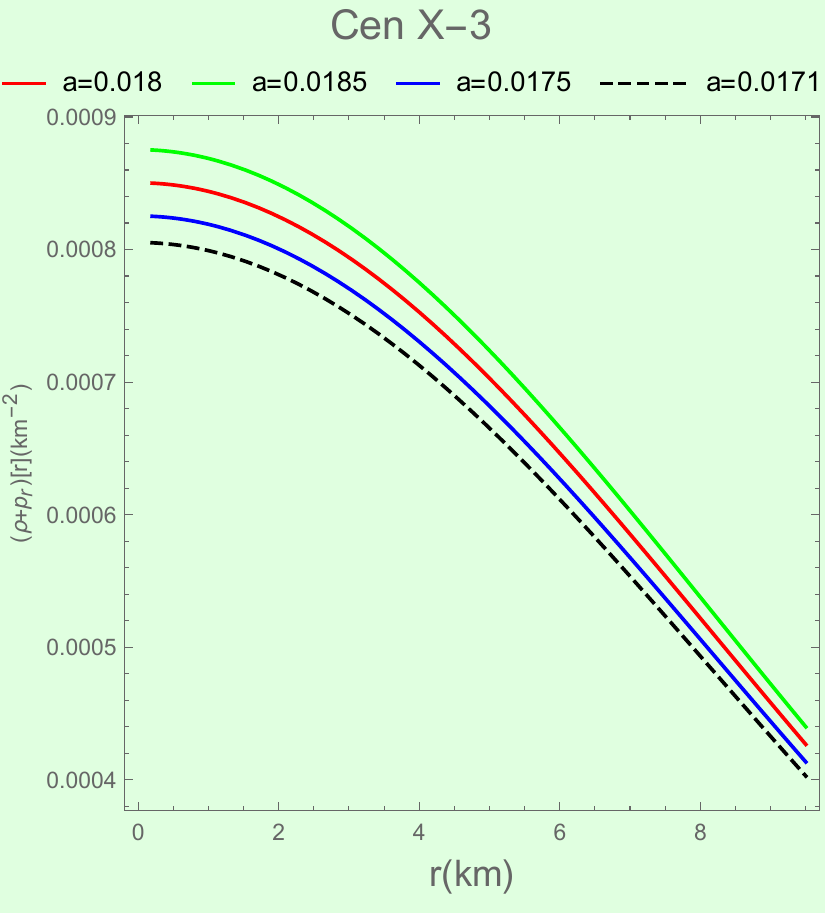}
        \small (c) Cen X-3
    \end{minipage}
    \hfill
    \begin{minipage}{0.24\textwidth}
        \centering
        \includegraphics[width=\linewidth]{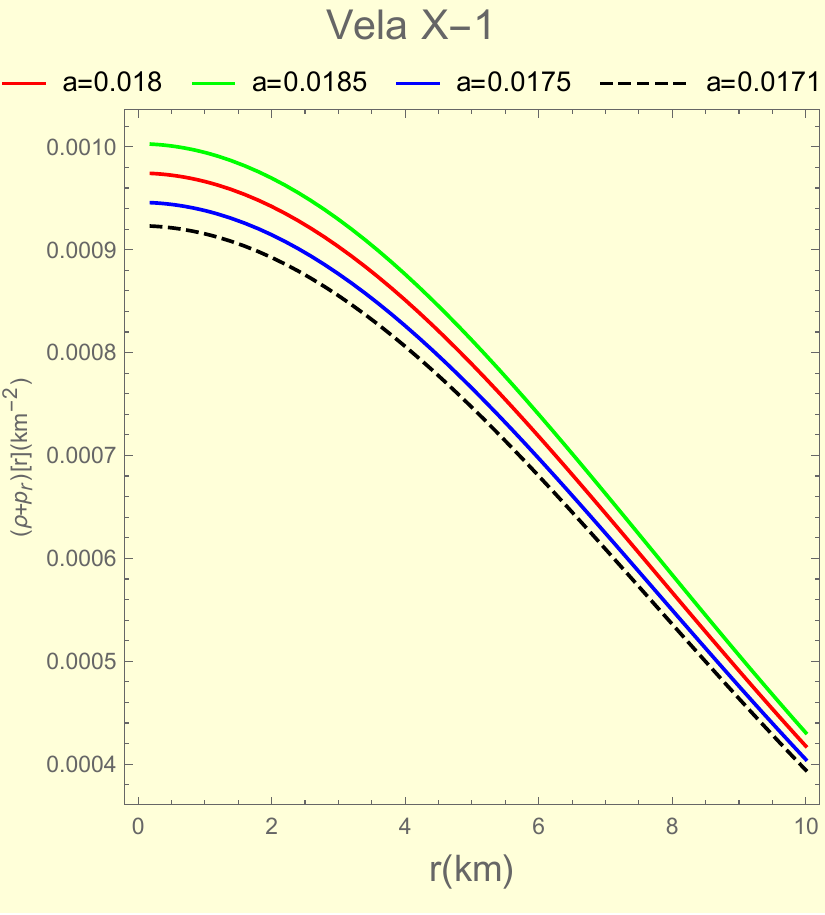}
        \small (d) Vela X-1
    \end{minipage}

    \caption{Evolution of $\rho + p_r$ with respect to the radial coordinate $r$ for (a) LMC X-4, (b) SMC X-4, (c) Cen X-3, and (d) Vela X-1.}
    
    \label{fig:sec1}
\end{figure}

\begin{figure}[!ht]
    \centering

    \begin{minipage}{0.24\textwidth}
        \centering
        \includegraphics[width=\linewidth]{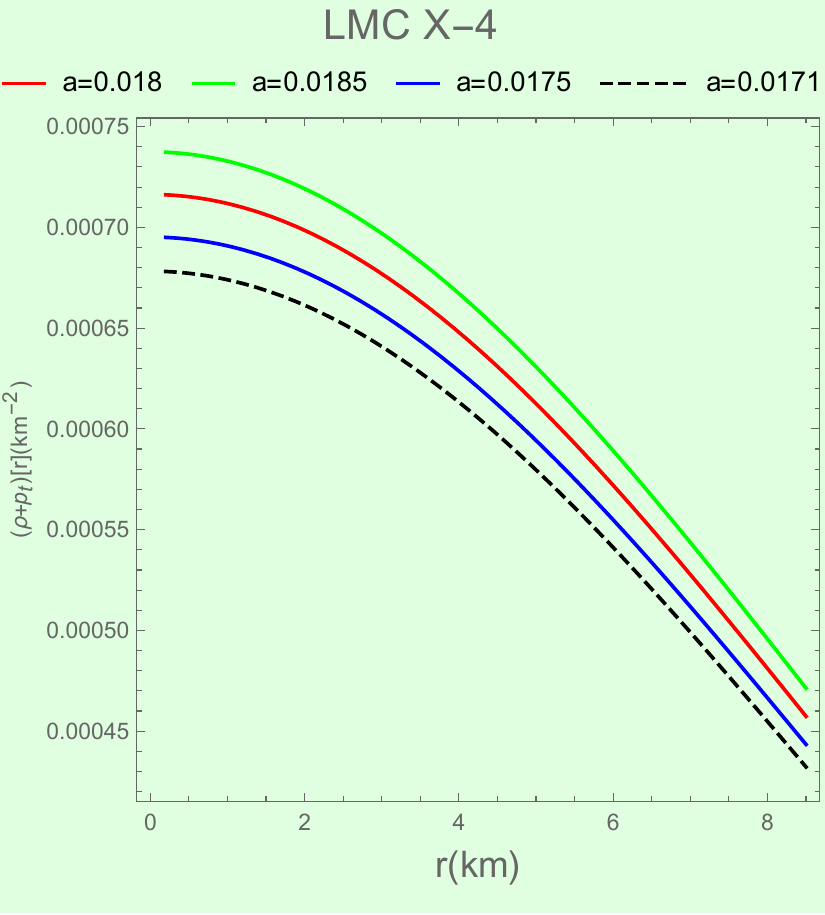}
        \small (a) LMC X-4
    \end{minipage}
    \hfill
    \begin{minipage}{0.24\textwidth}
        \centering
        \includegraphics[width=\linewidth]{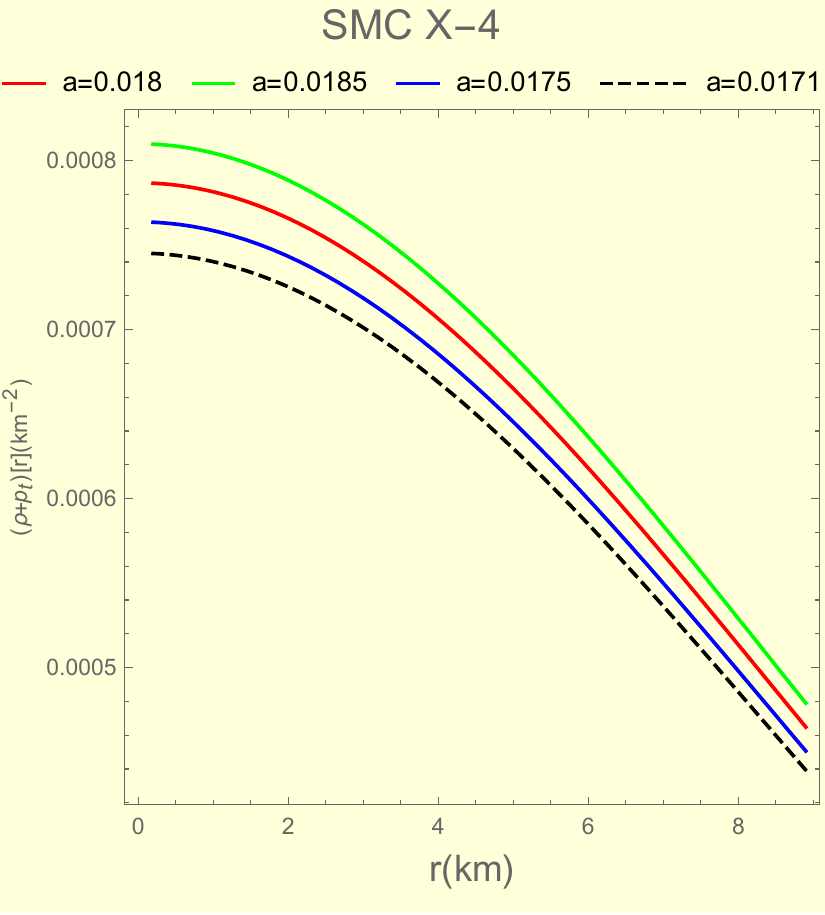}
        \small (b) SMC X-4
    \end{minipage}
    \hfill
    \begin{minipage}{0.24\textwidth}
        \centering
        \includegraphics[width=\linewidth]{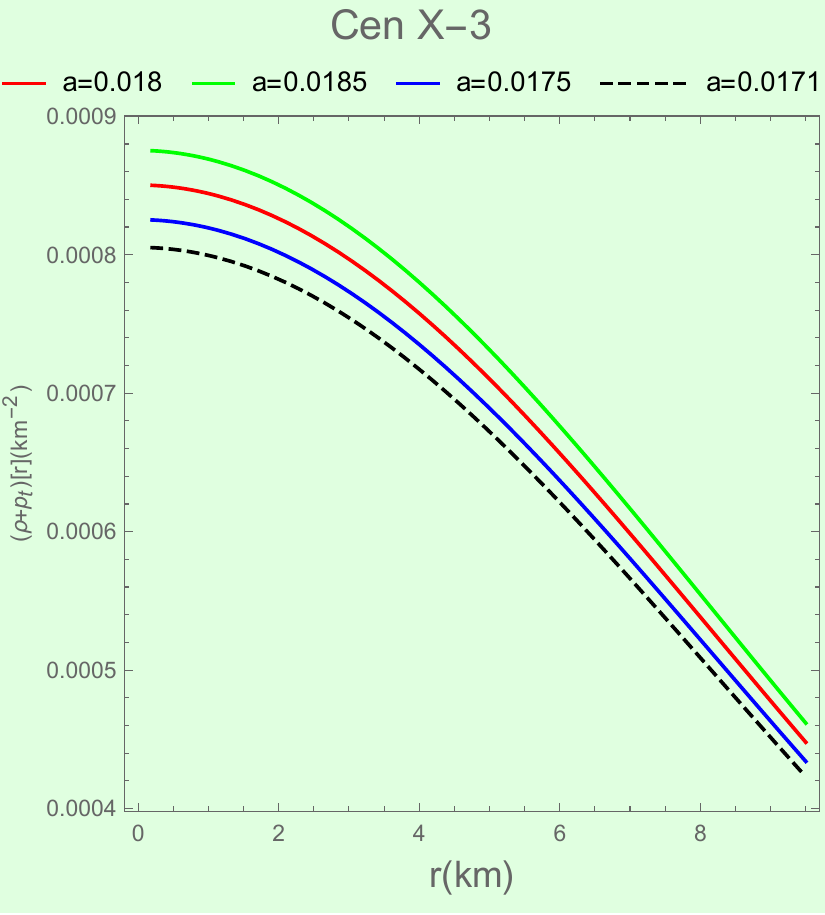}
        \small (c) Cen X-3
    \end{minipage}
    \hfill
    \begin{minipage}{0.24\textwidth}
        \centering
        \includegraphics[width=\linewidth]{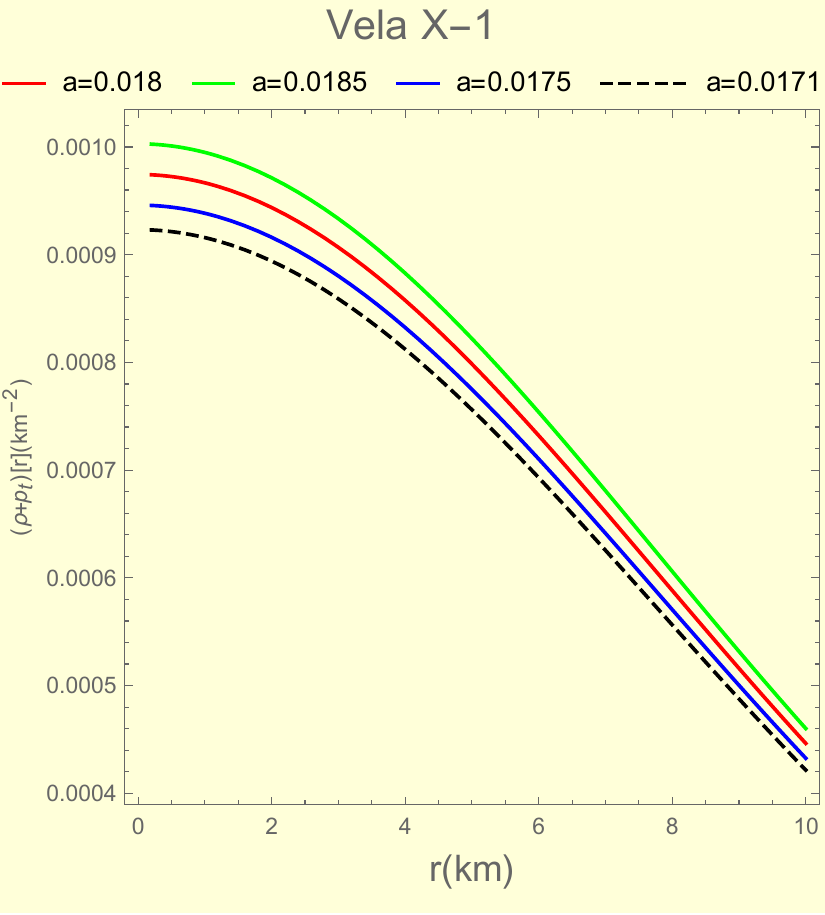}
        \small (d) Vela X-1
    \end{minipage}

    \caption{Evolution of $\rho + p_t$ with respect to the radial coordinate $r$ for (a) LMC X-4, (b) SMC X-4, (c) Cen X-3, and (d) Vela X-1.}
    
    \label{fig:sec2}
\end{figure}

\begin{figure}[!ht]
    \centering

    \begin{minipage}{0.24\textwidth}
        \centering
        \includegraphics[width=\linewidth]{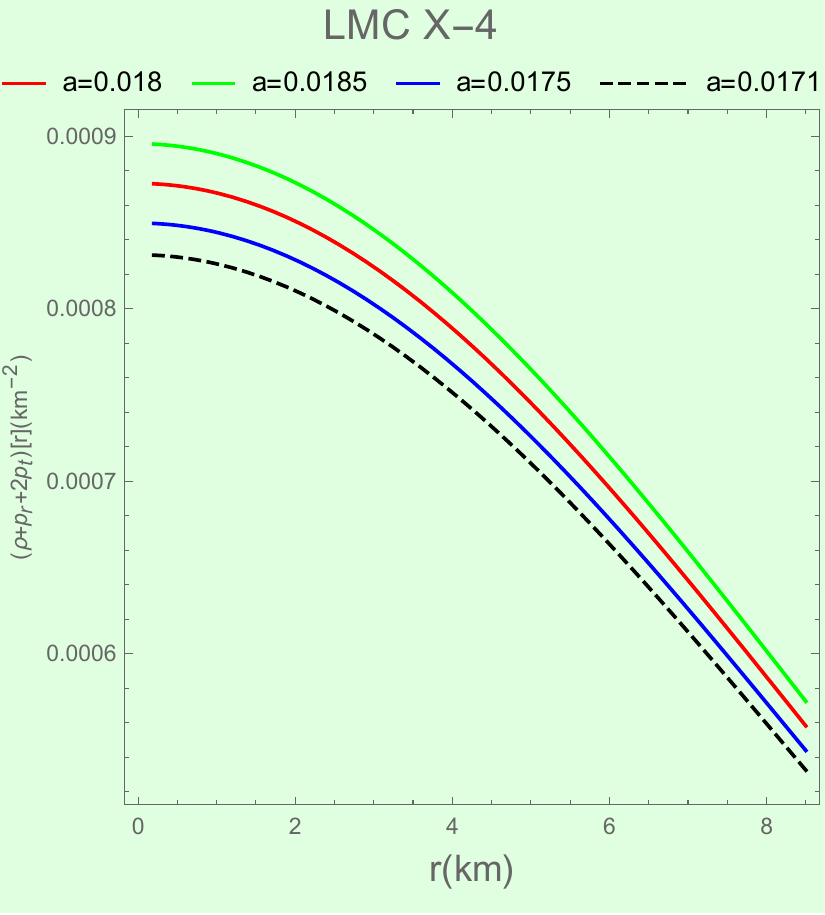}
        \small (a) LMC X-4
    \end{minipage}
    \hfill
    \begin{minipage}{0.24\textwidth}
        \centering
        \includegraphics[width=\linewidth]{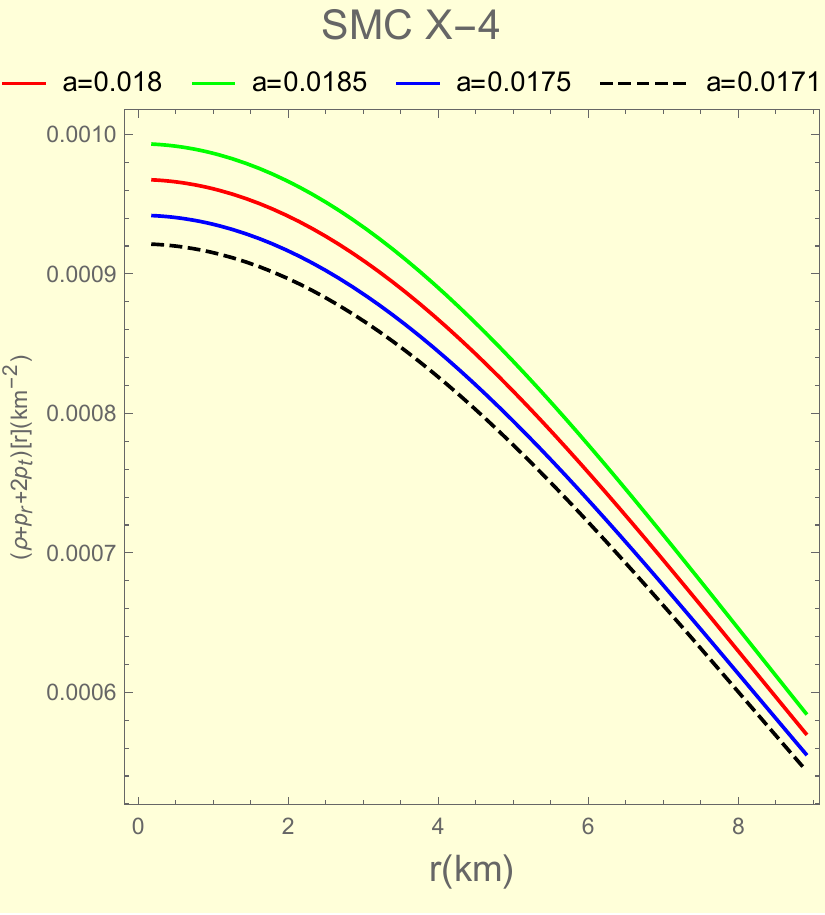}
        \small (b) SMC X-4
    \end{minipage}
    \hfill
    \begin{minipage}{0.24\textwidth}
        \centering
        \includegraphics[width=\linewidth]{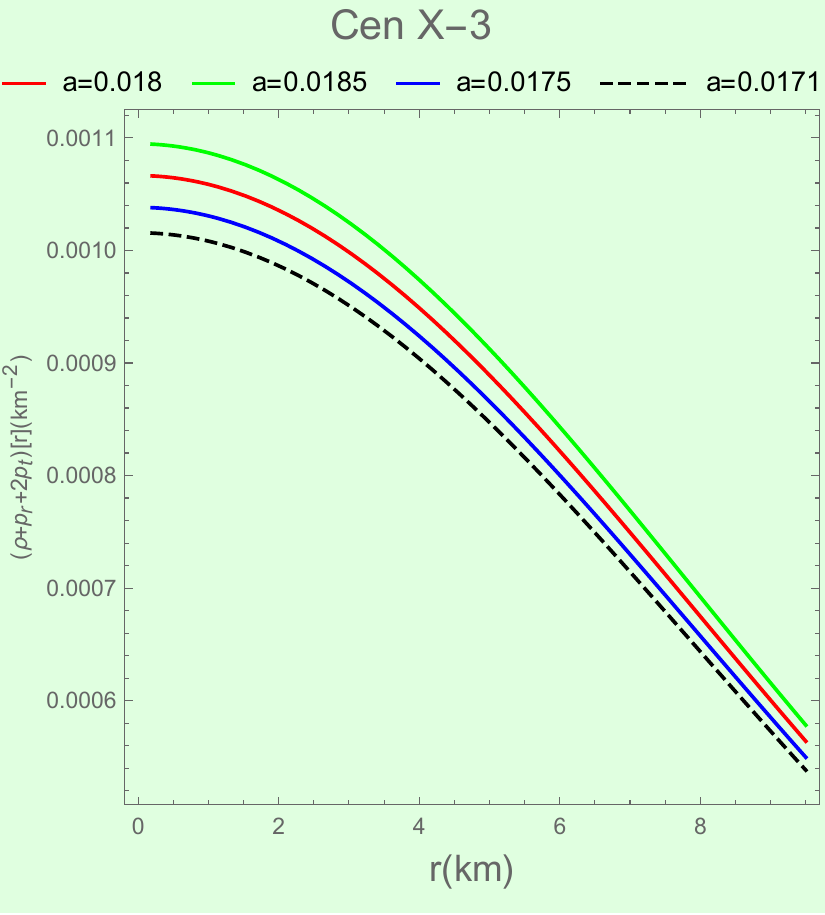}
        \small (c) Cen X-3
    \end{minipage}
    \hfill
    \begin{minipage}{0.24\textwidth}
        \centering
        \includegraphics[width=\linewidth]{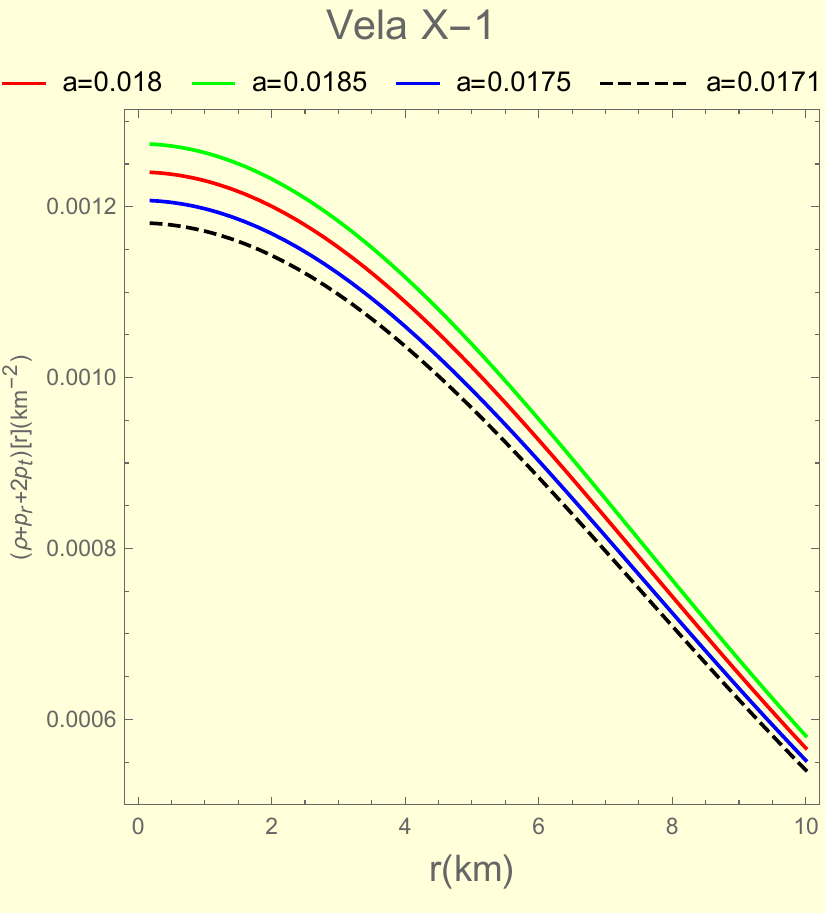}
        \small (d) Vela X-1
    \end{minipage}

    \caption{Evolution of $\rho + p_r + 2p_t$ with respect to the radial coordinate $r$ for (a) LMC X-4, (b) SMC X-4, (c) Cen X-3, and (d) Vela X-1.}
    
    \label{fig:sec3}
\end{figure}

Graphical representations for the energy conditions are depicted in Fig \ref{fig:sec1}, Fig \ref{fig:sec2}, and Fig \ref{fig:sec3}.

Fig \ref{fig:sec1}, Fig \ref{fig:sec2}, and Fig \ref{fig:sec3} show the graphical representation of Null Energy Condition (NEC) and Strong Energy Condition (SEC) for the pulsars LMC X-4, SMC X-4, Cen X-3, and Vela X-1. Here we can observe the energy conditions are preserved within the compact stars for our chosen value of $a$. This show that the proposed $f(Q)$ model is plausible.
By verifying the NEC it can be concluded that, an observer crossing a null diagram is implied by this energy condition to characterize the typical matter density as non-negative. According to SEC, the tidal tensor trace that the relating observers test is consistently positive.

\subsection{Equation of state parameter}
 
Additionally, we have examined the viability of charged matter and whether dark matter can be inferred from certain limitations that apply to EoS components such as $\omega_r$ and $\omega_t$ \cite{bib64,bib65,bib66,bib67}. By solving the field equations, the radial pressure and matter density are assumed to be linearly connected in the model; however, the transverse pressure and matter density relationship is still unknown.
The two dimensionless numbers known as the equation of state parameters, represented by $\omega_r$ and $\omega_t$, are utilized to describe the connection between pressure and matter density. 
Simplified form of EoS parameter for our study is
\begin{equation}
\omega _r=\frac{p_r}{\rho}=\frac{b e^{A r^2} r^2+a \left(2-2 e^{A r^2}+4 B r^2\right)}{-2 b e^{A r^2} r^2+4 a \left(-1+e^{A r^2}+2 A r^2\right)},
\label{E036}
\end{equation}
\begin{equation}
\omega _t=\frac{r^2 \left(-b e^{A r^2}+2 a \left(A+A B r^2-B \left(2+B r^2\right)\right)\right)}{2 b e^{A r^2} r^2-4 a \left(-1+e^{A r^2}+2
A r^2\right)}
\label{E037}
\end{equation}

\begin{figure}[!ht]
    \centering

    \begin{minipage}{0.24\textwidth}
        \centering
        \includegraphics[width=\linewidth]{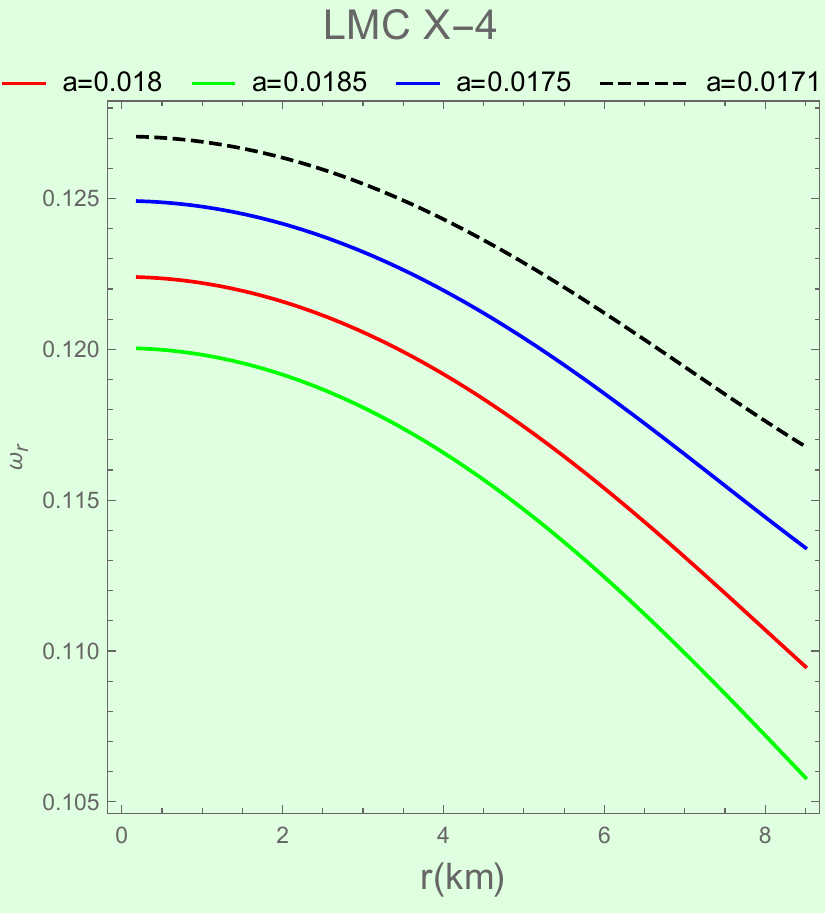}
        \small (a) LMC X-4
    \end{minipage}
    \hfill
    \begin{minipage}{0.24\textwidth}
        \centering
        \includegraphics[width=\linewidth]{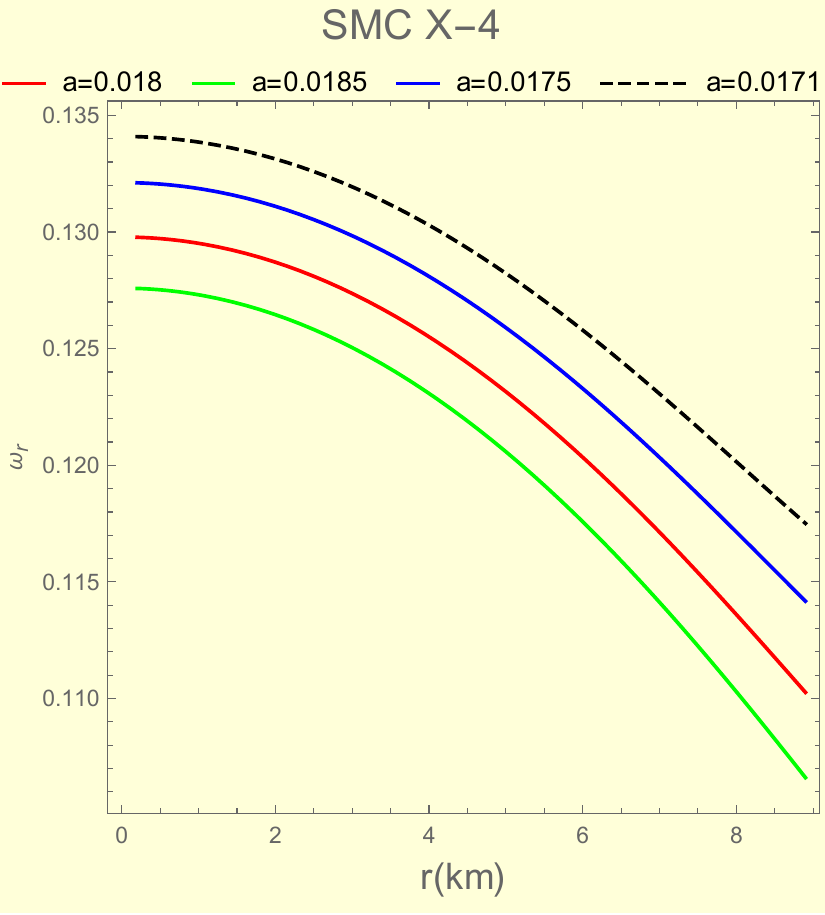}
        \small (b) SMC X-4
    \end{minipage}
    \hfill
    \begin{minipage}{0.24\textwidth}
        \centering
        \includegraphics[width=\linewidth]{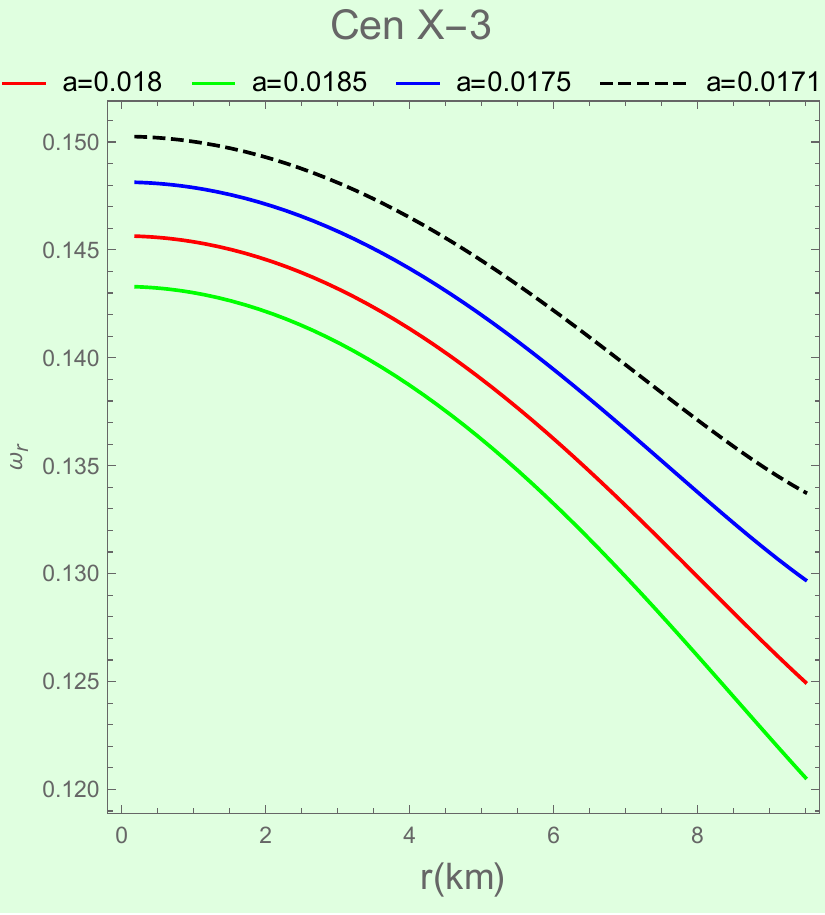}
        \small (c) Cen X-3
    \end{minipage}
    \hfill
    \begin{minipage}{0.24\textwidth}
        \centering
        \includegraphics[width=\linewidth]{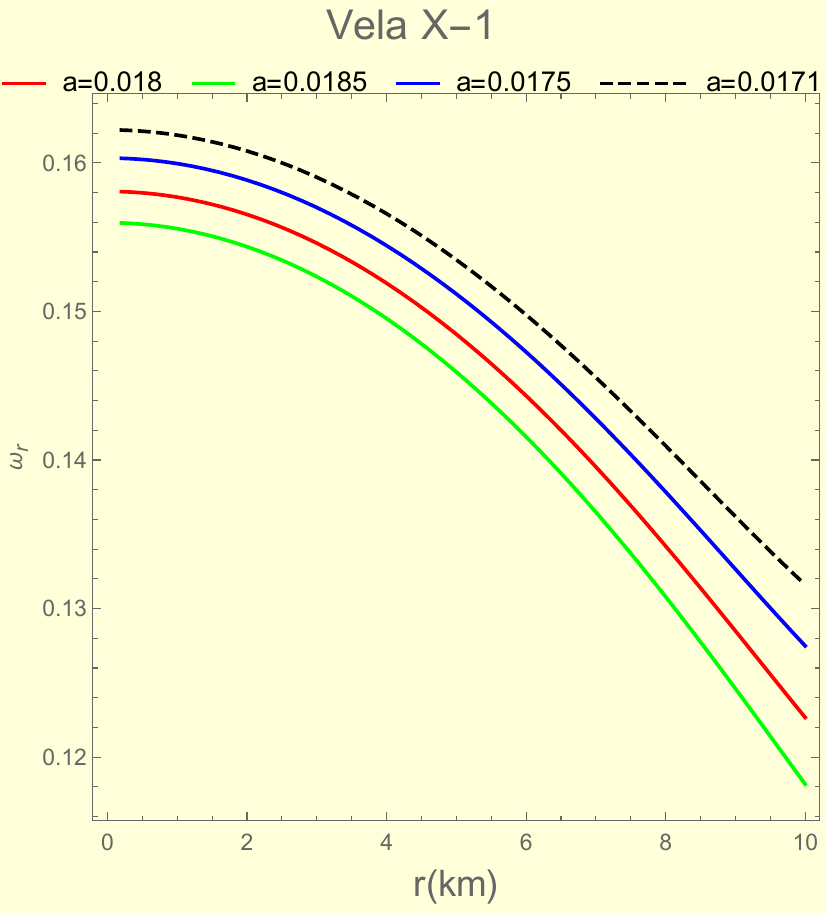}
        \small (d) Vela X-1
    \end{minipage}

    \caption{Evolution of the EoS parameter for the radial component with respect to the radial coordinate $r$ for (a) LMC X-4, (b) SMC X-4, (c) Cen X-3, and (d) Vela X-1.}  
    \label{fig:eosr}
\end{figure}

\begin{figure}[!ht]
    \centering

    \begin{minipage}{0.24\textwidth}
        \centering
        \includegraphics[width=\linewidth]{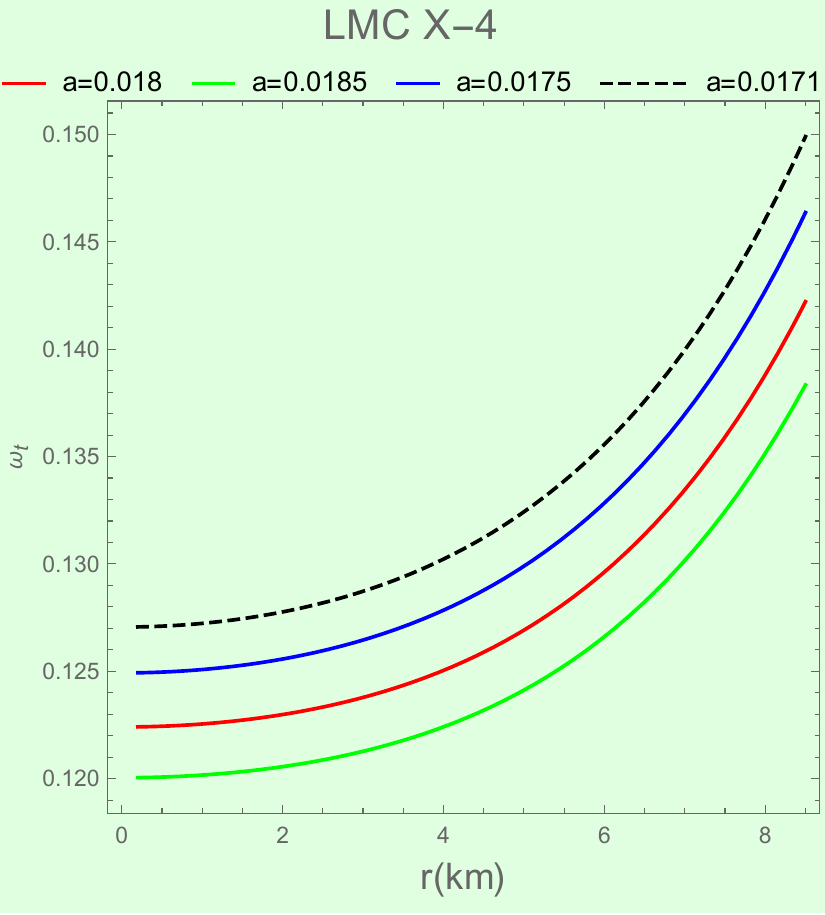}
        \small (a) LMC X-4
    \end{minipage}
    \hfill
    \begin{minipage}{0.24\textwidth}
        \centering
        \includegraphics[width=\linewidth]{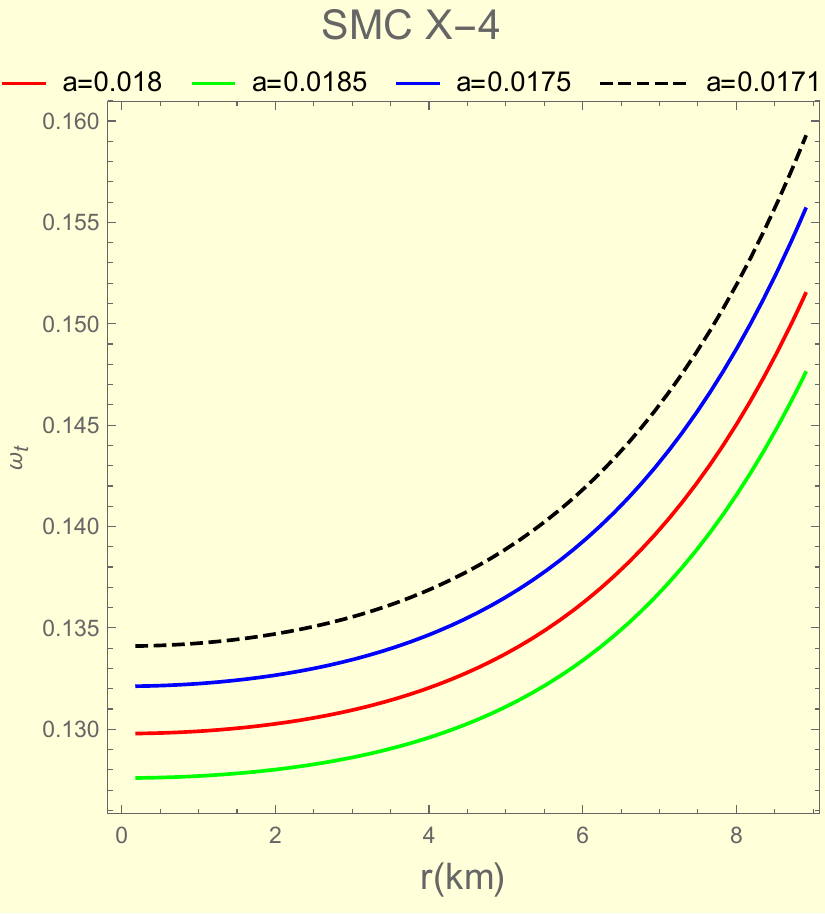}
        \small (b) SMC X-4
    \end{minipage}
    \hfill
    \begin{minipage}{0.24\textwidth}
        \centering
        \includegraphics[width=\linewidth]{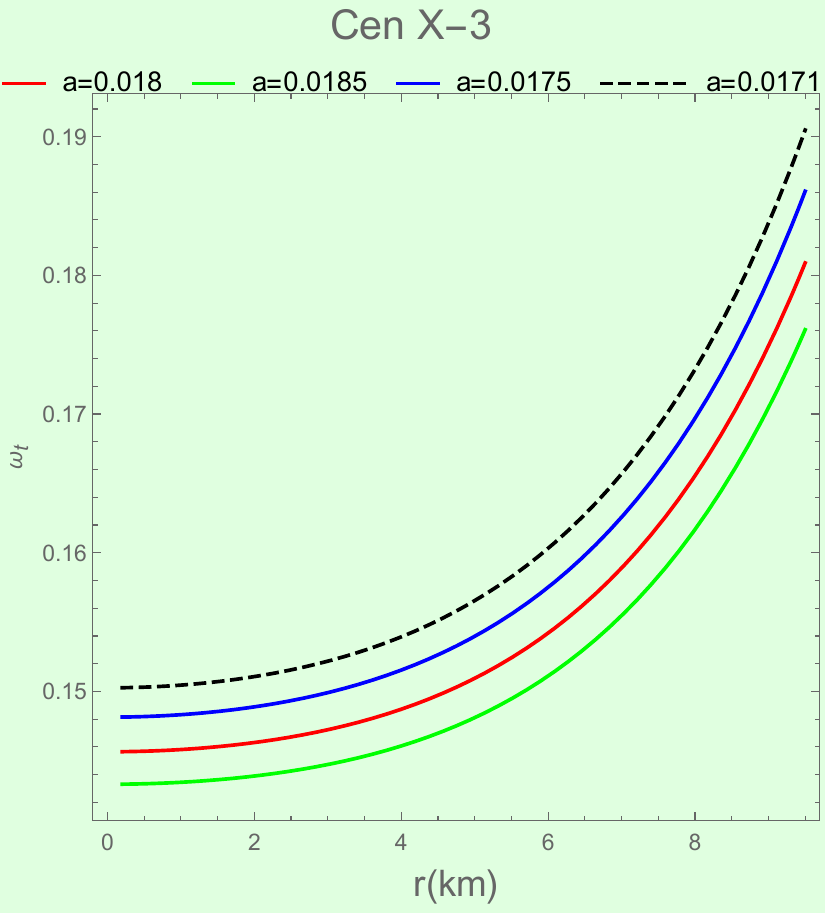}
        \small (c) Cen X-3
    \end{minipage}
    \hfill
    \begin{minipage}{0.24\textwidth}
        \centering
        \includegraphics[width=\linewidth]{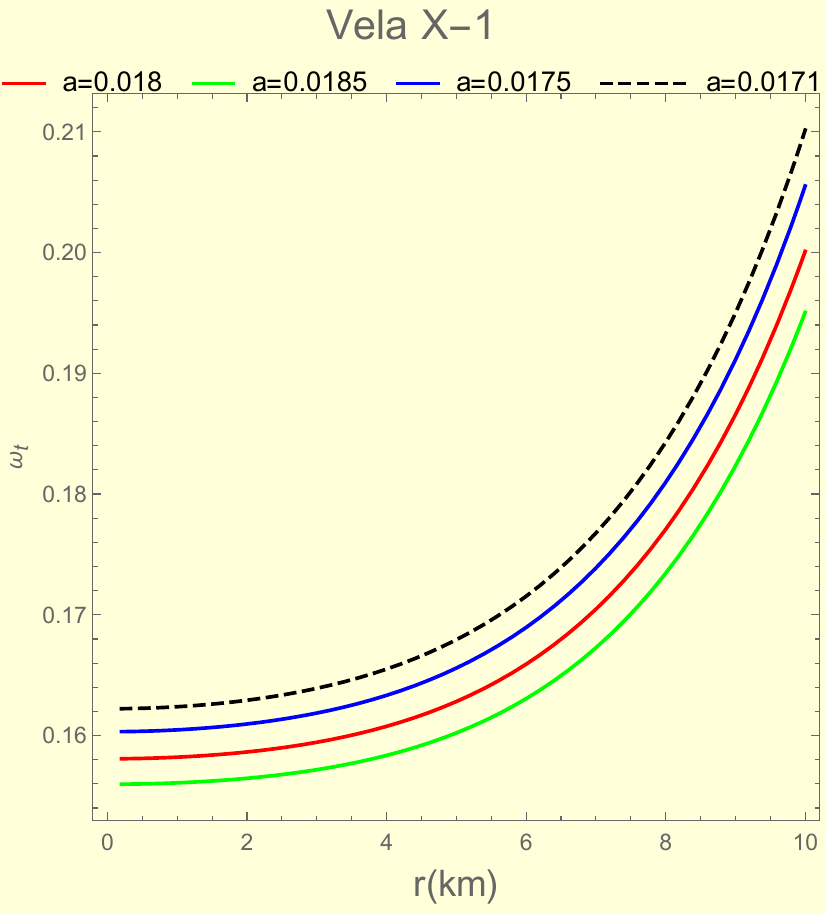}
        \small (d) Vela X-1
    \end{minipage}

    \caption{Evolution of the EoS parameter for the tangential component with respect to the radial coordinate $r$ for (a) LMC X-4, (b) SMC X-4, (c) Cen X-3, and (d) Vela X-1.}
    
    \label{fig:eost}
\end{figure}

Graphical representations of $\omega_r$ and $\omega_t$ from Eq. (\ref{E036}) and Eq. (\ref{E037}) respectively are depicted in Fig \ref{fig:eosr} and Fig \ref{fig:eost}.

Fig \ref{fig:eosr}, and Fig \ref{fig:eost} show the graphical representation of EoS parameter in radial and transverse direction for the pulsar LMC X-4, SMC X-4, Cen X-3, and Vela X-1. Anisotropic charged fluids have a positive equation of state that is contained inside $0< \omega_r , \omega_t < 1$, meaning that they are not exotic in any way. The Eos parameter decreases monotonically and vanishes at the star's surface \cite{bib17}. 

\subsection{Causality and Stability analysis}

The stability of a compact star model can be verified using various methods. Among them  $(i)$ the causality condition and $(ii)$ the adiabatic index has a great impact.
In astrophysical studies it is of much significance to analyze systems that are physically viable. The study of stability of stellar configurations are very critical in this context. In a notable study \cite{bib1} discussed the stability for anisotropic stellar configurations generated by gravitational decoupling with vanishing complexity factor in the framework of $f(Q)$ gravity. In this connection let us mention another study by \cite{bib22} where the stability condition  of hybrid star was investigated in the context of $f(Q)$ gravity. Given this works we have analyzed the stability of our model in a similar way and the same is demonstrated below.\\

\subsubsection{The Causality Condition}

Utilizing the "cracking" technique, which has to do with how stable anisotropic stars are under tiny radial perturbations The regions of an anisotropic star that are possibly stable are those where the radial speed of sound crosses the transverse speed of sound, as demonstrated by Herrera \cite{bib18} and Abreu et al. \cite{bib19}. They discovered that for configurations of anisotropic matter, changes in density alone do not trigger the system to lose equilibrium. Such departures are only induced by disturbances of both density and local anisotropy. The integration of Einstein's equation yielded the expansion, collapse, overturning, and cracking. Hence, the permissible limit for the square speed of sound is $0<v^2 _{sr},v^2 _{st}< 1$. Mathematical simplification for square speed of sound in radial and transverse direction for our model is
\begin{equation}
v_{sr}^2=\frac{{dp}_r}{{d\rho }}=\frac{1-e^{A r^2}+A \left(r^2+2 B r^4\right)}{2 \left(-1+e^{A r^2}-A r^2+2 A^2 r^4\right)}
\label{E038}
\end{equation} 
\begin{equation}
v_{st}^2=\frac{{dp}_t}{{d\rho }}=-\frac{r^4 \left(B^2+A^2 \left(1+B r^2\right)-A B \left(3+B r^2\right)\right)}{2 \left(-1+e^{A r^2}-Ar^2+2 A^2 r^4\right)}
\label{E039}
\end{equation} 
As the expression square speed of sound in radial and transversal direction can derived as $v_{sr}^2= \frac{dp_r}{dr}=\frac{dp_r}{dr} \frac{dr}{d\rho}$ and $v_{st}^2= \frac{dp_t}{dr}=\frac{dp_t}{dr} \frac{dr}{d\rho}$, we have evaluated Eq. (\ref{E038}) and Eq. (\ref{E039}) using  Eq. (\ref{E032}),  Eq. (\ref{E033}), and  Eq. (\ref{E034}).

\begin{figure}[!ht]
    \centering

    \begin{minipage}{0.42\textwidth}
        \centering
        \includegraphics[width=\linewidth]{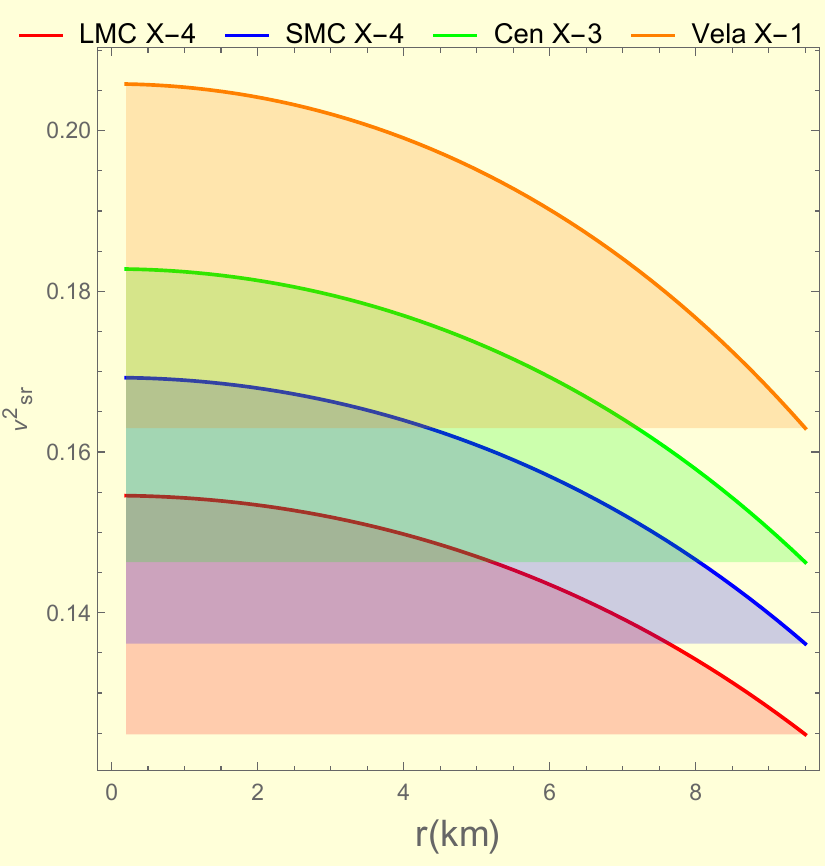}
        \small (a) Radial direction
    \end{minipage}
    \hfill
    \begin{minipage}{0.42\textwidth}
        \centering
        \includegraphics[width=\linewidth]{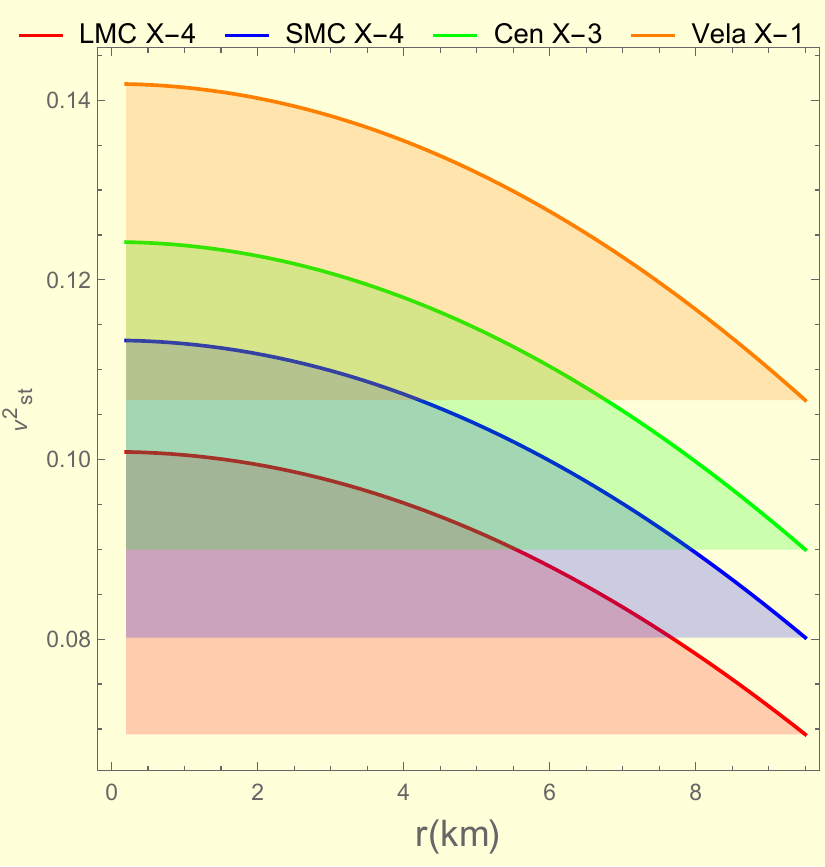}
        \small (b) Tangential direction
    \end{minipage}

    \caption{Evolution of the squared speed of sound with respect to the radial coordinate $r$: (a) radial direction and (b) tangential direction for LMC X-4, SMC X-4, Cen X-3, and Vela X-1.}
    
    \label{fig:sound}
\end{figure}

Here we can observe that the square speed of sound in radial and transverse direction is not affected by the integrating constant $a$ and $b$. 
Graphical representation of $v^2$ in radial and transverse direction from  Eq. (\ref{E038}) and Eq. (\ref{E039}) are depicted in Fig \ref{fig:sound}.

Fig \ref{fig:sound} shows the graphical representation of the square speed of sound in radial and transverse direction for the pulsar LMC X-4, SMC X-4, Cen X-3, and Vela X-1. We can observe for all four compact objects square speed of sound lies within the permissible range, hence our model does not violate the cracking criteria and is potentially stable under this background.

\subsubsection{Adiabatic Index}

The adiabatic index $\Gamma=\left(\frac{\rho +p_r}{p_r}\right)v_{sr}^2$, which is the ratio of the two specific temperatures, indicates the stiffness of the equation of state for a given density. The dynamical stability of the star structure under an infinitesimal radial adiabatic impact can be examined using this adiabatic index $\Gamma$, which should be greater than 4/3 in the interior region \cite{bib21}. This limiting value was created by Chandrashekhar to assess the dynamical stability of an isotropic relativistic star with spherical symmetry in the presence of a tiny radial adiabatic perturbation \cite{bib20}.
The mathematical form of the adiabatic index for our model is
\begin{equation}
\Gamma =\frac{\left(b e^{A r^2} r^2-2 a \left(-1+e^{A r^2}+4 A r^2+2 B r^2\right)\right) \left(-1+e^{A
r^2}-A \left(r^2+2 B r^4\right)\right)}{2 \left(-1+e^{A r^2}-A r^2+2 A^2 r^4\right) \left(b e^{A r^2} r^2+a \left(2-2 e^{A r^2}+4 B r^2\right)\right)}
\label{E040}
\end{equation}

\begin{figure}[!ht]
    \centering

    \begin{minipage}{0.24\textwidth}
        \centering
        \includegraphics[width=\linewidth]{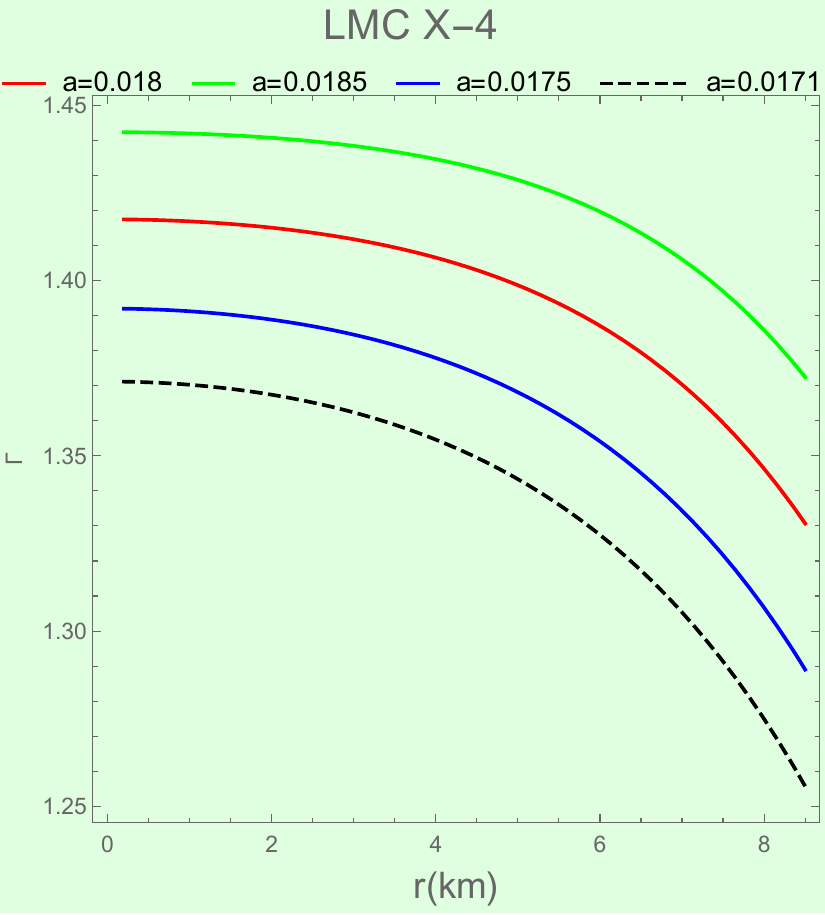}
        \small (a) LMC X-4
    \end{minipage}
    \hfill
    \begin{minipage}{0.24\textwidth}
        \centering
        \includegraphics[width=\linewidth]{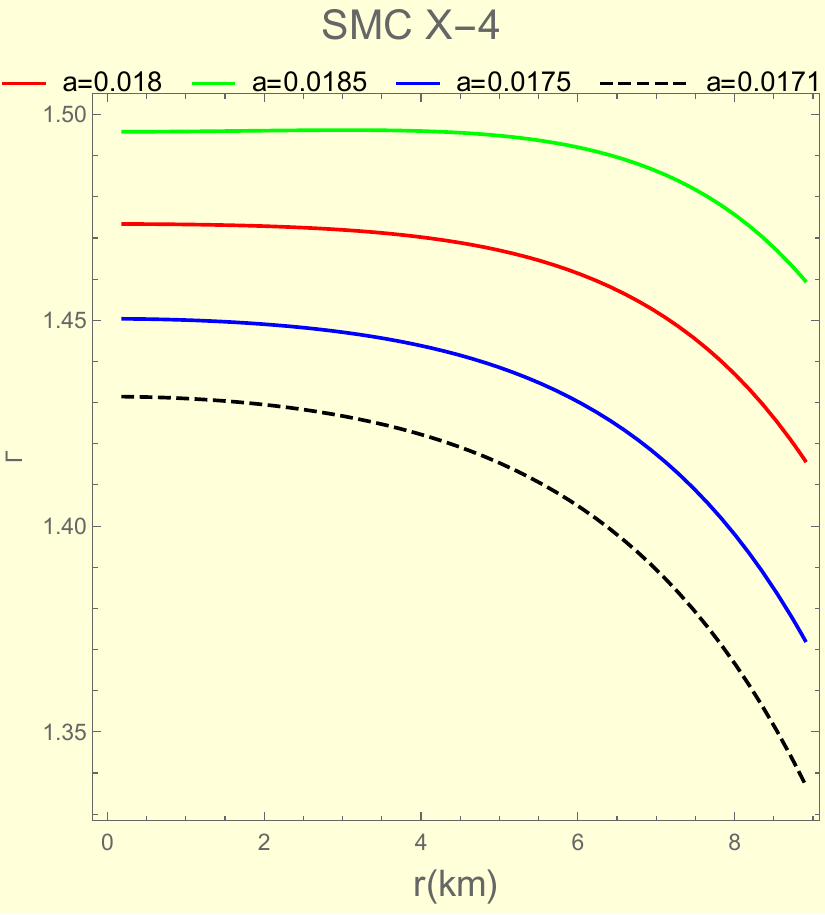}
        \small (b) SMC X-4
    \end{minipage}
    \hfill
    \begin{minipage}{0.24\textwidth}
        \centering
        \includegraphics[width=\linewidth]{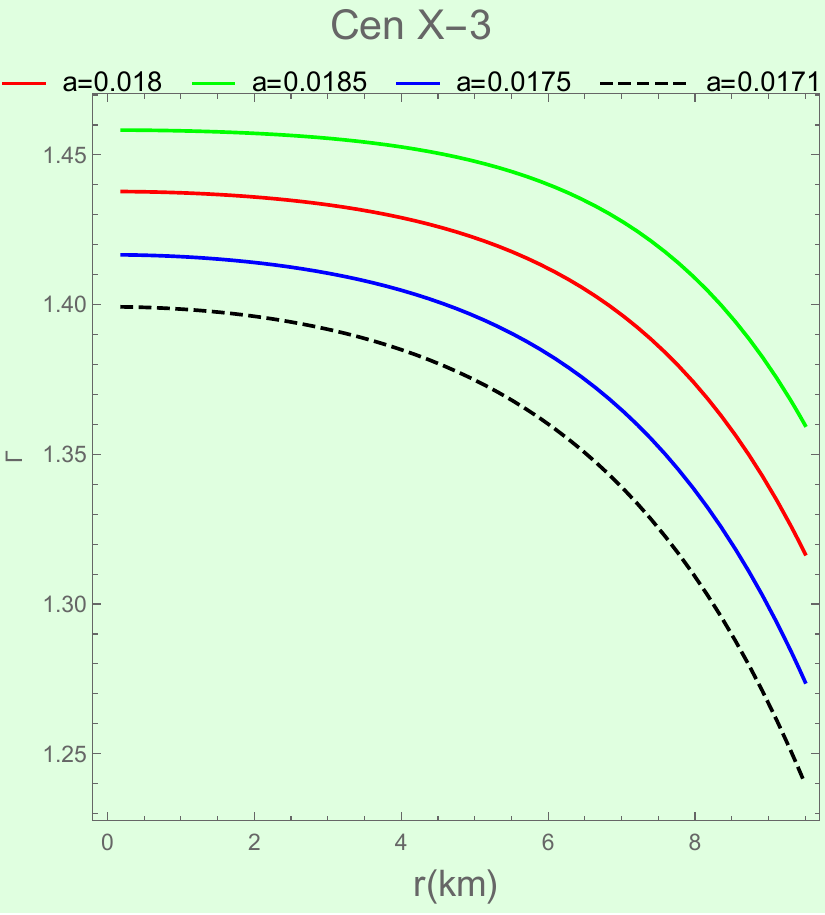}
        \small (c) Cen X-3
    \end{minipage}
    \hfill
    \begin{minipage}{0.24\textwidth}
        \centering
        \includegraphics[width=\linewidth]{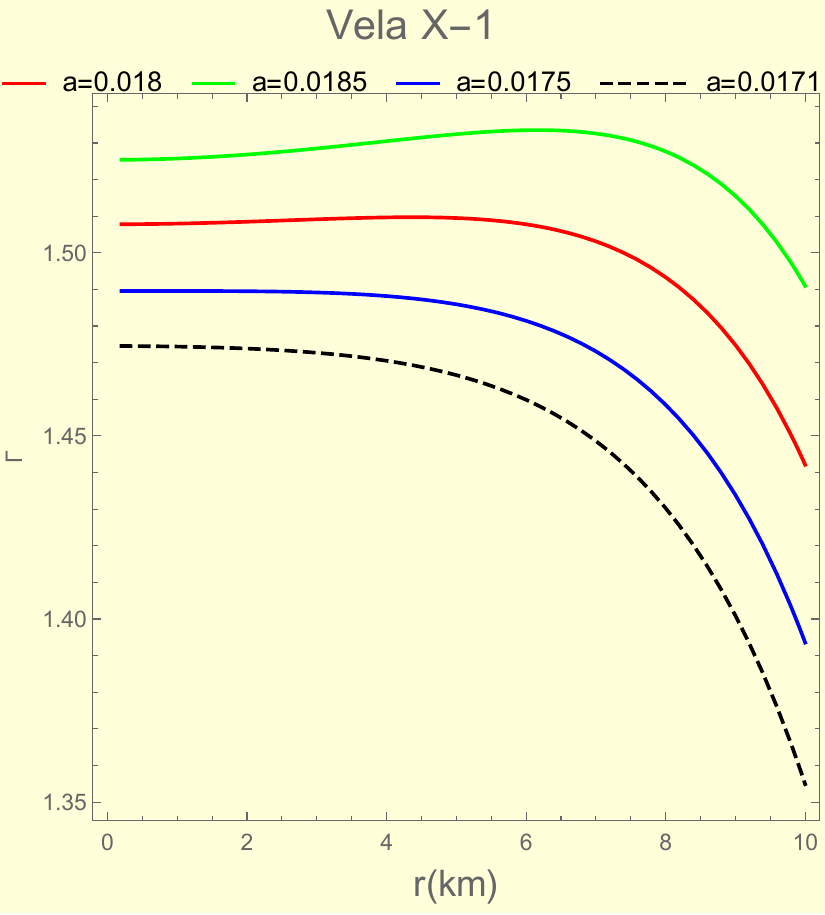}
        \small (d) Vela X-1
    \end{minipage}

    \caption{Evolution of the adiabatic index with respect to the radial coordinate $r$ for (a) LMC X-4, (b) SMC X-4, (c) Cen X-3, and (d) Vela X-1.}
    
    \label{fig:gamma}
\end{figure}

Graphical representation of $\Gamma$ for LMC X-4, SMC X-4, Cen X-3, and Vela X-1 for Eq. (\ref{E040}) is depicted in Fig \ref{fig:gamma}.

From Fig \ref{fig:gamma} it is shown that at the surface of perturbation, all the compact objects have an adiabatic index value larger than 4/3.
Hence, our model is stable by the causality condition as well as the dynamic stability condition for the compact objects.
In this connection let us note that according to Bondi \cite{revi2}, the model's instability is for $\Gamma $ less than $4/3$. Heintzmann and Hillebrandt \cite{revi3} investigated whether the compact object satisfies the relativistic stability condition when $\Gamma $ is more than 4/3 and there is an increasing, positive anisotropy factor. Hence the model demonstrated above is consistent with \cite{revi2}, \cite{revi3}.

\subsection{Equillibrium Condition}

The gravitational force, the hydrostatic force, and the anisotropic force are the three forces that the Tolmann- Oppenheimer- Volkoff (TOV) equation \cite{bib62,bib63} expresses in order to define the equilibrium condition of the stellar structure. We propose to determine if the three forces acting on our current star system—(i)hydrostatic force $F_h$, (ii)gravitational force $F_g$, and (iii)anisotropic force $F_a$—are in a stable equilibrium state using the TOV equation. The positive anisotropic factor in Fig \ref{fig:ani} induce positive repulsive anisotropic force. On the other hand, the negative gravitational force will be balanced by the combined action of the hydrostatic force and the anisotropic force. The TOV equation gives
\begin{equation}
-\frac{\mu '}{2} (\rho + p_{r}) - \frac{dp_{r}}{dr} + \frac{2}{r} (p_{t}-p_{r})=0 .
    \label{tov}
\end{equation}
In Eq. (\ref{tov}) the required force components are expressed as
\begin{equation} 
\begin{array}{cc}
    F_{g}=-\frac{\mu '}{2} (\rho + p_{r}) = \frac{A \left(-2 a+b r^2+e^{-A r^2} \left(2 a-4 a (2 A+B) r^2\right)\right)}{2 r} ; \\
    F_{h}=  - \frac{dp_{r}}{dr}= \frac{2 a \left(-1+e^{-A r^2} \left(1+A \left(r^2+2 B r^4\right)\right)\right)}{r^3} ;\\
    F_{a}= \frac{2}{r} (p_{t}-p_{r})= \frac{2 a e^{-A r^2} \left(-1+e^{A r^2}-A r^2+B (-A+B) r^4\right)}{r^3} .
\end{array}
    \label{force}
\end{equation}

\begin{figure}[!ht]
    \centering

    \begin{minipage}{0.24\textwidth}
        \centering
        \includegraphics[width=\linewidth]{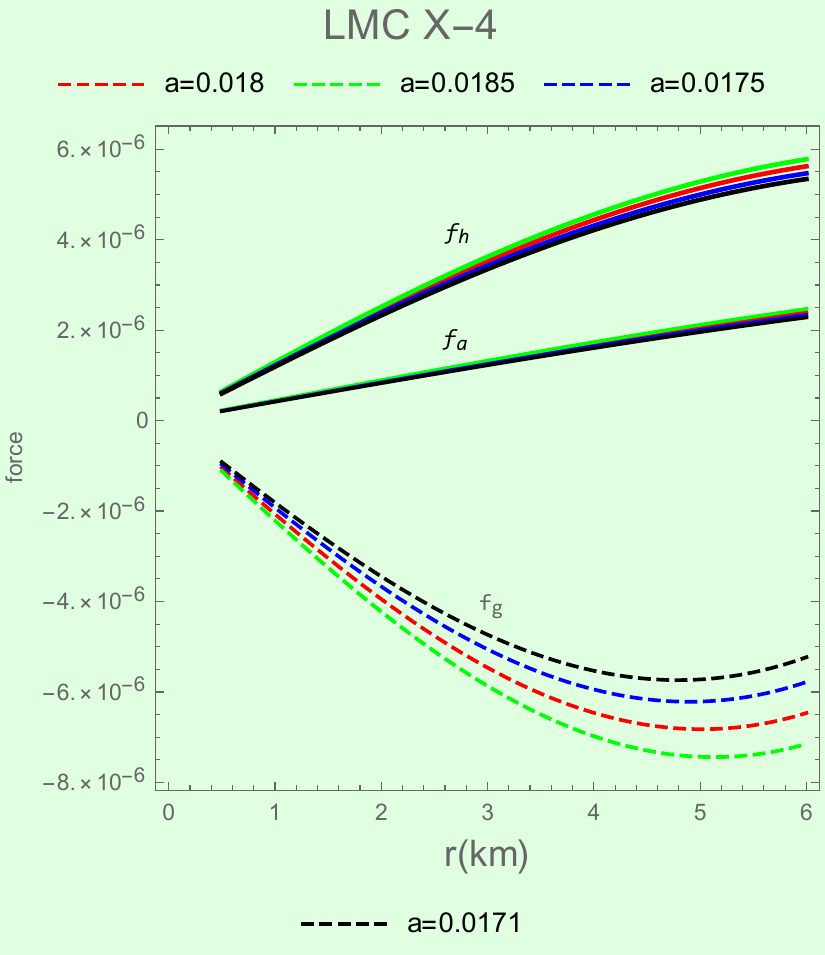}
        \small (a) LMC X-4
    \end{minipage}
    \hfill
    \begin{minipage}{0.24\textwidth}
        \centering
        \includegraphics[width=\linewidth]{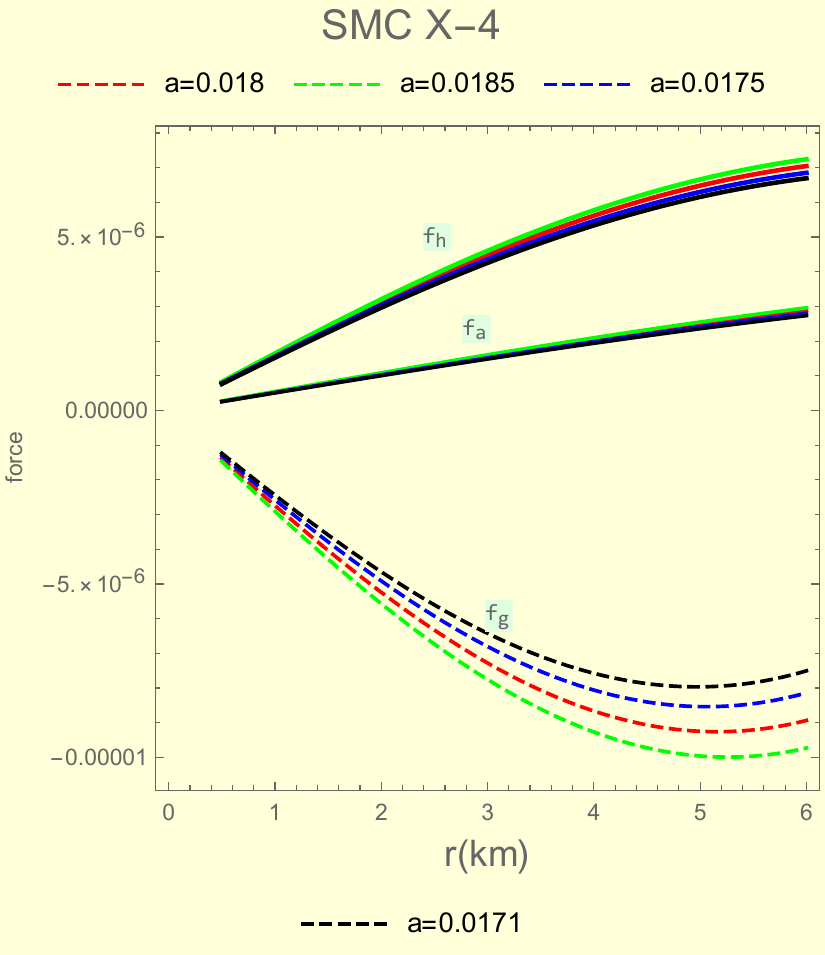}
        \small (b) SMC X-4
    \end{minipage}
    \hfill
    \begin{minipage}{0.24\textwidth}
        \centering
        \includegraphics[width=\linewidth]{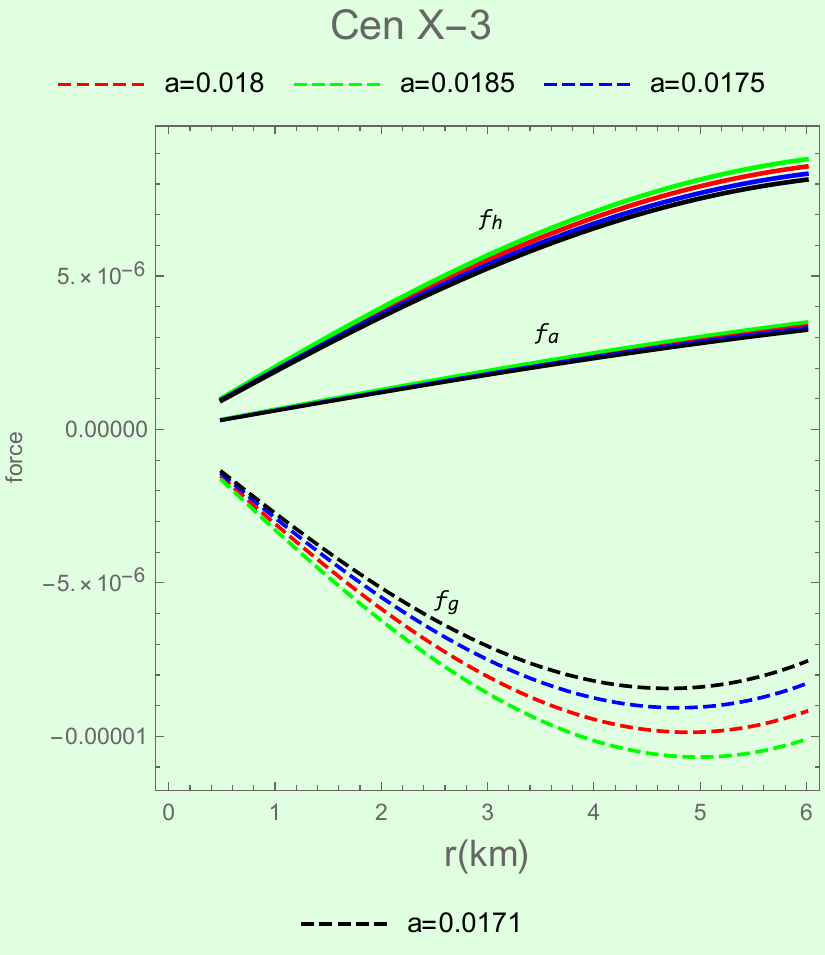}
        \small (c) Cen X-3
    \end{minipage}
    \hfill
    \begin{minipage}{0.24\textwidth}
        \centering
        \includegraphics[width=\linewidth]{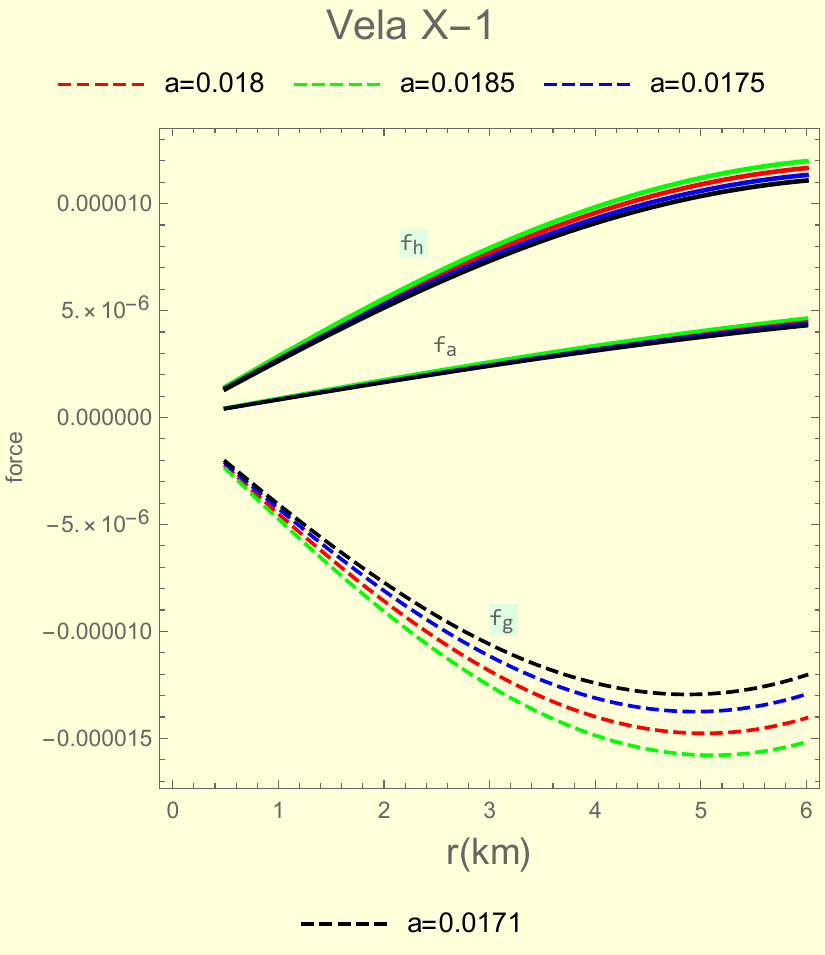}
        \small (d) Vela X-1
    \end{minipage}

    \caption{Evolution of the force components with respect to the radial coordinate $r$ for (a) LMC X-4, (b) SMC X-4, (c) Cen X-3, and (d) Vela X-1.}
    
    \label{fig:force}
\end{figure}

The equilibrium of the system under three different forces is shown in Fig. \ref{fig:force}. Here we can observe that the dotted lined expressed the anisotropic force, the solid bold lines represent the hydrostatic force and the solid lines represent the gravitational force. For different parametric values we have depicted the force components for four realistic compact objects. In each case we can observe that the hydrastatic force and the anisotropic force have positive inclination, whereas gravitational force is balancing the joint action of other two major forces.
Three forces work together to generate an equilibrium system for each compact object, even if the anisotropic force has less of an influence.

\section{Mass-Radius Relation}

Mass function of a neutron star is defined by \cite{bib22}
\begin{equation}
m=\int _0^r4 \pi  x^2\rho (x) dx.
\label{E041}
\end{equation}
Now from Eq. (\ref{E010}) and Eq. (\ref{E041}) mass function derived as
\begin{equation}
m=\frac{4}{3} \pi  r \left(a \left(6-6 e^{-A r^2}\right)-b r^2\right)
\label{E042}
\end{equation}
We have determined the mass and compared it with the observable mass for the aforementioned compact items by maintaining the constraints derived for $a$ and $b$ for distinct compact objects. The derived constraints have shown in Table \ref{Table:1}.

 \begin{table}[!h]
\begin{center}
\begin{tabular}{||c| c| c| c| c| c| c| c ||} 
 \hline
 Sl. No. & Star Model &  Estimated Radius & Estimated Mass & Constraint & Derived Mass & Compactness  & Surface Redshift  \\[0.5ex] 
   &   & R $(km)$ & $M(M_{\odot})$ &  & $M(M_{\odot})$ & & \\ 
 \hline\hline
 1 & LMC X-4 & 8.301 $\pm$ 0.2 & 0.95 - 1.13 & $a$=0.018 & 1.11001 & 0.137022 & 0.173666 \\ 
  & & & & $a$=0.0185 & 1.14641 & 0.141515 & 0.180999 \\
  & & & & $a$=0.0175 & 1.07361 & 0.132528 & 0.166468 \\
  & & & & $a$=0.0171 & 1.04449 & 0.128934 & 0.160805 \\
 \hline
1 & SMC X-4 & 8.831 $\pm$ 0.09 & 1.24 - 1.34 & $a$=0.018 & 1.43566 & 0.164244 & 0.220319 \\ 
  & & & & $a$=0.0185 & 1.48253 & 0.169607 & 0.230182 \\
  & & & & $a$=0.0175 & 1.38879 & 0.158882 & 0.210689 \\
  & & & & $a$=0.0171 & 1.35129 & 0.154592 & 0.203147 \\
 \hline
 3 & Cen X-3 & 9.178 $\pm$ 0.13 & 1.41 - 1.57 & $a$=0.018 & 1.62124 & 0.179182 & 0.248406 \\ 
  & & & & $a$=0.0185 & 1.67489 & 0.185112 & 0.260106 \\
  & & & & $a$=0.0175 & 1.56759 & 0.173252 & 0.237026 \\
  & & & & $a$=0.0171 & 1.52467 & 0.168509 & 0.228143 \\
 \hline
 4 & Vela X-1 & 9.56 $\pm$ 0.08 & 1.69 - 1.85 & $a$=0.018 & 1.97255 & 0.208075 & 0.308727 \\ 
  & & & & $a$=0.0185 & 2.03725 & 0.2149 & 0.324301 \\
  & & & & $a$=0.0175 & 1.90784 & 0.201249 & 0.293691 \\
  & & & & $a$=0.0171 & 1.85608 & 0.195789 & 0.282028 \\ [1ex] 
 \hline
\end{tabular}
\caption{\label{Table:2}Mass-Radius computation for different values of $a$ }
\end{center}
\end{table}

From Table \ref{Table:2}  we can observe the mass values for different values of $a$ for the compact objects LMC X-4, SMC X-4, Cen X-3, and Vela X-1 in $f(Q)$ modified gravity background.

\begin{figure}[!ht]
    \centering

    \begin{minipage}{0.24\textwidth}
        \centering
        \includegraphics[width=\linewidth]{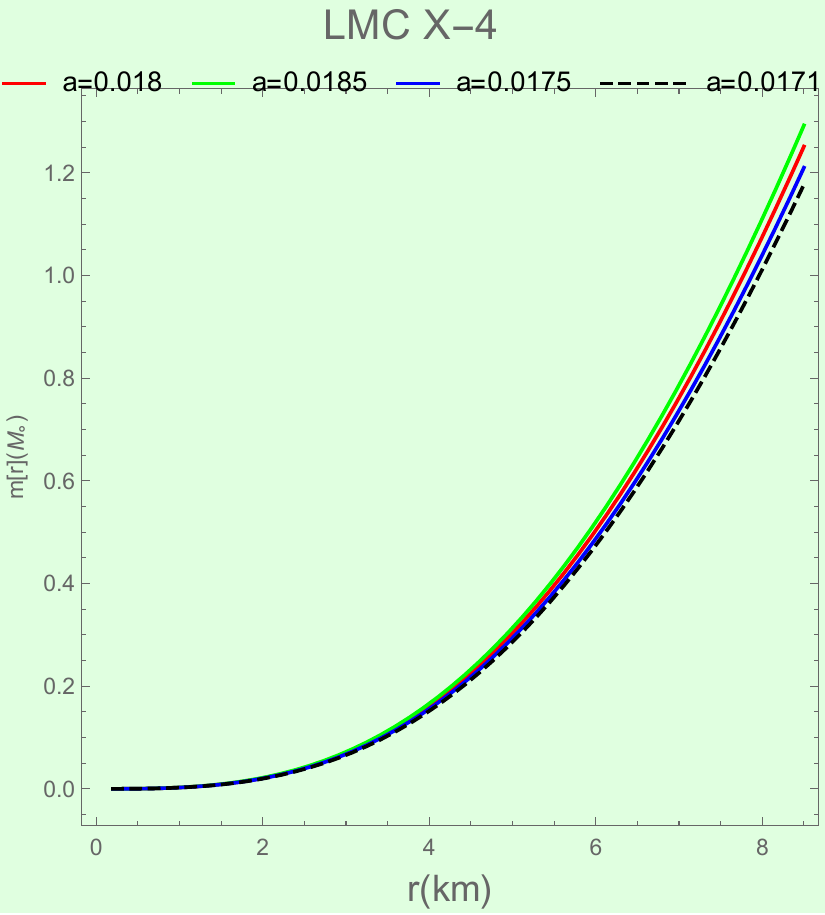}
        \small (a) LMC X-4
    \end{minipage}
    \hfill
    \begin{minipage}{0.24\textwidth}
        \centering
        \includegraphics[width=\linewidth]{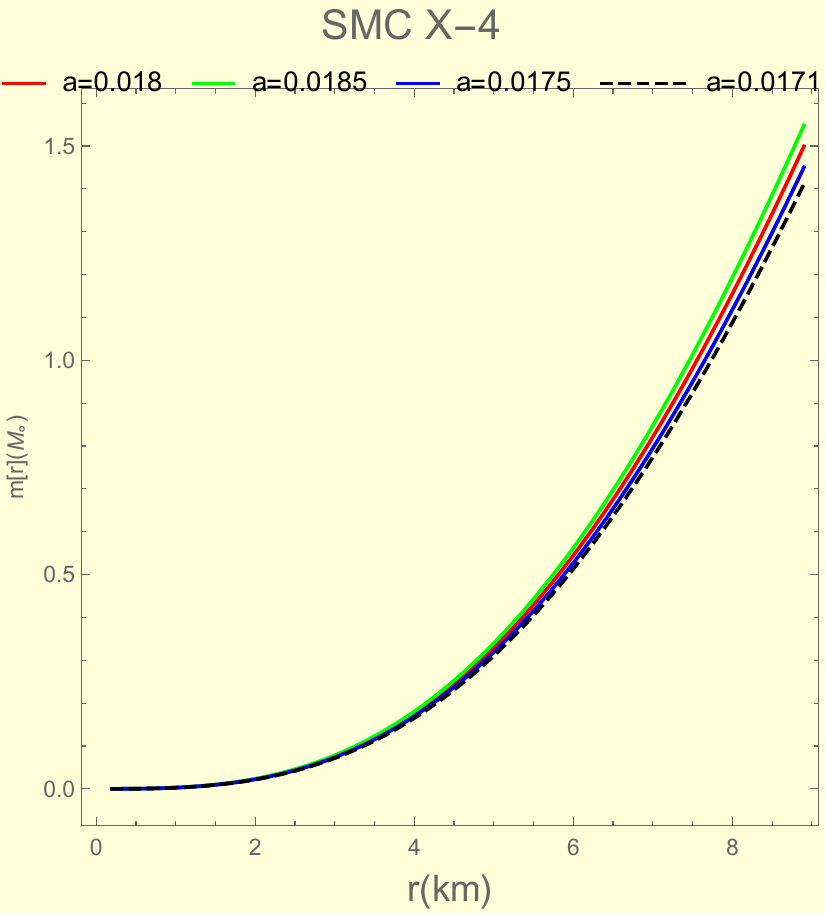}
        \small (b) SMC X-4
    \end{minipage}
    \hfill
    \begin{minipage}{0.24\textwidth}
        \centering
        \includegraphics[width=\linewidth]{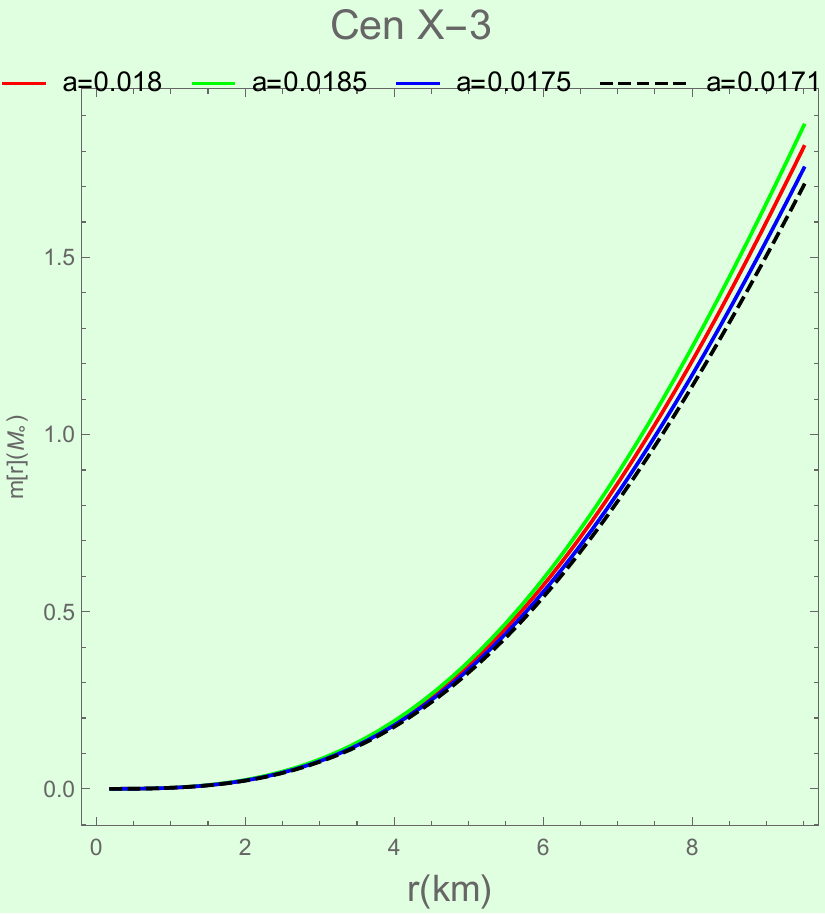}
        \small (c) Cen X-3
    \end{minipage}
    \hfill
    \begin{minipage}{0.24\textwidth}
        \centering
        \includegraphics[width=\linewidth]{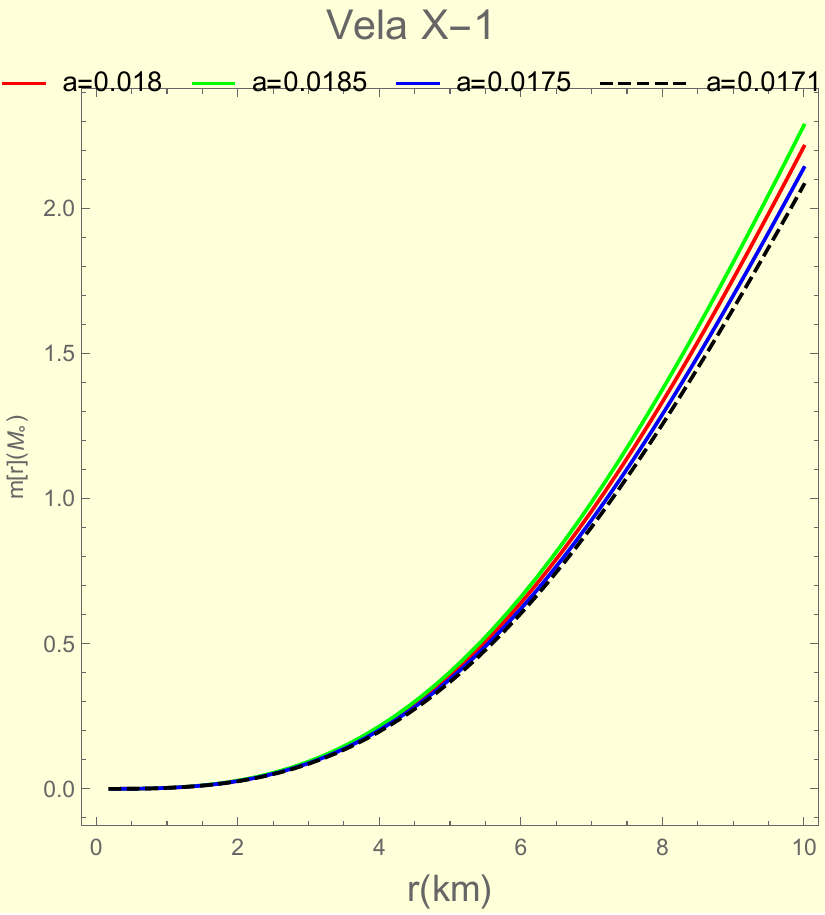}
        \small (d) Vela X-1
    \end{minipage}

    \caption{Evolution of the mass with respect to the radial coordinate $r$ for (a) LMC X-4, (b) SMC X-4, (c) Cen X-3, and (d) Vela X-1.}
    
    \label{fig:mass}
\end{figure}

Graphical representation of mass-radius for LMC X-4, SMC X-4, Cen X-3, and Vela X-1 for Eq. (\ref{E042}) is depicted in Fig \ref{fig:mass}.

Fig \ref{fig:mass} shows a mass-radius graph for the compact objects LMC X-4, SMC X-4, Cen X-3, and Vela X-1. Here we can observe mass shows an increasing pattern with respect to radius. But from Fig \ref{fig:density} we have seen density shows a decaying pattern. The explanation for the decreasing pattern of density despite rising mass could be due to an increasing rate of volume.

\begin{figure}[!ht]
    \centering

    \begin{minipage}{0.24\textwidth}
        \centering
        \includegraphics[width=\linewidth]{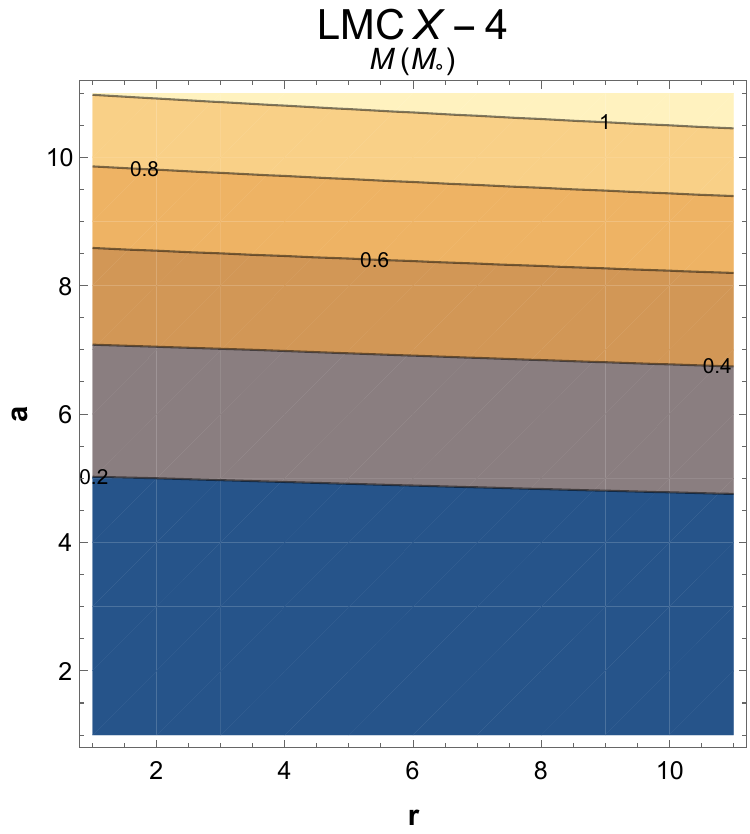}
        
    \end{minipage}
    \hfill
    \begin{minipage}{0.24\textwidth}
        \centering
        \includegraphics[width=\linewidth]{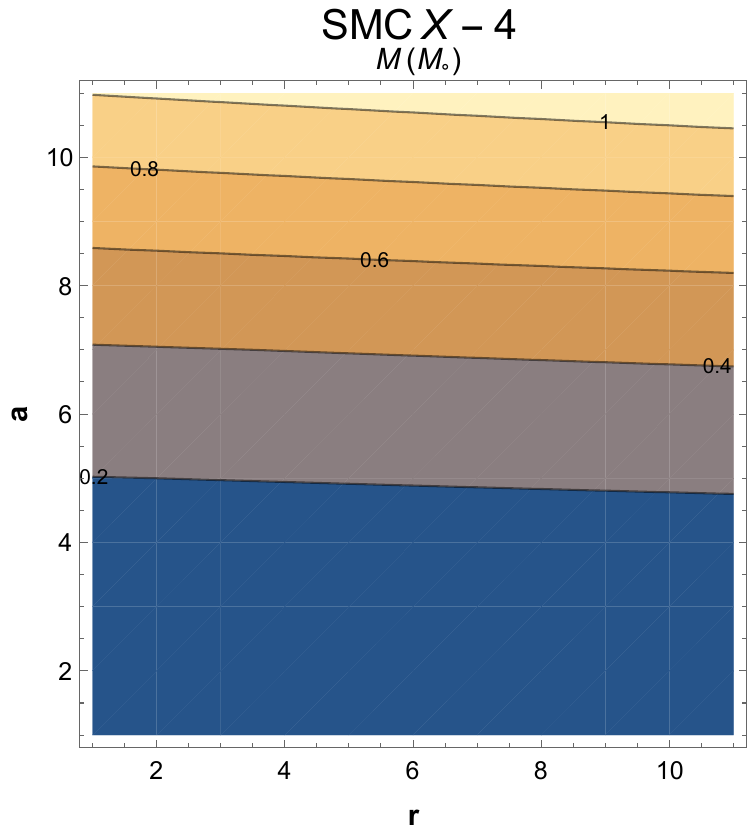}
        
    \end{minipage}
    \hfill
    \begin{minipage}{0.24\textwidth}
        \centering
        \includegraphics[width=\linewidth]{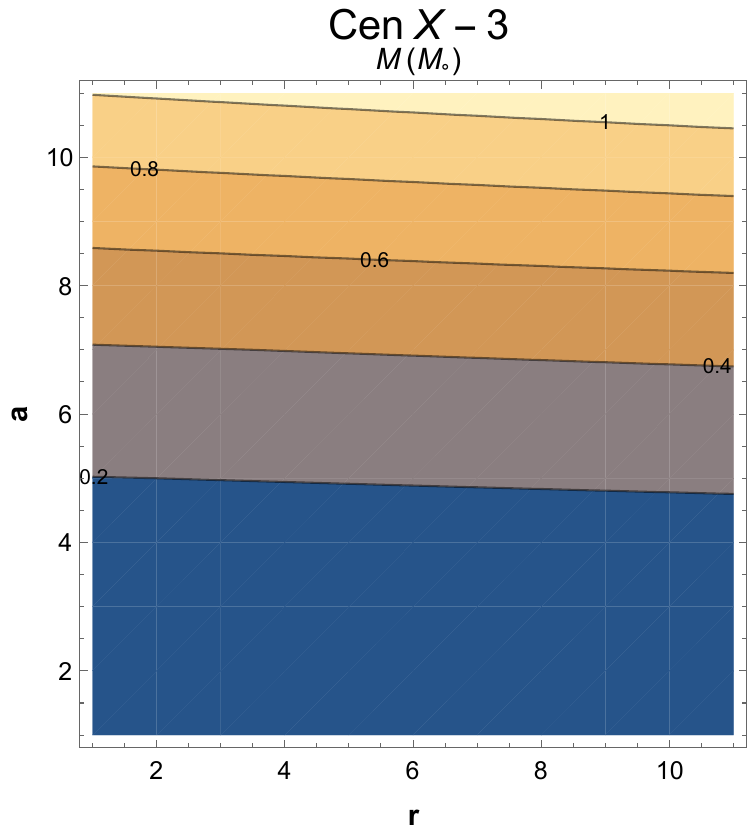}
        
    \end{minipage}
    \hfill
    \begin{minipage}{0.24\textwidth}
        \centering
        \includegraphics[width=\linewidth]{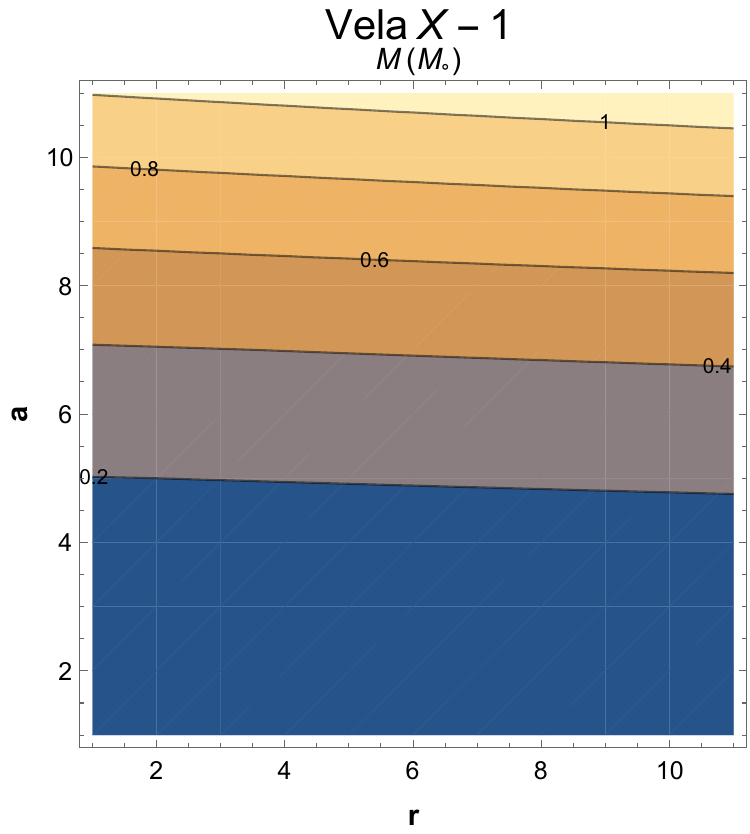}
        
    \end{minipage}

    \caption{Equi-mass diagram with respect to the radial coordinate $r$ for (a) LMC X-4, (b) SMC X-4, (c) Cen X-3, and (d) Vela X-1.}
    
    \label{fig:equi-mass}
\end{figure}
Fig \ref{fig:equi-mass} is showing the equi-mass diagram in $r-a$ plane by fixing the value $b$ at 0.0009. We can observe that if we fix $a$ and increase $r$ than the mass will be increasing. This implies the mass of the stellar objects depends on the parameter $a$.

\subsubsection{Hypothesis Testing}

In Table \ref{Table:2} we have represented four chosen values of $a$ and derived the model mass, compactness and the surface redshift for the said values. We have determined the model mass using thirty distinct measurements of $a$ to test the hypothesis. We opt for the Chi-Square test in this model.

The chi-square test is a statistical method for comparing predicted and actual results. This test aims to determine if a difference between observed and expected data is due to random variation or a relationship between the variables under investigation \cite{bib26}. We can assess the fairness, bias, and consistency between two entities with the aid of this test. The Chi-square test is a non-parametric statistic that is also referred to as a distribution-free test. Non-parametric tests ought to be applied when any of the following conditions apply to the data \cite{bib27}:

(i) A standard or ordinal range is used to evaluate each variable.

(ii) The sample sizes of the research groups are not the same; for the second, the groups may have equal or unequal sizes, although certain parametric tests need groups to have sizes that are equal or very close.

(iii) The levels (or categories) of the variables are exclusive of one another. Put otherwise, a given topic only corresponds to one level across all variables.

(iv) Each participant may enter information into a single matrix cell (two). For example, if the same people are tested again and comparisons are performed between identical participants at Time 1, Time 2, Time 3, etc., then the number two cannot be utilized.

Formula used to test this hypothesis is
\begin{equation}
    \chi^{2}= \sum \frac{(Observed  mass - Expected mass)^2}{Expected mass}.
    \label{E046}
\end{equation}
Utilizing information from Table \ref{Table:2}, we are utilizing this hypothesis to compare the observed mass and the expected mass for this model. Assume

$H_{0}: $  There is no discernible difference between the observed and model-generated masses.

$H_{1}: $  There is a substantial discrepancy between the observed and model-generated masses.

The degree of freedom for LMC X-4, SMC X-4, Cen X-3, and Vela X-1 is 29.
 For the compact object LMC X-4,  $ \chi^{2} _{cal} = 0.1717185$ and $\chi^{2} _{tab} = 42.557$ (we get this value from $\chi^{2}$ table for the degree of freedom 29). 
 For the compact object SMC X-4,  $ \chi^{2} _{cal} = 0.18394023$ and $\chi^{2} _{tab} = 42.557$ (we get this value from $\chi^{2}$ table for the degree of freedom 29). 
 For the compact object Vela X-1,  $ \chi^{2} _{cal} = 0.09133$ and $\chi^{2} _{tab} = 42.557$ (we get this value from $\chi^{2}$ table for the degree of freedom 29).
  For the compact object Cen X-3,  $ \chi^{2} _{cal} = 0.05043$ and $\chi^{2} _{tab} = 42.557$ (we get this value from $\chi^{2}$ table for the degree of freedom 29).

  For all four compact objects $\chi^{2} _{cal} < \chi^{2} _{tab}$, hence null hypothesis ($H_{0} $ ) is accepted. So, there is no discernible difference between the observed and model-generated masses.

\subsection{Compactness}

Compactness of a Stellar object is defined as $u(r)=m/r$. Compactness is a very crucial observation. The compactness of stellar objects is categorised as follows:\\
$(i)$ Regular Star :($u \sim 10^{-5}$),
$(ii)$ White Dwarf : ($u \sim 10^{-3}$),
$(iii)$ Neutron Star : ($0.1<u<0.25$),
$(iv)$ Ultra Compact Object : ($0.25<u<0.5$),
$(v)$ Black Hole : ($u=0.5$) \cite{bib22}.
The simplified expression of compactness for our model using Eq. (\ref{E042}) is
\begin{equation}
u=\frac{4}{3} \pi  \left(a \left(6-6 e^{-A r^2}\right)-b r^2\right)
\label{E043}
\end{equation}
For model verification the compactness should lie within Buchdahl's limit which is $u<4/9$ \cite{bib23}.

\begin{figure}[!ht]
    \centering

    \begin{minipage}{0.24\textwidth}
        \centering
        \includegraphics[width=\linewidth]{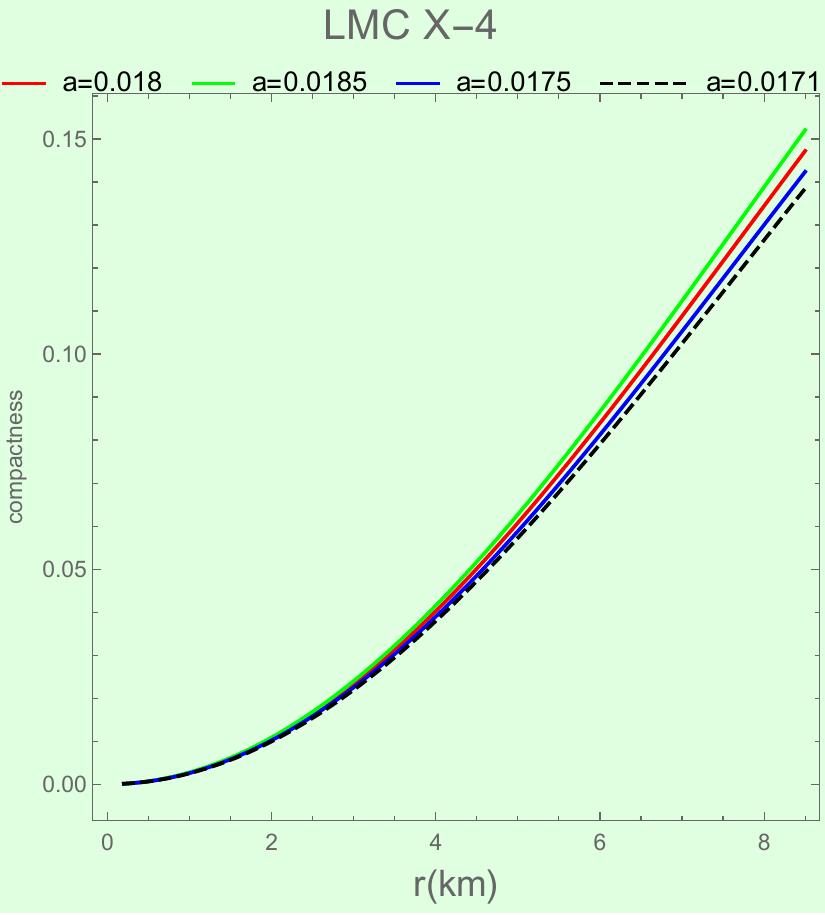}
        \small (a) LMC X-4
    \end{minipage}
    \hfill
    \begin{minipage}{0.24\textwidth}
        \centering
        \includegraphics[width=\linewidth]{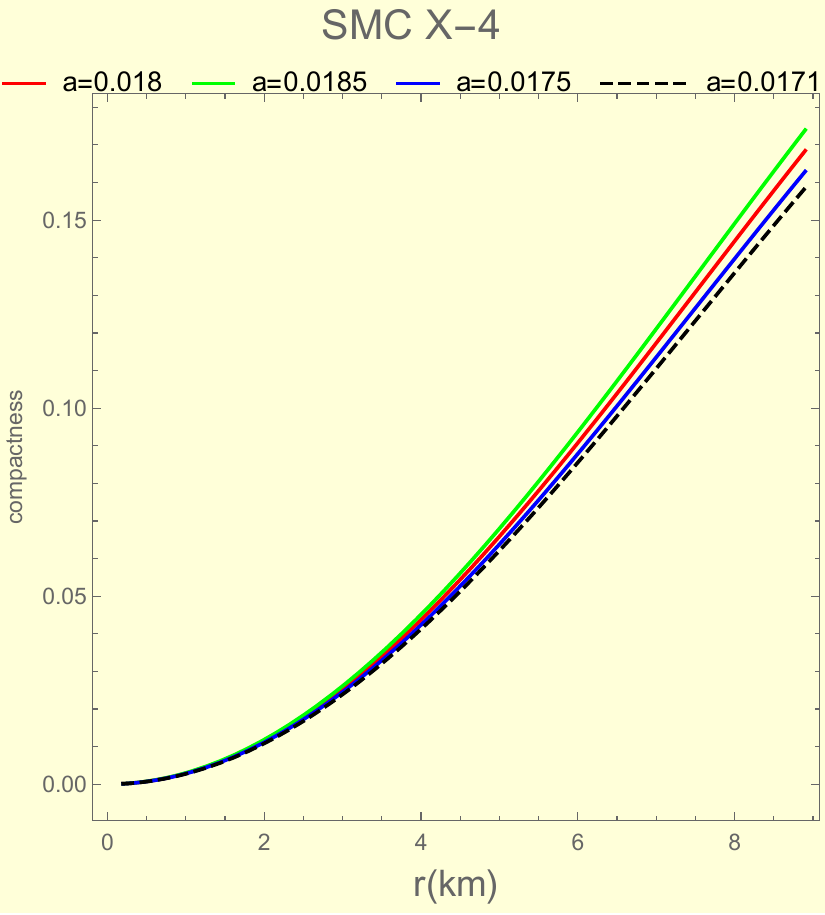}
        \small (b) SMC X-4
    \end{minipage}
    \hfill
    \begin{minipage}{0.24\textwidth}
        \centering
        \includegraphics[width=\linewidth]{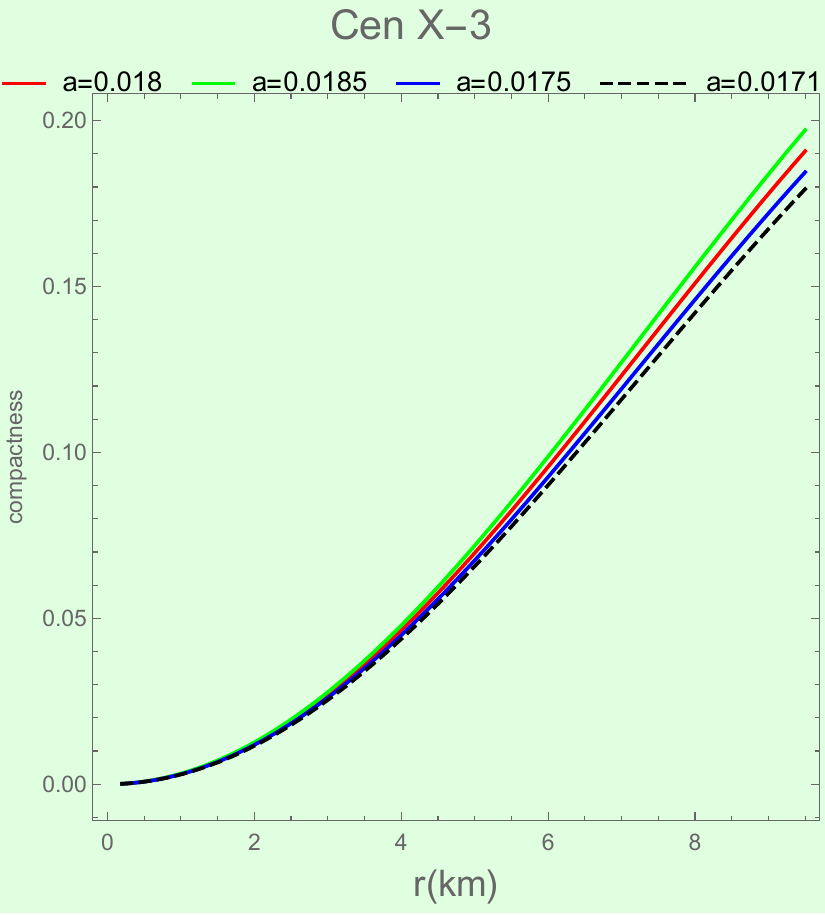}
        \small (c) Cen X-3
    \end{minipage}
    \hfill
    \begin{minipage}{0.24\textwidth}
        \centering
        \includegraphics[width=\linewidth]{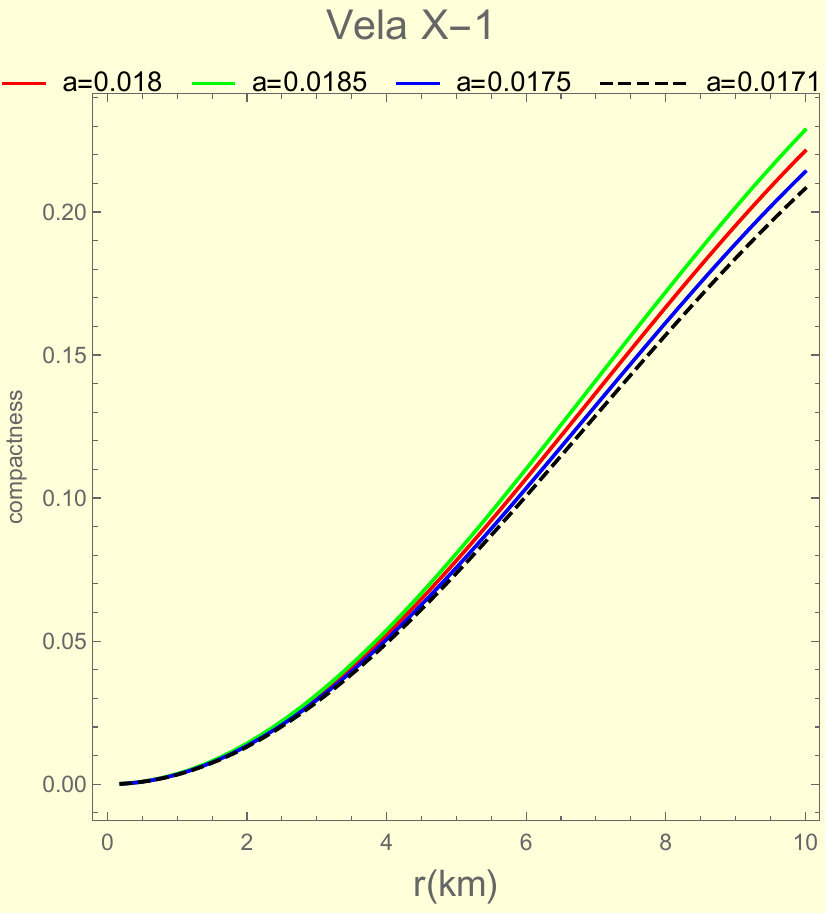}
        \small (d) Vela X-1
    \end{minipage}

    \caption{Evolution of the compactness with respect to the radial coordinate $r$ for (a) LMC X-4, (b) SMC X-4, (c) Cen X-3, and (d) Vela X-1.}
    
    \label{fig:compact}
\end{figure}

Graphical representation of compactness from Eq. (\ref{E043}) is depicted in Fig \ref{fig:compact}.

From Table \ref{Table:2}  we can observe that for different values of $a$ the compactness ($u$) of the pulsars LMC X-4, SMC X-4, Cen X-3, and Vela X-1 lies within the Buchdahl's limit and it also indicates that these are Neutron stars in $f(Q)$ modified gravity background as $0.1<u<0.25$. Fig \ref{fig:compact} shows the graphical representation of the compactness evolution of four compact objects with respect to $r$.

\subsection{Surface Redshift}

The surface redshift is evaluated by $z_s=(1-2u)^{-1/2} -1$. In physics and general relativity, gravitational redshift is the phenomenon of electromagnetic waves or photons leaving a gravitational well (also known as the Einstein shift in earlier literature) \cite{bib24}. A redshift, often referred to as a drop in wave frequency and an increase in wavelength, is the outcome of this energy loss. For compatibility the surface redshift lies within ($z_{s}$) $\leq$ 5.211 \cite{bib25}.
\begin{equation}
z=-1+\frac{1}{\sqrt{1-\frac{8}{3} \pi  \left(a \left(6-6 e^{-A r^2}\right)-b r^2\right)}}
\label{E045}
\end{equation}

\begin{figure}[!ht]
    \centering

    \begin{minipage}{0.24\textwidth}
        \centering
        \includegraphics[width=\linewidth]{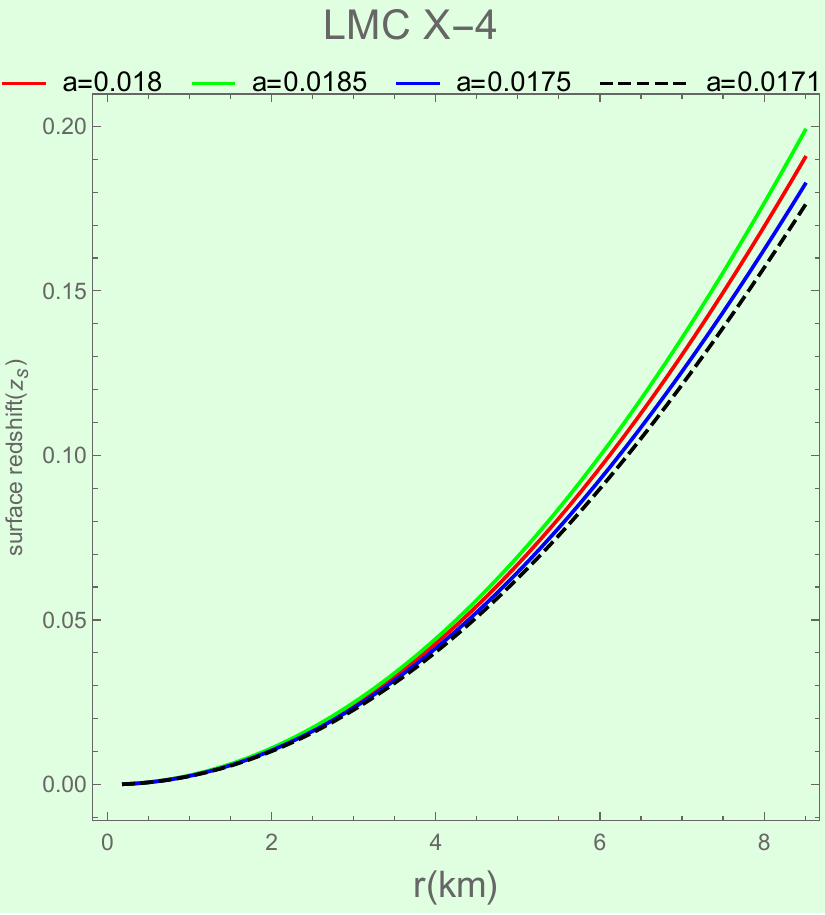}
        \small (a) LMC X-4
    \end{minipage}
    \hfill
    \begin{minipage}{0.24\textwidth}
        \centering
        \includegraphics[width=\linewidth]{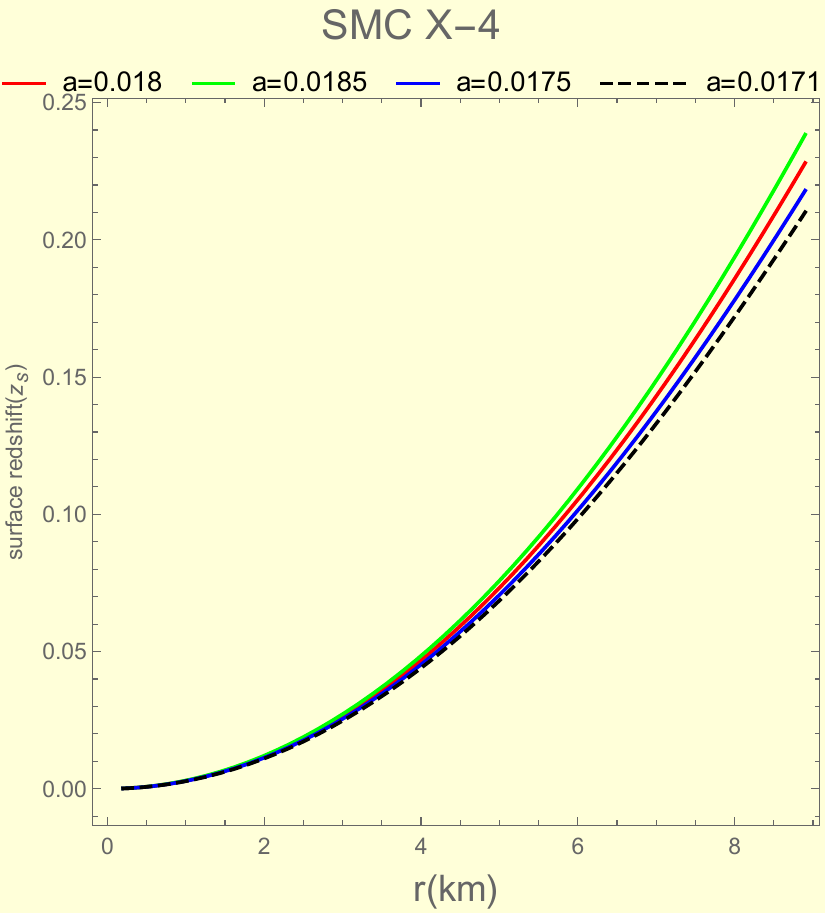}
        \small (b) SMC X-4
    \end{minipage}
    \hfill
    \begin{minipage}{0.24\textwidth}
        \centering
        \includegraphics[width=\linewidth]{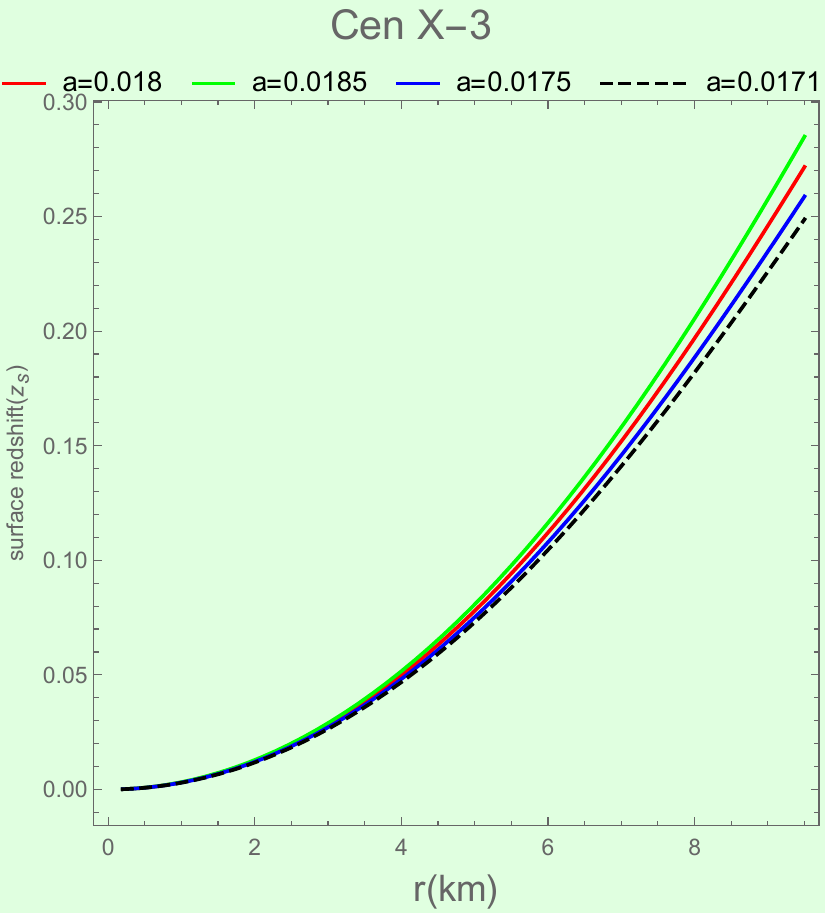}
        \small (c) Cen X-3
    \end{minipage}
    \hfill
    \begin{minipage}{0.24\textwidth}
        \centering
        \includegraphics[width=\linewidth]{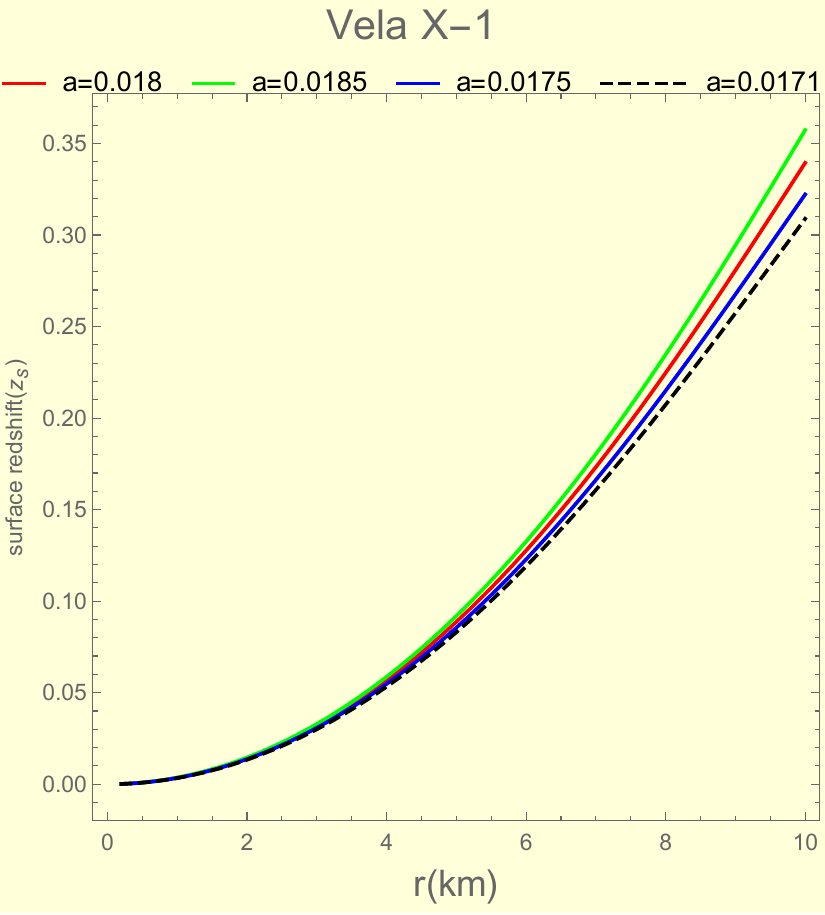}
        \small (d) Vela X-1
    \end{minipage}

    \caption{Evolution of the surface redshift with respect to the radial coordinate $r$ for (a) LMC X-4, (b) SMC X-4, (c) Cen X-3, and (d) Vela X-1.}
    
    \label{fig:redshift}
\end{figure}

Graphical representation of compactness from Eq. (\ref{E045}) is depicted in Fig \ref{fig:redshift}.

From Table \ref{Table:2}  we can observe that for different values of $a$ the surface redshift of LMC X-4, SMC X-4, Cen X-3, and Vela X-1 lies within the permissible limit in $f(Q)$ modified gravity background.
Fig \ref{fig:redshift} shows the graphical evolution of surface redshift for four compact objects with respect to $r$.

The true reason for the rising pattern of surface redshift within compact objects is that a photon loses a great deal of energy and disperses more widely while traveling from the star's center to its surface over a longer and denser region. Rather than altering speed, this energy must be lost by altering frequency. The frequency of a photon decreases as its energy decreases. The photon either travels toward the red end of the electromagnetic spectrum or its wavelength grows as a result. A photon that bursts from near the surface, on the other hand, must go through a less dense area and on a shorter journey, resulting in less dispersion and energy loss. When the radius rises slightly and the surface gravity increases with mass, the surface redshift is greatest near the surface and diminishes toward the core. The image also makes it simple to confirm that there isn't a singularity of any kind in any part of its arrangement.

\section{concluding remarks}
A few realistic compact objects are analyzed in this study and are evaluated in the presence of a gravitational interaction between two particles characterized by a nonmetricity $Q$. We have chosen anisotropic equation of motion in $f(Q)$ gravity framework and deduced $f(Q)$ as a linear function of nonmetricity $Q$. We have chosen to use the Krori-Barua metric in our study to assess the field equations. 
The vacuum Schwarzchild solution has been used to characterize the external spacetime and develop the boundary condition accordingly. 
After assessing the density, pressure in both radial and angular direction we have evaluated the restriction in the integrating constant $a$ and $b$ for the compact models. In Table \ref{Table:1} we have presented the derived condition for $a$ and $b$ for the compact objects LMC X-4, SMC X-4, Cen X-3 and Vela X-1. Density, radial pressure and transversal pressure have been graphically depicted in Fig \ref{fig:density}, Fig \ref{fig:radial}, and Fig \ref{fig:tangential} by choosing $a$=0.018,0.0185,0.0175 and 0.0171. These figures show that our model does not have any singularity and has the maximum density around the centre the star.
We obtained the anisotropic factor for the four compact objects. There can be two cases:  $p_{t} > p_{r}$ and  $p_{t} < p_{r}$.
In the first case the system encounters a repulsive force that diminishes the gravitational gradient. In contrast, in the second case, the force transmitted by anisotropy adds to the gravitational force squeezing the star. If the pressure from the nuclear force is strong enough to overcome gravity, the structure will eventually continue to collapse until it reaches its Schwarzschild radius. The first case is accepted for our model since the Fig. \ref{fig:ani} in our study demonstrates that the anisotropic component is positive and increases monotonically. Therefore, it is interpreted that the gravitational attraction can be resisted by the nuclear force.
Fig \ref{fig:drho}, Fig \ref{fig:dradial}, and Fig \ref{fig:dtangential} show the graphical evolution of the gradient of density, radial pressure and transversal pressure. All gradients have been observed to be of negative value.

The energy conditions have been verified for our model and graphically demonstrated in Fig \ref{fig:sec1}, Fig \ref{fig:sec2}, and Fig \ref{fig:sec3}. From the graphical representation we can say that the interior of the compact objects does not include any exotic matters. 
Two Eos parameters define the spectrum of realistic and normal distribution of matter by  $\omega_r$ and $\omega_t$. Fig \ref{fig:eosr}, and Fig \ref{fig:eost} represent the graphical demonstrations of EoS parameter and it decreases monotonically and vanishes at the surface of the star. The EoS parameters demonstrated in the above-mentioned figures are confined within $0< \omega_r , \omega_t < 1$, which imply the present fluid is non-exotic, similar to the work \cite{bib17}. It is necessary to verify the stability analysis and the causality criterion. We have utilized the cracking method \cite{bib18} to evaluate the stability of the model for the said compact objects.  Fig \ref{fig:sound} and Fig \ref{fig:gamma} demonstrate the stability analysis and causality criteria. According to Herrera \cite{bib18} and Abreu et al. \cite{bib19}, the potentially stable zones of an anisotropic star are those where the radial speed of sound and the transverse speed of sound, confirming $0<v^2 _{sr},v^2 _{st}< 1$, which we have verified in Fig \ref{fig:sound}. It is possible to use the adiabatic index $\Gamma$ to investigate the dynamical stability of the star structure under an infinitesimal radial adiabatic impact; it should be greater than 4/3 in the interior area \cite{bib21}. The graphical representation in Fig \ref{fig:gamma} verify the last mentioned condition.
Through the definition of three distinct force components, which are graphically depicted in  Fig \ref{fig:force}, we have verified the TOV equation for the equilibrium of the system. Anisotropic force has less of an impact, yet the three forces remain in balance as seen in the graph. As a result, we deduce that our model is in equilibrium under various force perturbations.

The mass-radius relation has been established and graphically depicted in Fig \ref{fig:mass}. We have determined the model mass and conducted the Chi-Square test to check whether there is any discernible difference between the observed and model-generated masses by setting thirty different values of $a$. We found that the null hypothesis has been accepted for LMC X-4, SMC X-4, Cen X-3, and Vela X-1 from Chi-Square test. Hence, there is no substantial discrepancy between the observed and model-generated masses. For the compact objects, we have also verified the compactness and the surface redshift. For neutron stars the compactness $u=m/r$ must be confined within (0.1,0.25) \cite{bib22} and surface redshift must be $(z_{s})$ $\leq$ 5.211 \cite{bib25}. Table \ref{Table:2} enables us to confirm that the pulsars LMC X-4, SMC X-4, Cen X-3, and Vela X-1 have compactness values less than 0.25 and that their surface redshifts are less than 5.211. The graphical representation of compactness and surface redshift have been demonstrated in Fig \ref{fig:compact} and Fig \ref{fig:redshift} as well.

While concluding, let us comment on our work with respect to the very relevant study of \cite{lin}, where the authors investigated the nonrotating neutron stars in $f(T)$ gravity. They \cite{lin}  utilized the SLy and BSk family of equations of state for perfect fluid to describe the neutron stellar matter and search for the effects of the $f (T )$ modification on the models of neutron stars. In our case, we looked into the density, radial pressure, and transversal pressure of a neutron star husing the Krori-Barua metric potential and obtained a boundary condition for the integrating constants $a$ and $b$ by applying the matching condition from Schwarzschild solution and the non-singularity condition of neutron star density. Some other noteworthy works in this context include \cite{Odi1,Odi2,Odi3,Odi4}. In \cite{Odi1}, the authors showed that a neutron star with an observed mass can be consistently explained with the mass-radius relation obtained by extended theories of Gravity. Furthermore, \cite{Odi1} demonstrated the equations of state for their consistency with LIGO observational constraints. In this context, we would like to mention that it is noteworthy to extend our study by checking for consistency of our results in other modified gravity frameworks with LIGO observational constraints.

\section{Acknowlegment}
The authors acknowlege the insiteful comments from the anonymous reviewers.
The Inter-University Centre for Astronomy and Astrophysics (IUCAA), Pune, India, graciously hosted the first author (SD) on her October–November 2023 scientific visit, and this is where part of the work was done. IUCAA's visiting associateship is acknowledged by the second author (SC).

\section{Data Availability Statement}
No Data are associated with the manuscript.

\end{document}